        \def\dotdeg{\hbox{$^\circ$\hskip-3pt .}}

        \documentclass[10pt]{article}
        \usepackage[dvips]{graphicx,epsfig}
        \usepackage[usenames,dvipsnames]{color}
        \usepackage{natbib}
        \usepackage{times,mathptm,mathptmx}
        \usepackage{bm}
        \usepackage{geometry}
        \usepackage{pdflscape}
\begin{document}

\vspace*{1.0in}

\begin{center}
			Do Tides Destabilize Trojan Exoplanets?

\vspace*{1.0in}

                        Anthony R. Dobrovolskis

                        SETI Institute

                        245-3 NASA Ames Research Center

                        Moffett Field, CA 94035-1000

                        Email: anthony.r.dobrovolskis@nasa.gov

\vspace*{0.5in}

			Jack J. Lissauer [0000-0001-6513-1659]

			Space Science \& Astrobiology Division

			245-3 NASA Ames Research Center

			Moffett Field, CA 94035-1000

\vspace*{1.0in}
                                2022 April 28
\end{center}

\vspace*{1.0in}

37 pages \\

9 figures (including 1 with color) \\

4 tables

\newpage

\vspace*{1.0in}

Proposed running head:

\begin{center}
			Do Tides Destabilize Trojan Exoplanets?
\end{center}

\vspace*{1.0in}

Correspondence:

\begin{center}
                        Anthony R. Dobrovolskis

                        245-3 NASA Ames Research Center

                        Moffett Field CA 94035-1000

                        anthony.r.dobrovolskis@nasa.gov

			Mobile:  (669) 290-3437

%                        'Phone: (650) 604-4194

%                        FAX: (650) 604-6779
\end{center}

\vspace*{0.3in}

\underline{Key Words}: \\

Celestial mechanics \\

Extra-solar planets \\

Planetary dynamics \\

Tides, solid body \\

Resonances, orbital

\newpage

\begin{center}
				HIGHLIGHTS
\end{center}

- Tides may prevent the survival of co-orbital exoplanets \\

- We generalize the theory of tides to include interactions among multiple perturbers \\

- We solve analytically for the motion of Trojan planets with tides \\

- We find that tidal dissipation pumps up small tadpole librations exponentially \\

- Our numerical simulations verify that tides remove Trojan planets 

% \vspace{0.5in}

\begin{center}
                                ABSTRACT
\end{center}

One outstanding problem in extrasolar planet studies is why no co-orbital exoplanets have been found, 
despite numerous searches among the many known planetary systems, many of them in other 
mean-motion resonances.  Here we examine the hypothesis that dissipation of energy 
by tides in Trojan planets is preventing their survival.  

The Appendix of this paper generalizes the conventional theory 
of tides to include tidal forces independent of dissipation, 
as well as the effects of one body on tides raised by another.  
The main text applies this theory to a model system 
consisting of a primary of stellar mass, a secondary 
of sub-stellar mass in a circular orbit about the primary, 
and a much lighter Trojan planet librating with small amplitude 
about an equilateral point of the system.  

Next, we linearize the equations of motion about the Trojan points, 
including the tidal forces, and solve for the motion of the Trojan.  
The results indicate that tides damp out the Trojan's motion perpendicular to the orbital plane 
of the primary and secondary, as well as its epicycles due to its eccentricity; 
but they pump up the amplitude of its tadpole librations exponentially.  
We then verify our analytic solutions by integrating 
the non-linearized equations of motion numerically for several sample cases.  
In each case, we find that the librations grow until the Trojan escapes its 
libration, which leads to a close encounter with either the primary or the secondary.

\section{Introduction}

Our Solar System contains numerous examples of ``co-orbitals'', 
that is, objects sharing nearly the same orbital period about the same primary.  
The best known are the several thousand Trojan asteroids librating in ``tadpole'' orbits 
about either of Jupiter's equilateral Lagrange points, like twin Sargassos of space; 
but Neptune also has numerous Trojan companions which are stable for Gyr timescales 
({\it e.g.}, Sheppard and Trujillo, 2006).  
Small numbers of Trojans, all with large-amplitude librations, have been found for the planets 
Uranus, Mars, Earth, and Venus as well; but the orbits of these bodies are relatively short-lived, 
and imply that their occupants were captured into libration relatively recently in Solar System history.  

In addition, each of Saturn's moons Tethys and Dione has one small satellite at its leading Lagrange point L4, 
and another at its trailing Lagrange point L5.  In the Circular Restricted Three-Body Problem (CR3BP), 
Trojans are stable for ratios $M_1/M_2$ of the primary mass to the secondary mass greater than 
$2/(1-\sqrt{23/27}) = (27+\sqrt{23\times 27})/2 \approx$ 25.9599 ({\it e.g.}, Dobrovolskis, 2013); 
for example, Pluto's moon Charon, with $M_1/M_2 \approx$ 8.2, cannot have Trojan companions.  

Besides Trojans in tadpole orbits, other types of co-orbital companions are possible as well.  
For example, Saturn's small moons Janus and Epimetheus are in a ``horseshoe'' resonance, 
such that Epimetheus librates about a wide arc enclosing both Janus' L4 and L5 points.  
Horse-shoe resonances are stable for mass ratios $M_1/M_2 >\sim$ 1200 (\'{C}uk {\it et al.}, 2012); 
for example, with $M_1/M_2 \approx$ 3499, Saturn could support long-term horseshoe companions, 
while with $M_1/M_2 \approx$ 1047, Jupiter is not expected to have any.  

``Quasi-satellites'' appear to orbit a moon or planet 
in the retrograde direction (contrary to its orbital motion) 
and outside its Hill sphere (sometimes called its sphere of influence); 
but they are really in eccentric prograde orbits about the primary.  
Several quasi-satellites of Venus, Earth, and Neptune are known.  In the CR3BP, 
quasi-satellites are stable for $M_1/M_2 >\sim$ 21.0 (H\'{e}non and Guyot, 1970).  

Furthermore, numerous asteroids are in hybrid co-orbital resonances, 
which alternate among tadpole, horseshoe, and quasi-satellite states.  
The Earth quasi-satellite 469219 Kamo'oalewa (a.k.a. 2016 HO3) is of this type, 
and is the target of a planned sample return mission 
by the China National Space Administration.  

Finally, ``counter-orbital'' resonances are possible, when two objects orbit with the same period, 
but in opposite directions (Dobrovolskis, 2012; Morais \& Namouni, 2013, 2016).  
To date only one counter-orbital companion of Jupiter has been found 
(Wiegert {\it et al.}, 2017): asteroid 2015 BZ 509, also known as (514107) Ka'epaoka'awela.  
Counter-orbitals seem to be stable for mass ratios $M_1/M_2 >\sim$ 7; 
so even Charon could support counter-orbitals (Dobrovolskis, 2012).  

In the past three decades, several thousands of extra-solar planets have been discovered and confirmed 
({\it e.g.}, Rowe {\it et al.}, 2014; Guerrero {\it et al.}, 2021), including hundreds of multi-planet systems, 
several of them with resonances among their orbital periods (Lissauer {\it et al.}, 2011).  
Yet despite numerous directed searches for satellites or Trojan companions of exoplanets 
({\it e.g.}, Caton {\it et al}., 1999; Davis {\it et al.}, 2001; Ford \& Gaudi, 2006; 
Ford \& Holman, 2007; Narita {\it et al.}, 2007; Madhusudhan \& Winn, 2009; Janson, 2013; 
Lillo-Box {\it et al.}, 2018a, 2018b; Kipping, 2020), not one exomoon or co-orbital exoplanet has yet been found. 

Hippke \& Angerhousen (2015) searched for statistical evidence for Trojan companions 
by stacking lightcurves of $\sim$4000 {\it Kepler} planet candidates.  
They did not find evidence for such Trojans in their bulk analysis of the entire population; 
% and obtained a $2\sigma$ upper bound of 6.6 $\times 10^5$ km$^2$ for the average cross-section.  
but they claimed a statistically significant detection of a Trojan population 
in the sub-sample of planet candidates with periods exceeding 60 days.  

It remains a puzzle that no co-orbital exoplanets have been found yet, 
so some physics may well be preventing their formation, survival, or detection.  
Although our own Solar System contains no Trojan planets, theories of planet formation 
often predict that extrasolar planets should be accompanied by Trojan companions 
({\it e.g.}, Laughlin \& Chambers, 2002; Nauenberg, 2002; Go\'zdziewski \& Konacki, 2006; 
Beaug\'e {\it et al.}, 2007; Giuppone {\it et al.}, 2012).  
Shared orbital periods can cause misinterpretation of astrometry and radial velocity signals, 
but usually should not confuse transit signatures and prevent detection (Dobrovolskis, 2013, 2015).  
Therefore we suspect that Trojan exoplanets may be unstable, 
on timescales shorter than the ages of most of the exoplanets 
identified by {\it Kepler}, {\it TESS}, and ground-based transiting planet searches.  

The most likely explanation for the lack of confirmed moons of extrasolar planets 
is that they raise tides in their parent planets which drive evolution of the moon's orbits, 
and ultimately destabilize most such exomoons large enough to be observable.  Similarly, 
it is possible that tides raised by the primary star in Trojan exoplanets may destabilize their orbits.  
See Ferraz-Mello (2022) for a recent review of tidal effects on exoplanets in the two-body case.  

In the two-body case, when the secondary already has been tidally despun to synchronous rotation, 
tides in the secondary always damp both its orbital eccentricity and semi-major axis 
({\it e.g.}, Murray \& Dermott, 1999).  In the classic CR3BP, 
the equilateral Lagrange points L4 and L5 are known to be maxima of the energy integral 
({\it e.g.}, Murray \& Dermott, 1999).  Thus it is plausible that energy dissipation 
may increase the amplitude of tadpole librations, and ultimately may destabilize Trojan orbits.  

Many papers have addressed the effects of various types of dissipation on Trojans in our Solar System, 
such as drag by nebular gas and dust (Jeffreys, 1929; Greenberg, 1978; Yoder, 1979; Peale, 1993; Murray, 1994; 
Leleu {\it et al.}, 2019), radiation pressure and its associated Poynting-Robinson drag (Colombo et al., 1966; 
Schuerman, 1980; Simmons et al., 1985), and even torques from planetary rings (Lissauer et al., 1985).  
These studies show that energy dissipation can either damp or pump orbital eccentricity and libration, 
depending on the particular functional form of the drag (Yoder et al, 1983; Murray, 1994).  

However, the Trojan problem with tidal dissipation in the third body has been treated very little.  
Caudal (2013) found that tides raised by Saturn on Janus and Epimetheus tend to de-stabilize 
their mutual horseshoe orbit; but that torques from Saturn's rings tend to stabilize it, 
so that they may be evolving toward mutual tadpole orbits.  In the same year, 
Rodr\'{i}guez {\it et al.} (2013) used Mignard's (1979) formulation of viscous-type tides 
numerically to integrate the planar motion of Trojan pairs with equal masses, 
and found them unstable to collision in every case.  

Recently, Couturier {\it et al.} (2021) have used a Hamiltonian formulation 
of the planar problem to reach similar conclusions.  
An even more recent preprint by the same authors (Couturier {\it et al.}, 2022) finds 
that yet another massive planet, exterior to the co-orbitals but in mean-motion resonance 
with both of them, may significantly delay the onset of this instability, but cannot prevent it.  
This is intriguing because planets on nearby orbits typically destabilize co-orbital companions, 
as is the case for Trojan asteroids of Saturn (de la Barre {\it et al.}, 1996) 
and for multiple co-orbital planets on closely-spaced nested orbits (Smith and Lissauer, 2010).  

This paper re-examines the hypothesis that tidal dissipation 
may destabilize Trojan exoplanets and prevent their long-term survival.  
The Appendix derives a theory of tidal forces generalized for systems with more than two bodies.  
Section 2 of the main text generalizes the CR3BP to include tidal forces on the third body, 
while Section 3 linearizes the equations of motion for small departures from L4 or L5.  
Section 4 analytically solves the linearized equation of vertical motion, 
while Section 5 solves the linearized equations of horizontal motion analytically.  
For comparison, Section 6 displays particular numerical solutions of the non-linearized equations; 
some readers may wish to skip directly to this section.  
Finally, Section 7 discusses our conclusions.

\newpage

\section{Model}

For this study, we envision a primary of stellar mass $M_1$ (a star or a compact object) 
orbited by a secondary of sub-stellar mass $M_2 < M_1$ (a giant planet or a brown dwarf), 
and also by a third body of small mass $m_o << M_2$ (a terrestrial planet or a super-Earth), 
so that it does not significantly affect the motions of $M_1$ or $M_2$.  
All three objects are assumed to be essentially spherical, and the secondary $M_2$ is assumed to be 
in a fixed orbit of semi-major axis $\sigma$ and negligible eccentricity about the primary $M_1$; 
while the tertiary body $m_o$ is assumed already to be despun to synchronous rotation by tides, 
and librating about the secondary's Trojan point L4 or L5.  
Then the question is whether this libration is damped or pumped by the action of tides in the tertiary.  
Note that we neglect tides in the primary and secondary, because tidal dissipation in stars 
and giant planets is normally weak compared to that in solid/liquid bodies such as terrestrial planets 
(but see Lainey {\it et al}, 2020).  

We apply the machinery of the Circular Restricted Three-Body Problem (CR3BP) to the above model; 
symbols are defined in Tables 1 and 2.  As usual, we define a synodic Cartesian coordinate system 
$X$, $Y$, $Z$ (with corresponding unit vectors $\bf I$, $\bf J$, $\bf K$) 
where both $M_1$ and $M_2$ are fixed, with its origin at their mutual center of mass.  Let this system 
rotate counter-clockwise about its $+Z$ axis with angular velocity equal to that of $M_2$ about $M_1$, 
of magnitude $n = \sqrt{G[M_1+M_2]/\sigma^3}$, where $G$ is Newton's constant of universal gravitation.  
For convenience, we choose units of time, length, and mass such that $n$, $\sigma$, and $[M_1+M_2]$ 
all equal one; then $G$ also becomes unity.  

Furthermore, let the $X$ axis of these coordinates pass through both $M_1$ and $M_2$, 
so that the primary is fixed at point $(X,Y,Z) = (-M_2,0,0)$, 
while the secondary resides at point $(M_1,0,0)$.  
Finally, the $Y$ axis completes a right-handed triad.  
Then the leading equilateral point L4 lies at $(1/2-M_2,+\sqrt{3/4},0)$ 
while the trailing equilateral point L5 lies at $(1/2-M_2,-\sqrt{3/4},0)$.  

In this synodic frame of reference, the equations of motion for a particle of negligible mass $m_o$ become 
\begin{equation}
                        \ddot{X} =  2\dot{Y} +X +f_X +\partial U/\partial X ,
\end{equation}
\begin{equation}
                        \ddot{Y} = -2\dot{X} +Y +f_Y +\partial U/\partial Y ,
\end{equation}
\begin{equation}
                        {\rm and} \; \; \; \ddot{Z} = f_Z +\partial U/\partial Z .
\end{equation}
Here $f_X$, $f_Y$, and $f_Z$ are the components of $\bf f$, the acceleration of $m_o$ 
due to tides, in the $+X$, $+Y$, and $+Z$ directions, respectively; 
formulae for the corresponding tidal forces ${\bf F} = m_o {\bf f}$ are developed in the Appendix.  

In System (1) through (3) above, $U$ is just minus the usual gravitational potential:  
\begin{equation}
                                U = M_1/r_1 +M_2/r_2 . 
\end{equation}
Here $r_1$ and $r_2$ are the distances of $m_o$ from the centers of $M_1$ and $M_2$ respectively, given by 
\begin{equation}
                        r_1 = \sqrt{ (X +M_2)^2 +Y^2 +Z^2 }
                        \; \; \; {\rm and} \; \; \;
                        r_2 = \sqrt{ (X -M_1)^2 +Y^2 +Z^2 } ; 
\end{equation}
note that $U$ is always positive.  Then differentiating Formula (4) gives 
\begin{equation}
                        \partial U/\partial X = -M_1 [X +M_2] r_1^{-3} -M_2 [X -M_1] r_2^{-3} ,
\end{equation}
\begin{equation}
                        \partial U/\partial Y = -M_1 Y r_1^{-3} -M_2 Y r_2^{-3} ,
\end{equation}
\begin{equation}
                        \partial U/\partial Z = -M_1 Z r_1^{-3} -M_2 Z r_2^{-3} .
\end{equation}

\newpage

\begin{center}
					Table 1.  Roman Symbols.  
\vspace*{0.1in}

\begin{tabular}{|c|c|}

%\hline
%		Symbol		&			Meaning					\\
% \hline
% \hline
%		$\bf A$		&		tidal acceleration of $m_2$			\\
% \hline
%		$a_0$		&		orbital semi-major axis of $m_0$		\\
%		$a_2$		&		orbital semi-major axis of $m_2$		\\
%		$a_3$		&		orbital semi-major axis of $m_3$		\\
% \hline
%		$A$, $B$	&	complex amplitudes of horizontal motion			\\
%		$A$, $B$	&	Fourier coefficients of forced vertical motion		\\
% \hline
% 		$b$		&			radius of $m_0$				\\
% \hline

\hline
		$A_N$		&		real part of $\alpha_N$				\\
		$a_n$		&		imaginary part of $\alpha_N$			\\
		$B_N$		&		real part of $\beta_N$				\\
		$b_n$		&		imaginary part of $\beta_N$			\\
\hline
		$C_J$		&			Jacobi constant				\\
%		$C'_J$		&		modified Jacobi constant			\\
		$C$		&	cosine coefficient of free vertical motion		\\
		$c$		&	cosine coefficient of forced vertical motion		\\
		$D$, $E$	&	complex amplitudes of free horizontal motion		\\
\hline
		$e$		&			orbital eccentricity			\\
\hline
		$\bf F$		&	= $m_o {\bf f}$ = tidal force on $m_o$			\\
		$\bf f$		&		tidal acceleration of $m_o$			\\
\hline
		$G$		&	Newton's constant of universal gravitation		\\
		$g$		&		surface gravity on $m_o$			\\
\hline
		$H$		&		complex amplitude of free vertical motion	\\
		$h$		&		second-degree height Love number		\\
                $i$             &       orbital inclination from the $XY$ plane         	\\
		$j$		&		imaginary unit; $j^2 = -1$			\\
                $k$             &               second-degree potential Love number     	\\
\hline
      $\bf I$, $\bf J$, $\bf K$ &	unit vectors in directions of increasing $X$, $Y$, $Z$  \\
\hline
		L4		&		leading equilateral Lagrange point		\\
		L$'4$		&		tidally shifted location of L4			\\
		L5		&		trailing equilateral Lagrange point		\\
		L$'5$		&		tidally shifted location of L5			\\
		$\ell_N$	&		angular frequencies of horizontal motion	\\
\hline
%		$M$		&		mass of perturbed body				\\
		$M$		&		mass of tide-raising body			\\
		$M_1$		&			mass of primary				\\
		$M_2$		&		mass of secondary ($< M_1$)			\\
		$m$		&		mass of perturbed body				\\
                $m_o$           &               mass of tertiary ($<< M_2$)                     \\

\hline
		$N$		&			integer index				\\
		$n$		&	mean motion = $\sqrt{G[M_1+M_2]/a_2^3}$			\\
\hline
		$P$		&		period of tides					\\
		$Q$		&		tidal quality factor				\\
		$q_N$		&		complex quotient $\beta_N/\alpha_N$		\\
% hline
%		$\bf R$		&		radial force due to tides			\\
\hline
		$R$		&		radius of $m_o$					\\
		$R_1$		&		radius of $M_1$					\\
		$\bf r$		&	current location of tide-raising body relative to $m_o$	\\
		$\bf r'$	&   time-lagged location of tide-raising body relative to $m_o$	\\
                $r_o$           &               distance from center of $m_o$           	\\
		$r_1$		&		distance between $m_o$ and $M_1$		\\
		$r_2$		&		distance between $m_o$ and $M_2$		\\
\hline
		$S$		&		sine coefficient of free vertical motion	\\
		$s$		&		sine coefficient of forced vertical motion	\\
\hline
%		$\bf s$		&		apparent transverse velocity			\\
%		$\bf T$		&		transverse force due to tides			\\
%		$\bf t$		&		transverse velocity vector			\\
% \hline
		$t$		&			time					\\
%		$t_z$		&		decay time of vertical motion			\\
\hline
		$U$		&		(minus) gravitational potential			\\
		$V_1$		&		tidal potential from $M_1$			\\
		$V_2$		&		tidal potential from $M_2$			\\
		$V'$		&		lagged tidal potential				\\
		$W_1$		&		quadrupole potential raised by $M_1$		\\
		$W_2$		&		quadrupole potential raised by $M_2$		\\
\hline
		$X,Y,Z$		&		global Cartesian coordinates			\\
		$x,y,z$		&		local Cartesian coordinates			\\
		$x_0,y_0,z_0$	&	initial values of $x-x'$, $y-y'$, and $z$, respectively	\\
		$x'$, $y'$	&	horizontal offsets of equilibria from L4 or L5		\\
\hline

\end{tabular}

\newpage
                                        Table 2.  Greek Symbols.
\vspace*{0.1in}

\begin{tabular}{|c|c|}

\hline
		$\alpha_N$	&		complex coefficients in Formula (75)		\\
		$\beta_N$	&		complex coefficients in Formula (76)		\\
\hline
		$\Gamma$	&			defined by Formula (74)			\\
		$\gamma_N$	&		complex roots of biquadratic Eq. (55)		\\
		$\Delta$	&			defined by Formula (73)			\\
		$\delta_N$	&			$\lambda_N -\gamma_N$			\\
\hline
		$\epsilon$	&  angle between $\bf K$ and $\boldsymbol \omega$ in the sidereal frame	\\
\hline
		$\zeta$		&		e-folding rate of free vertical motion			\\
		$\eta$		&		angular frequency of free vertical motion		\\
\hline
		$\theta_1$	&		angular distance from sub-$M_1$ point on $m_o$		\\
		$\theta_2$	&		angular distance from sub-$M_2$ point on $m_o$		\\
		$\Theta$	&		angular distance between $M_1$ and $M_2$ at $m_o$	\\
\hline
		$\kappa$	&		tidal constant defined by Formula (26)		\\
		$\kappa_c$	&		critical value of $\kappa$			\\
\hline
		$\Lambda_N$	&		e-folding rates of horizontal motion			\\
		$\lambda_N$	&		complex rate constants of horizontal motion		\\
%		$\mu$		&		mass of tertiary ($<< M_2$)				\\
		$\nu$		&		complex rate constant of free vertical motion		\\
\hline
		$\rho$		&			mean density of $m_o$				\\
\hline
%		$\sigma_0$	&		orbital semi-major axis of $m_0$ about $m_1$		\\
		$\sigma$	&		orbital semi-major axis of $M_2$ about $M_1$		\\
\hline
		$\tau$		&				tidal time lag				\\
	$\boldsymbol \omega$	&			spin angular velocity of $m_o$			\\
	$\boldsymbol \varpi$	&  apparent transverse velocity ${\bf r} \times {\boldsymbol \omega}$	\\
\hline

\end{tabular}

\end{center}

When dissipative tides are included, 
System (1) through (3) does not conserve energy, 
and does not possess any isolating integrals.  However, 
if $f$ is neglected, System (1) through (3) is conservative, 
and possesses a single isolating integral 
\begin{equation}
                C_J = 2U +X^2 +Y^2 -\dot{X}^2 -\dot{Y}^2 -\dot{Z}^2 +M_1 M_2 
\end{equation}
called the Jacobi constant (also known as the Tisserand parameter), 
equal to --2 times the total energy per unit mass of $\mu$ 
(gravitational potential +centrifugal potential +specific kinetic energy).  
% \footnote{In the intermediate case $\kappa \ne 0 = \tau$, 
% when tides are present but without dissipation, 
% the tidal terms in System (1) through (3) 
% can be integrated to give a modified Jacobi constant 
% $C'_J = C_J -\kappa[m_1 -m_2][x-x'] \pm\kappa\sqrt{3}[11m_1 m_2/4 -1][y-y']$ 
% to second degree in small quantities $x$, $y$, $z$, and $\kappa$. } 
The arbitrary constant of integration has been set to $M_1 M_2$ by convention,
so that $C_J$ = 3 at L4 and L5 when $\dot{X} = \dot{Y} = \dot{Z} = 0$.

\section{Linearization}

In order to study the stability of small tadpole orbits, 
we linearize the equations of motion about the Trojan points.  
First, define local displacements $(x,y,z)$ about these points, 
parallel to the global coordinates $(X,Y,Z)$; but please note that 
this convention is {\it opposite} to that of Murray and Dermott (1999, p. 85).  
Then near the leading Trojan point L4, 
\begin{equation}
                X = x +1/2 -M_2 = x +M_1 -1/2 , \; \; Y = y +\sqrt{3/4} , \; \; {\rm and} \; \; Z = z ; 
\end{equation}
while near the trailing Trojan point L5, 
\begin{equation}
                X = x +1/2 -M_2 = x +M_1 -1/2, \; \; Y = y -\sqrt{3/4} , \; \; {\rm and} \; \; Z = z .  
\end{equation}
Note the sign difference in the $Y$ coordinate;  note also that 
$\dot{X} = \dot{x}$, $\ddot{X} = \ddot{x}$, {\it etc.}, and similarly for $Y$ and $Z$.  

In these local coordinates, System (1) through (3) becomes 
\begin{equation}
                        \ddot{x} =  2\dot{y} +x +1/2 -M_2 +f_X +\partial U/\partial x ,
\end{equation}
\begin{equation}
                        \ddot{y} = -2\dot{x} +y \pm\sqrt{3/4} +f_Y +\partial U/\partial y ,
\end{equation}
\begin{equation}
                        {\rm and} \; \; \; \ddot{z} = f_Z +\partial U/\partial z .
\end{equation}
In Eq. (13), and until further notice, the upper sign applies at the leading Trojan point L4, 
while the lower sign applies at the trailing Trojan point L5.

\subsection{Potential}

                        In System (12) through (14) above, 
\begin{equation}
                        \partial U/\partial x = \partial U/\partial X
                        = -M_1 [X +M_2] r_1^{-3} -M_2 [X -M_1] r_2^{-3}
                        = -M_1 [x +1/2] r_1^{-3} -M_2 [x -1/2] r_2^{-3} ,
\end{equation}
\begin{equation}
                        \partial U/\partial y = \partial U/\partial Y
                        = -M_1 Y r_1^{-3} -M_2 Y r_2^{-3}
                        = -M_1 [y \pm\sqrt{3/4}] r_1^{-3} -M_2 [y \pm\sqrt{3/4}] r_2^{-3} ,
\end{equation}
\begin{equation}
                        {\rm and} \; \; \; \partial U/\partial z = \partial U/\partial Z
                        = -M_1 Z r_1^{-3} -M_2 Z r_2^{-3} = -M_1 z r_1^{-3} -M_2 z r_2^{-3} .
\end{equation}

		Linearizing Formulae (15) through (17) above is relatively easy.  From Eqs. (5), 
\begin{equation}
r_1 = \sqrt{(x+1/2)^2 +(y \pm\sqrt{3/4})^2 +z^2} \approx \sqrt{1 +x \pm y\sqrt{3}} \approx 1 +x/2 \pm y\sqrt{3/4} , 
\end{equation}
			while 
\begin{equation}
r_2 = \sqrt{(x-1/2)^2 +(y \pm\sqrt{3/4})^2 +z^2} \approx \sqrt{1 -x \pm y\sqrt{3}} \approx 1 -x/2 \pm y\sqrt{3/4} , 
\end{equation}
			to first degree in the local coordinates.  

			Substituting Formulae (18) and (19) above for $r_1$ and $r_2$ into Eqs. (15) through (17) gives 
\begin{equation}
                        \partial U/\partial x = -M_1 [x +1/2] r_1^{-3} -M_2 [x -1/2] r_2^{-3} 
\end{equation} \[
			\approx -M_1 [x +1/2] [1 -3x/2 \mp 3y\sqrt{3/4}] -M_2 [x -1/2] [1 +3x/2 \mp 3y\sqrt{3/4}] 
\] \[
			\approx -M_1 [1/2 +x/4 \mp 3\sqrt{3} y/4] -M_2 [-1/2 +x/4 \pm 3\sqrt{3} y/4] 
\] \[
				= -(M_1 -M_2) [1/2 \mp 3\sqrt{3} y/4] -(M_1 +M_2) x/4 
\] \[
				= -(1 -2M_2) [1/2 \mp 3\sqrt{3} y/4] -x/4 , 
\]
\begin{equation}
                        \partial U/\partial y = -M_1 [y \pm\sqrt{3/4}] r_1^{-3} -M_2 [y \pm\sqrt{3/4}] r_2^{-3} 
\end{equation} \[
	\approx -M_1 [y \pm\sqrt{3/4}] [[1 -3x/2 \mp 3y\sqrt{3/4}] -M_2 [y \pm\sqrt{3/4}] [1 +3x/2 \mp 3y\sqrt{3/4}] 
\] \[
		\approx -M_1 [\pm\sqrt{3/4} \mp3\sqrt{3}x/4 -5y/4] -M_2 [\pm\sqrt{3/4} \pm3\sqrt{3}x/4 -5y/4] , 
\] \[
				= -(M_1 +M_2) [\pm\sqrt{3/4} -5y/4] \pm 3\sqrt{3} (M_1 -M_2) x/4 
\] \[
				= \mp\sqrt{3/4} +5y/4 \pm 3\sqrt{3} (1 -2M_2) x/4 , 
\]
\begin{equation}
                        {\rm and} \; \; \; \partial U/\partial z = -M_1 z r_1^{-3} -M_2 z r_2^{-3} 
			\approx -M_1 z -M_2 z = -(M_1 +M_2)z = -z , 
\end{equation}
			again to first degree in the local coordinates.  

		Then substituting Formulae (20) through (22) above into System (12) through (14) and simplifying gives 
\begin{equation}
                        \ddot{x} = +2\dot{y} +3x/4 \pm 3\sqrt{3} (1 -2M_2) y/4 +f_X , 
\end{equation}
\begin{equation}
                        \ddot{y} = -2\dot{x} +9y/4 \pm 3\sqrt{3} (1 -2M_2)x/4 +f_Y , 
\end{equation}
\begin{equation}
	                        {\rm and} \; \; \; \ddot{z} = -z +f_Z .  
\end{equation}
Note that Eqs. (23) and (24) reduce to Eqs. (3.101) and (3.140) of Murray \& Dermott (1999) 
when the tidal terms are neglected.  

% COMPARE WITH MURRAY AND DERMOTT - AGREES - YAY ! 

% ABOVE ALL CHECKS OUT 

% APPENDIX CHECKS TOO ! 

\subsection{Tidal terms}

The Appendix gives vectorial formulae for the tidal forces $\bf F$ on $m_o$, in dimensional units.  
Now we must express these as dimensionless accelerations $\bf f$, in Cartesian coordinates.  
To convert the tidal force $\bf F$ into the corresponding acceleration $\bf f$, 
divide the force by the mass $m_o$ of the tertiary body.  
This presents no difficulties even if $m_o$ approaches zero, because 
each of the four force terms ${\bf F}_{11}$, ${\bf F}_{12}$, ${\bf F}_{21}$, and ${\bf F}_{22}$ 
contains a factor of $kR^5$, 
% the fifth power of the radius of the tertiary body; 
while its mass $m_o = 4\pi \rho R^3/3$ scales only as the cube of its radius 
(for a given mean density $\rho$).  Then the net acceleration $\bf f$ scales as $kR^2$.  
% the square of the radius of the tertiary body.  

In order to render $\bf f$ dimensionless in the context of the Three-Body Problem, 
we also must divide it by the constant $\sigma n^2 = G[M_1+M_2]/\sigma^2$, with dimensions of acceleration.  
                        For convenience, we define the constant coefficient 
\begin{equation}
                        \kappa \equiv \frac{3kGR^5}{m_o \sigma n^2}
                        = \frac{3kGR^5}{4\pi \rho R^3 \sigma n^2/3}
                        = \frac{9kGR^2}{4\pi \rho \sigma n^2}
                        = \frac{9kGR^2 \sigma^2}{4\pi \rho G[M_1+M_2]}
                        = \frac{9kR^2 \sigma^2}{4\pi \rho [M_1+M_2]} ,
\end{equation}
with dimensions of Length$^7$/Mass$^2$. 

If $m_o$ has a similar density to the primary $M_1$, then 
$\kappa$ is of order $R^2 R_1^3/\sigma^5$ or less in dimensionless units, 
where $R_1$ is the radius of the primary.  
Thus for most astrophysical situations, $\kappa$ is much less than unity.  
For the Earth-Moon system, for example, $R \approx$ 1738 km, $R_1 \approx$ 6378 km, 
and $\sigma \approx$ 384 000 km, so $\kappa$ is on the order of $10^{-10}$ or less; 
while for the Sun-Earth system, $R \approx$ 6378 km, $R_1 \approx$ 696 000 km, 
and $\sigma \approx 150 \times 10^6$ km, so $\kappa < \sim 2 \times 10^{-16}$ !  

The formulae for $\bf F$ in the Appendix already have been linearized in the tidal time lag $\tau$, 
but now they must be linearized in the local coordinates $x,y,z$ as well.  
Because $M_1$ and $M_2$ are fixed in the synodic frame, only the coordinates of $m_o$ vary.  
Then the expressions of the Appendix can be written as in Table 3, 
to first degree in $x$, $y$, and $z$.  
Note from this table that $\dot{\bf r}_1$ and $\dot{\bf r}_2$ are equal; 
furthermore, ${\bf r}_1 \bullet \dot{\bf r}_2 \approx \dot{r}_1$ 
and          ${\bf r}_2 \bullet \dot{\bf r}_1 \approx \dot{r}_2$.  
These result in some welcome simplifications to the formulae for $\bf F$ and $\bf f$.  
An even greater simplification is also possible, as follows.  

We define a planet's obliquity $\epsilon$ as the angle between its orbit normal $\bf K$ 
and its rotational angular velocity $\boldsymbol \omega$, in the sidereal frame.  
Most planets and asteroids in our Solar System start with a wide range of obliquities, 
and rotation periods of about half a day to one day.  However, planets massive enough 
and close enough to their parent star to be affected significantly by solar tides 
despin to low obliquities and slow rotations within the first $10^9$ years 
({\it e.g.}, Dobrovolskis, 2007).  

In our own Solar System, Mercury has been captured into a spin-orbit resonance 
such that it rotates three times during every two orbits, because of its permanent 
quadrupole moment and its relatively high orbital eccentricity $e \approx$ 0.206 
({\it e.g.}, Noyelles {\it et al.}, 2014); 
while Venus presumably has been caught in a balance between gravitational tides in its interior 
and thermal tides in its massive atmosphere ({\it e.g.}, Ingersoll and Dobrovolskis, 1978).  

Most regular satellites of the Sun's planets are locked into synchronous rotation, 
so that their spin periods exactly match their orbit periods.  
Furthermore, their obliquities with respect to their orbits are small; 
thus they always keep nearly the same hemisphere facing toward their parent planets, 
as our Moon always presents roughly the same face to the Earth.  
This appears to be the most likely outcome for despun solid exoplanets 
with low orbital eccentricities, as well.  
Fluid exoplanets may despin to a ``pseudo-synchronous'' state 
with $\omega/n \approx (1+5e^2)/(1-e^2)$ (Dobrovolskis, 2007; see also Hut, 1981).  

Henceforth we assume synchronous rotation of $m_o$, with low obliquity $\epsilon$.  
Then in the synodic frame where $M_1$ and $M_2$ both are fixed, 
the $Z$ component of $\boldsymbol \omega$ vanishes to first degree in $\epsilon$, leaving 
\begin{equation}
                {\boldsymbol \omega} = -{\bf I}\epsilon\sin(t) -{\bf J}\epsilon\cos(t) .
\end{equation}
Without loss of generality, here we have chosen the origin of time 
when $\boldsymbol \omega$ points in the $-Y$ direction; 
if $m_o$ were in a Keplerian orbit lying in the $XY$ plane, 
this would correspond to its northern vernal equinox.  

\newpage

\begin{center}
			Table 3.  Linearized expressions.  
\vspace*{0.1in}

% \begin{tabular}{|c|c||c|c|}

% \hline
%                Expression                      &               Linearization                   &	Expression	&	Linearization			\\
% \hline
% \hline
%                ${\bf r}_1$                     &       $(-x -1/2, -y \mp\sqrt{3/4}, -z)$       &	$r_1$		&	$1 +x/2 \pm y\sqrt{3/4}$	\\
%                ${\bf r}_2$                     &       $(-x +1/2, -y \mp\sqrt{3/4}, -z)$       &	$r_2$		&	$1 -x/2 \pm y\sqrt{3/4}$	\\
% \hline
%                $\dot{\bf r}_1$                 &       $(-\dot{x}, -\dot{y}, -\dot{z})$        &	$\dot{r}_1$	&   $+\dot{x}/2 \pm\dot{y}\sqrt{3/4}$	\\
%                $\dot{\bf r}_2$                 &       $(-\dot{x}, -\dot{y}, -\dot{z})$        &	$\dot{r}_2$	&   $-\dot{x}/2 \pm\dot{y}\sqrt{3/4}$	\\
% \hline
%        ${\bf r}_1 \bullet {\bf r}_2$           &               $1/2 \pm y\sqrt{3}$             &	$r_1 r_2$	&	$1 \pm y\sqrt{3}$		\\
%        ${\bf r}_1 \bullet \dot{\bf r}_2$       &       $+\dot{x}/2 \pm\dot{y}\sqrt{3/4}$       &	$r_1^2$		&	$1 +x \pm y\sqrt{3}$		\\
%        ${\bf r}_2 \bullet \dot{\bf r}_1$       &       $-\dot{x}/2 \pm\dot{y}\sqrt{3/4}$       &	$r_2^2$		&	$1 -x \pm y\sqrt{3}$		\\
% \hline

% \end{tabular}

\begin{tabular}{|c|c|}

\hline
			Expression			&			Linearization			\\
\hline
\hline
			${\bf r}_1$			& $(-x -1/2){\bf I} +(-y \mp\sqrt{3/4}){\bf J} -z{\bf K}$ \\
			${\bf r}_2$			& $(-x +1/2){\bf I} +(-y \mp\sqrt{3/4}){\bf J} -z{\bf K}$ \\
\hline
                	$r_1^2$                         &               $1 +x \pm y\sqrt{3}$            	\\
                	$r_2^2$                         &               $1 -x \pm y\sqrt{3}$            	\\
\hline
                        $r_1$                           &               $1 +x/2 \pm y\sqrt{3/4}$                \\
                        $r_2$                           &               $1 -x/2 \pm y\sqrt{3/4}$                \\
			$r_1 r_2$                       &               $1 \pm y\sqrt{3}$			\\
\hline
                        $\dot{r}_1$                     &               $+\dot{x}/2 \pm\dot{y}\sqrt{3/4}$       \\
                        $\dot{r}_2$                     &               $-\dot{x}/2 \pm\dot{y}\sqrt{3/4}$       \\
\hline
			$\dot{\bf r}_1$			&   $-\dot{x}{\bf I} -\dot{y}{\bf J} -\dot{z}{\bf K}$	\\
			$\dot{\bf r}_2$			&   $-\dot{x}{\bf I} -\dot{y}{\bf J} -\dot{z}{\bf K}$	\\
\hline
		${\bf r}_1 \bullet {\bf r}_2$		&		$1/2 \pm y\sqrt{3}$			\\
		${\bf r}_1 \bullet \dot{\bf r}_2$	&		$+\dot{x}/2 \pm\dot{y}\sqrt{3/4}$	\\
		${\bf r}_2 \bullet \dot{\bf r}_1$	&		$-\dot{x}/2 \pm\dot{y}\sqrt{3/4}$	\\
\hline
							& $(z\omega_y -y\omega_z \mp\sqrt{3/4}\omega_z){\bf I}$	\\
${\boldsymbol \varpi}_1 = {\bf r}_1 \times \boldsymbol \omega$	& $-(z\omega_x -x\omega_z -\omega_z/2){\bf J}$	\\
					& $+(y\omega_x \pm\sqrt{3/4}\omega_x -x\omega_y -\omega_y/2){\bf K}$	\\
\hline
							& $(z\omega_y -y\omega_z \mp\sqrt{3/4}\omega_z){\bf I}$	\\
${\boldsymbol \varpi}_2 = {\bf r}_2 \times \boldsymbol \omega$	& $-(z\omega_x -x\omega_z +\omega_z/2){\bf J}$	\\
					& $+(y\omega_x \pm\sqrt{3/4}\omega_x -x\omega_y +\omega_y/2){\bf K}$	\\
\hline

\end{tabular}

\end{center}

From Table 3 and Formula (27) above, the transverse velocities ${\boldsymbol \varpi}_1$ and ${\boldsymbol \varpi}_2$ 
then reduce to vertical velocities $\varpi_1{\bf K}$ and $\varpi_2{\bf K}$, respectively, where 
\begin{equation}
		\varpi_1 = \pm\sqrt{3/4}\omega_x -\omega_y/2 = \epsilon[\mp\sqrt{3}\sin(t) +\cos(t)]/2 
\end{equation}
				and 
\begin{equation}
                \varpi_2 = \pm\sqrt{3/4}\omega_x +\omega_y/2 = \epsilon[\mp\sqrt{3}\sin(t) -\cos(t)]/2 , 
\end{equation}
to first degree in the small quantities $x,y,z$, and $\epsilon$.

From Formulae (26) and (28), and Formula (104) of the Appendix, 
% and neglecting $\bf s$, 
the primary tidal term ${\bf f}_{11} = {\bf F}_{11}/m_o$ can be expressed as 
\begin{equation}
{\bf f}_{11} \approx \kappa M_1^2 r_1^{-8} \{ [1 +2\dot{r}_1\tau/r_1]{\bf r}_1 +[\dot{\bf r}_1 +{\boldsymbol \varpi}_1]\tau \} 
\end{equation} \[
	\approx \kappa M_1^2 [1 -4x \mp4y\sqrt{3}] \{ [1 +(\dot{x} \pm\dot{y}\sqrt{3})\tau] 
[-(x+1/2){\bf I} -(y\pm\sqrt{3/4}){\bf J} -z{\bf K}] -(\dot{x}{\bf I} +\dot{y}{\bf J} +[\dot{z} -\varpi_1]{\bf K})\tau \} 
\] \[
        \approx \kappa M_1^2 \{ [-\frac{1}{2} +x \pm2y\sqrt{3} -\frac{3}{2}\dot{x}\tau \mp\sqrt{\frac{3}{4}}\dot{y}\tau]{\bf I} 
                +[\mp\sqrt{\frac{3}{4}} \pm2x\sqrt{3} +5y \mp\sqrt{\frac{3}{4}}\dot{x}\tau -\frac{5}{2}\dot{y}\tau]{\bf J} 
%	\approx \kappa m_1^2 \{ [-1/2 +x \pm2y\sqrt{3} -3\dot{x}\tau/2 \mp\dot{y}\tau\sqrt{3/4}]{\bf I} 
%		+[\mp\sqrt{3/4} \pm2x\sqrt{3} +5y \pm\dot{x}\tau\sqrt{3/4} -5\dot{y}\tau/2]{\bf J} 
				-[z +\dot{z}\tau -\varpi_1\tau]{\bf K} \} . 
\]
% Here $\bf I$, $\bf J$, $\bf K$ are the unit vectors in the directions of increasing $X$, $Y$, $Z$, respectively.

Similarly, linearizing Formula (105) of the Appendix gives the secondary tidal term 
\begin{equation}
{\bf f}_{22} \approx \kappa M_2^2 r_2^{-8} \{ [1 +2\dot{r}_2\tau/r_2]{\bf r}_2 +[\dot{\bf r}_2 +{\boldsymbol \varpi}_2]\tau \} 
\end{equation} \[
        \approx \kappa M_2^2 [1 +4x \mp4y\sqrt{3}] \{ [1 +(-\dot{x} \pm\dot{y}\sqrt{3})\tau] 
[-(x-1/2){\bf I} -(y\pm\sqrt{3/4}){\bf J} -z{\bf K}] -(\dot{x}{\bf I} +\dot{y}{\bf J} +[\dot{z} -\varpi_2]{\bf K})\tau \}
\] \[
        \approx \kappa M_2^2 \{ [+\frac{1}{2} +x \mp2y\sqrt{3} -\frac{3}{2}\dot{x}\tau \pm\sqrt{\frac{3}{4}}\dot{y}\tau]{\bf I} 
                +[\mp\sqrt{\frac{3}{4}} \mp2x\sqrt{3} +5y \pm\sqrt{\frac{3}{4}}\dot{x}\tau -\frac{5}{2}\dot{y}\tau]{\bf J} 
%	\approx \kappa m_2^2 \{ [+1/2 +x \mp2y\sqrt{3} -3\dot{x}\tau/2 \pm\dot{y}\tau\sqrt{3/4}]{\bf I} 
%		+[\mp\sqrt{3/4} \mp2x\sqrt{3} +5y \pm\dot{x}\tau\sqrt{3/4} -5\dot{y}\tau/2]{\bf J} 
				-[z +\dot{z}\tau -\varpi_2\tau]{\bf K} \} . 
\]
Note the pattern of sign differences between Formulae (30) and (31) above.  

\newpage

Linearizing Formula (106) of the Appendix gives the first mixed term:  
\[
		{\bf f}_{12} \approx \kappa M_1 M_2 r_1^{-5} r_2^{-7} \{ [5({\bf r}_2 \bullet {\bf r}_1)^2 
                -10({\bf r}_2 \bullet {\bf r}_1)[{\bf r}_2 \bullet (\dot{\bf r}_1 +{\boldsymbol \varpi}_1)]\tau 
                        -r_1^2 r_2^2 -3r_2^2 r_1 \dot{r}_1\tau]{\bf r}_2/2 
\] \[
                -[{\bf r}_2 \bullet ({\bf r}_1 -[\dot{\bf r}_1 +{\boldsymbol \varpi}_1]\tau)]r_2^2 {\bf r}_1 
                        +[{\bf r}_2 \bullet {\bf r}_1]r_2^2 (\dot{\bf r}_1 +{\boldsymbol \varpi}_1)\tau 
                                +5[5({\bf r}_2 \bullet {\bf r}_1)^2 {\bf r}_2/2 
                        -({\bf r}_2 \bullet {\bf r}_1)r_2^2{\bf r}_1]\dot{r}_1\tau/r_1 \} 
\]
\[
		\approx \kappa M_1 M_2 [1 -5x/2 \mp5y\sqrt{3/4}][1 +7x/2 \mp7y\sqrt{3/4}] 
	\{ [5(1/2 \pm y\sqrt{3})^2 -10(1/2 \pm y\sqrt{3})(-\dot{x}/2 \pm \dot{y}\sqrt{3/4} -z\varpi_1)\tau 
\] \[
				-(1 +x \pm y\sqrt{3})(1 -x \pm y\sqrt{3}) 
		-3(1 -x \pm y\sqrt{3})(1 +x/2 \pm y\sqrt{3/4})(+\dot{x}/2 \pm\dot{y}\sqrt{3/4})\tau]
			[(-x +1/2){\bf I} -(y \pm\sqrt{3/4}){\bf J} -z{\bf K}]/2 
\] \[
			-[(1/2 \pm y\sqrt{3}) -(-\dot{x}/2 \pm\dot{y}\sqrt{3/4} -z\varpi_1)\tau]
		(1 -x \pm y\sqrt{3})[(-x -1/2){\bf I} -(y \pm\sqrt{3/4}){\bf J} -z{\bf K}] 
\] \[
	+[1/2 \pm y\sqrt{3}](1 -x \pm y\sqrt{3})[-\dot{x}{\bf I} -\dot{y}{\bf J} +(\varpi_1 -\dot{z}){\bf K}]\tau 
\] \[
		+5[5(1/2 \pm y\sqrt{3})^2((-x+1/2){\bf I} -(y\pm\sqrt{3/4}){\bf J} -z{\bf K})/2 
	-(1/2 \pm y\sqrt{3})(1 -x \pm y\sqrt{3})((-x-1/2){\bf I} -(y \pm\sqrt{3/4}){\bf J} -z{\bf K})]
\] \[
			(1 -x/2 \mp y\sqrt{3/4})(\dot{x}/2 \pm\dot{y}\sqrt{3/4})\tau \} 
\]
\begin{equation}
%			\approx \kappa m_1 m_2 \{ {\bf I}[5/16 +x/16 +(45/32)\dot{x}\tau 
        	\approx \kappa M_1 M_2 \{ {\bf I}[\frac{5}{16} +\frac{7}{16}x +\frac{45}{32}\dot{x}\tau 
			\mp\frac{3\sqrt{3}}{8}y \pm\frac{5\sqrt{3}}{32}\dot{y}\tau] 
%                        \approx \kappa m_1 m_2 \{ {\bf I}[5/16 +7x/16 +(45/32)\dot{x}\tau 
%				+(\mp3\sqrt{3}/8)y +(\pm5\sqrt{3}/32)\dot{y}\tau] 
\end{equation} 
\[
%	+{\bf J}[\pm3\sqrt{3}/16 \mp\sqrt{3}x/16 \mp(5\sqrt{3}/32)\dot{x}\tau +3y/2 +(41/32)\dot{y}\tau] 
+{\bf J}[\pm\frac{3\sqrt{3}}{16} \mp\frac{\sqrt{3}}{16}x \mp\frac{5\sqrt{3}}{32}\dot{x}\tau -3y +\frac{41}{32}\dot{y}\tau] 
                        +{\bf K}[\frac{3}{8}z +\frac{1}{2}(\varpi_1 -\dot{z})\tau] \} . 
%        +{\bf J}[\pm3\sqrt{3}/16 \mp\sqrt{3}x/16 \mp(5\sqrt{3}/32)\dot{x}\tau  -3y  +(41/32)\dot{y}\tau] 
%				+{\bf K}[3z/8 +(\varpi_1 -\dot{z})\tau/2] \} . 
\]

% OK, THE THING TO DO HERE IS TO RE-DERIVE EVERYTHING AS $f_{21}$ 
% AND COMPARE WITH $f_{12}$ AS A CHECK; HERE GOES 

Then swapping subscripts in Formula (32) above gives the second mixed term:  
\[
                {\bf f}_{21} \approx \kappa M_1 M_2 r_2^{-5} r_1^{-7} \{ [5({\bf r}_1 \bullet {\bf r}_2)^2 
                -10({\bf r}_1 \bullet {\bf r}_2)[{\bf r}_1 \bullet (\dot{\bf r}_2 +{\boldsymbol \varpi}_2)]\tau 
                        -r_1^2 r_2^2 -3r_1^2 r_2 \dot{r}_2\tau]{\bf r}_1/2 
\] \[
                -[{\bf r}_1 \bullet ({\bf r}_2 -[\dot{\bf r}_2 +{\boldsymbol \varpi}_2]\tau)]r_1^2 {\bf r}_2 
                        +[{\bf r}_1 \bullet {\bf r}_2]r_1^2 (\dot{\bf r}_2 +{\boldsymbol \varpi}_2)\tau 
                                +5[5({\bf r}_1 \bullet {\bf r}_2)^2 {\bf r}_1/2 
                        -({\bf r}_1 \bullet {\bf r}_2)r_1^2{\bf r}_2]\dot{r}_2\tau/r_2 \} 
\]

\[
		\approx \kappa M_1 M_2 (1 +5x/2 \mp5y\sqrt{3/4})(1 -7x/2 \mp7y\sqrt{3/4}) 
	\{ [5(1/2 \pm y\sqrt{3})^2 -10(1/2 \pm y\sqrt{3})(+\dot{x}/2 \pm\dot{y}\sqrt{3/4} -z\varpi_2)\tau 
\] \[
	-(1 \pm y\sqrt{3})^2 -3(1 \pm y\sqrt{3})(1 +x/2 \pm y\sqrt{3/4})(-\dot{x}/2 \pm\dot{y}\sqrt{3/4})\tau]
			[(-x -1/2){\bf I} -(y \pm\sqrt{3/4}){\bf J} -z{\bf K}]/2
\] \[
		-[(1/2 \pm y\sqrt{3}) -(+\dot{x}/2 \pm\dot{y}\sqrt{3/4} -z\varpi_2)\tau](1 +x \pm y\sqrt{3})
			[(-x +1/2){\bf I} -(y \pm\sqrt{3/4}){\bf J} -z{\bf K}] 
\] \[
	+[1/2 \pm y\sqrt{3}](1 +x \pm y\sqrt{3})[-\dot{x}{\bf I} -\dot{y}{\bf J} +(\varpi_2 -\dot{z}){\bf K}]\tau 
\] \[
		+5[5(1/2 \pm y\sqrt{3})^2[(-x -1/2){\bf I} -(y \pm\sqrt{3/4}){\bf J} -z{\bf K}]/2 
	-(1/2 \pm y\sqrt{3})(1 +x \pm y\sqrt{3})[(-x +1/2){\bf I} -(y \pm\sqrt{3/4}){\bf J} -z{\bf K}]]
\] \[
			(1 +x/2 \mp y\sqrt{3/4})(-\dot{x}/2 \pm\dot{y}\sqrt{3/4})\tau \} 
\]

\begin{equation}
% \approx \kappa m_1 m_2 \{ {\bf I}[-5/16 +x/16 +(45/32)\dot{x}\tau +(\pm3\sqrt{3}/8)y +(\mp5\sqrt{3}/32)\dot{y}\tau]  
\approx \kappa M_1 M_2 \{ {\bf I}[-\frac{5}{16}+\frac{7}{16}x +\frac{45}{32}\dot{x}\tau \pm\frac{3\sqrt{3}}{8}y 
												\mp\frac{5\sqrt{3}}{32}\dot{y}\tau] 
%  \approx \kappa m_1 m_2 \{ {\bf I}[-5/16+7x/16 +(45/32)\dot{x}\tau +(\pm3\sqrt{3}/8)y +(\mp5\sqrt{3}/32)\dot{y}\tau] 
\end{equation}
\[
%	+{\bf J}[\pm3\sqrt{3}/16 +(\pm\sqrt{3}/16)x +(\pm5\sqrt{3}/32)\dot{x}\tau +(3/2)y +(41/32)\dot{y}\tau] 
+{\bf J}[\pm\frac{3\sqrt{3}}{16} \pm\frac{\sqrt{3}}{16}x \pm\frac{5\sqrt{3}}{32}\dot{x}\tau -3y +\frac{41}{32}\dot{y}\tau] 
                                +{\bf K}[\frac{3}{8}z +\frac{1}{2}(\varpi_2 -\dot{z})\tau] \} . 
%	+{\bf J}[\pm3\sqrt{3}/16 +(\pm\sqrt{3}/16)x +(\pm5\sqrt{3}/32)\dot{x}\tau   -3y   +(41/32)\dot{y}\tau] 
%				+{\bf K}[(3/8)z +(\varpi_2 -\dot{z})\tau/2] \} . 
\]

% EQS. (32) AND (33) ABOVE ARE CONSISTENT NOW 

Note the pattern of sign differences between Formulae (32) and (33) above, 
like that between Formulae (30) and (31).  This leads to substantial simplifications 
when Formulae (32) and (33) are added together into a joint mixed term:  
\begin{equation}
%		{\bf f}_{12} +{\bf f}_{21} \approx \kappa m_1 m_2 \{ {\bf I}[x/8 +45\dot{x}\tau/16] 
%		+{\bf J}[\pm3\sqrt{3}/8 +3y +41\dot{y}\tau/16] +{\bf K}[3z/4 -\dot{z}\tau] \} . 
	{\bf f}_{12} +{\bf f}_{21} \approx \kappa M_1 M_2 \{ {\bf I}[\frac{7}{8}x +\frac{45}{16}\dot{x}\tau] 
+{\bf J}[\pm\frac{3\sqrt{3}}{8} -6y +\frac{41}{16}\dot{y}\tau] +{\bf K}[\frac{3}{4}z +\frac{1}{2}(\varpi_1 +\varpi_2)\tau -\dot{z}\tau] \} . 
%                {\bf f}_{12} +{\bf f}_{21} \approx \kappa m_1 m_2 \{ {\bf I}[7x/8 +45\dot{x}\tau/16] 
%	+{\bf J}[\pm3\sqrt{3}/8 -6y +41\dot{y}\tau/16] +{\bf K}[3z/4 +(\varpi_1 +\varpi_2)\tau/2 -\dot{z}\tau] \} . 
\end{equation}
Note also that $(\varpi_1 +\varpi_2)$ reduces to just 
$\pm\sqrt{3}\omega_x = \mp\epsilon\sqrt{3}\sin t$ from Formulae (28) and (29).  

Finally, Formulae (30), (31), and (34) above all can be combined 
to give the net tidal accelerations for System (23) through (25):  

\[
%  f_X = \kappa [(m_2^2 -m_1^2)/2 +(m_1^2 +m_2^2  +m_1 m_2/8)x +(45m_1 m_2/16 -3m_1^2/2 -3m_2^2/2)\dot{x}\tau 
   f_X = \kappa [(M_2^2 -M_1^2)/2 +(M_1^2 +M_2^2 +7M_1 M_2/8)x +(45M_1 M_2/16 -3M_1^2/2 -3M_2^2/2)\dot{x}\tau 
\] \[
			\pm2\sqrt{3}(M_1^2 -M_2^2)y \pm\sqrt{3}(M_2^2 -M_1^2)\dot{y}\tau/2] 
\] \[
%		= \kappa [(m_2 -m_1)/2 +(1 -15m_1 m_2/8)x +(-3/2 +93m_1 m_2/16)\dot{x}\tau 
%		= \kappa [(m_2 -m_1)/2 +(1  -9m_1 m_2/8)x +(-3/2 +63m_1 m_2/16)\dot{x}\tau 
%		= \kappa [(m_2 -m_1)/2 +(1  -9m_1 m_2/8)x +(-3/2 +93m_1 m_2/16)\dot{x}\tau
		= \kappa [(M_2 -M_1)/2 +(1  -9M_1 M_2/8)x +(93M_1 M_2/16 -3/2)\dot{x}\tau
\] \begin{equation}
%			\pm2\sqrt{3}(m_1 -m_2)y \pm\sqrt{3}(m_2 -m_1)\dot{y}\tau/2] , 
				\pm\sqrt{3}(M_1 -M_2)(2y -\dot{y}\tau/2)] ,
\end{equation}

\[
 	               f_Y = \kappa [\pm\sqrt{3}(3M_1 M_2/8 -M_1^2/2 -M_2^2/2) 
		\pm2\sqrt{3}(M_1^2 -M_2^2)x \mp\sqrt{3}(M_1^2 -M_2^2)\dot{x}\tau/2 
\] \[
%		+(5m_1^2 +5m_2^2 +3m_1 m_2)y +(41m_1 m_2/16 -5m_1^2/2 -5m_2^2/2)\dot{y}\tau] 
		+(5M_1^2 +5M_2^2 -6M_1 M_2)y +(41M_1 M_2/16 -5M_1^2/2 -5M_2^2/2)\dot{y}\tau]
\]  \[
%				= \kappa [\pm\sqrt{3}(-1/2 +11m_1 m_2/8) 
				= \kappa [\pm\sqrt{3}(11M_1 M_2/8 -1/2)
%			\pm2\sqrt{3}(m_1 -m_2)x \mp\sqrt{3}(m_1 -m_2)\dot{x}\tau/2 
				\pm\sqrt{3}(M_1 -M_2)(2x -\dot{x}\tau/2)
\] \begin{equation}
%			+(5 -7m_1 m_2)y +(-5   +121m_1 m_2/16)\dot{y}\tau , 
%			+(5-16m_1 m_2)y +(-5/2 +121m_1 m_2/16)\dot{y}\tau] ,
			+(5-16M_1 M_2)y +(121M_1 M_2/16 -5/2)\dot{y}\tau] ,
\end{equation}

\[
               	f_Z = -\kappa [(M_1^2 +M_2^2 -3M_1 M_2/4)z +(M_1^2 +M_2^2 +M_1 M_2)\dot{z}\tau 
			+(M_1^2 s_1 +M_2^2 s_2 +M_1 M_2 (s_1 +s_2)/2)\tau] 
\] \begin{equation}
			= -\kappa[(1 -11M_1 M_2/4)z +(1 -M_1 M_2)\tau\dot{z} 
		+(M_1^2 \varpi_1 +M_2^2 \varpi_2 +M_1 M_2 [\varpi_1 +\varpi_2]/2)\tau] . 
\end{equation}
Here we have used $M_1 +M_2 = 1$ to simplify Formulae (35) through (37) above slightly.

\section{Vertical Motion}

Substituting Formulae (35) through (37) into Eqs. (23) through (25), respectively, 
leaves Eqs. (23) and (24) a coupled system in $x$, $y$, and their derivatives; 
but Eq. (25) in $z$, $\dot{z}$, and $\ddot{z}$ remains decoupled from System (23) and (24).  
Therefore we begin by solving Eq. (25) for the vertical motion of $m_o$.  

Subsituting Formulae (28), (29), and (37) into Eq. (25) gives 
\[
			\ddot{z} -\kappa\tau[1 -M_1 M_2]\dot{z} +(1 +\kappa[1 -11M_1 M_2/4])z 
			= -\kappa\tau[M_1^2 \varpi_1 +M_2^2 \varpi_2 +M_1 M_2[\varpi_1 +\varpi_2]/2]
\] \[
%	= -\kappa\tau\epsilon[\mp\sqrt{3}(m_1^2 +m_2^2 +m_1 m_2/2)\sin(t) +(m_1^2 -m_2^2)\cos(t)]/2 
	= -\kappa\tau\epsilon[\mp\sqrt{3}(M_1^2 +M_2^2 +2M_1 M_2)\sin(t) +(M_1^2 -M_2^2)\cos(t)]/2 
\] 
\begin{equation}
%		= -\kappa\tau\epsilon[\mp\sqrt{3}(1 -3m_1 m_2/2)\sin(t) +(m_1 -m_2)\cos(t)]/2 , 
		= -\kappa\tau\epsilon[\mp\sqrt{3}\sin(t) +(M_1 -M_2)\cos(t)]/2 , 
\end{equation}
with the homogeneous part on the left-hand side, and the forcing terms on the right.

\subsection{Forced solution}

The general solution to Eq. (38) above consists of a free part and a forced part.  
Because the forcing terms are sinusoidal, the forced solution is also sinusoidal, 
of the form $z_{forced} = s\sin(t) +c\cos(t)$, where the coefficients $s$ and $c$ are real constants.  
Substituting this into Eq. (38) gives 
\[
			-s\sin(t) -c\cos(t) -\kappa\tau[1 -M_1 M_2][s\cos(t) -c\sin(t)] 
				+(1 +\kappa[1 -11M_1 M_2/4])[s\sin(t) +c\cos(t)] 
\]
\begin{equation}
%		= -\kappa\tau\epsilon[\mp\sqrt{3}(1 -3m_1 m_2/2)\sin(t) +(m_1 -m_2)\cos(t)]/2 . 
			= -\kappa\tau\epsilon[\mp\sqrt{3}\sin(t) +(M_1 -M_2)\cos(t)]/2 . 
\end{equation}
Then equating sine terms gives 
\begin{equation}
%		\kappa[1 -11m_1 m_2/4]A +\kappa\tau[1 -m_1 m_2]B = \pm\kappa\tau\epsilon\sqrt{3/4}(1 -3m_1 m_2/2) , 
		\kappa[1 -11M_1 M_2/4]s +\kappa\tau[1 -M_1 M_2]c = \pm\kappa\tau\epsilon\sqrt{3/4} , 
\end{equation}
while equating cosine terms gives 
\begin{equation}
			-\kappa\tau[1 -M_1 M_2]s +\kappa[1 -11M_1 M_2/4]c = -\kappa\tau\epsilon[M_1 -M_2]/2 . 
\end{equation}

Simultaneously solving Eqs. (40) and (41) above gives 
\begin{equation}
%		A \approx \frac{\epsilon}{2}\left[\frac{m_1 -m_2}{1 -m_1 m_2}\right] \; \; 
%			A \approx \pm\tau\epsilon\sqrt{3}(2 -3m_1 m_2)/(4 -11m_1 m_2) \; \; 
%			s \approx \pm2\sqrt{3}\tau\epsilon/(4 -11m_1 m_2) \; \; 
			s \approx \frac{\pm2\sqrt{3}\tau\epsilon}{4 -11M_1 M_2} \; \; 
{\rm and}
%		B \approx \pm\epsilon\sqrt{3/4}\left[\frac{1 -3m_1 m_2/2}{1 -m_1 m_2}\right] , 
%			\; \; c \approx -2\tau\epsilon[m_1 -m_2]/(4 -11m_1 m_2) . 
			\; \; c \approx \frac{-2\tau\epsilon[M_1 -M_2]}{4 -11M_1 M_2} . 
\end{equation}
Here we have neglected terms of order $\tau^2$ and higher, 
for consistency with the development of the Appendix.
% to first degree in $\tau$.  
% (provided that $\kappa \ne 0$).  
% To this order, note that $\kappa$ and $\tau$ drop out of the forced solution.  
Solution (42) above represents $m_o$ moving in a certain ``preferred'' orbital plane, 
tilted from the $XY$ plane by an inclination angle 
\begin{equation}
%		i = \sqrt{A^2 +B^2} = \epsilon\sqrt{1 -13m_1 m_2/4 +27m_1^2 m_2^2/16}/(1 -m_1 m_2) . 
%		i = \sqrt{A^2 +B^2} = \tau\epsilon\sqrt{16 -52m_1 m_2 +27m_1^2 m_2^2}/(4 -11m_1 m_2) . 
%		i = \sqrt{s^2 +c^2} = 4\tau\epsilon\sqrt{1 -m_1 m_2}/(4 -11m_1 m_2) . 
		i = \sqrt{s^2 +c^2} = \frac{2\tau\epsilon}{4 -11M_1 M_2}\sqrt{3 +[M_1 -M_2]^2} 
			= \frac{4\tau\epsilon\sqrt{1 -M_1 M_2}}{4 -11M_1 M_2} . 
\end{equation}
Note that Formulae (42) and (43) above are independent of $\kappa$, as long as $\kappa \ne 0$; 
however, when $\kappa = 0$, the tides vanish, $s$, $c$, and $i$ are arbitrary, and there is no preferred plane.  

Figure 1 graphs $|s|$, $|c|$, and $i$ as functions of $M_2$ (or of $M_1$) 
from Formulae (42) and (43), for $M_2$ up to 1/2, for completeness.  
In the two-body case when $M_1$ = 1, so $M_2$ vanishes (left-hand axis), then 
$s = \pm\tau\epsilon\sqrt{3/4} \approx \pm$0.866 025 $\tau\epsilon$, $c = -\tau\epsilon/2$, and $i = \tau\epsilon$.  
% $A = \epsilon/2$, $B= \pm\sqrt{3/4}\epsilon \approx \pm0.866025 \; \epsilon$, and $i = \epsilon$.  
% As expected, a bit of trigonometry reveals that in this case, 
% the preferred plane coincides with the equator plane of $m_0$, perpendicular to its spin axis.  
When $M_1 = 1/2 +\sqrt{5/44} \approx$ 0.837 100 and $M_2 = 1/2 -\sqrt{5/44} \approx$ 0.162 900, 
$c$ reaches a shallow minimum of $-4\tau\epsilon/\sqrt{55} \approx$ --0.539 360 $\tau\epsilon$, 
while $s = \pm4\sqrt{3}\tau\epsilon/5 \approx \pm$1.385 641 $\tau\epsilon$ 
and $i = 8\sqrt{19/22} \; \tau\epsilon/5 \approx$ 1.486 913 $\tau\epsilon$.  
% 0.162900060       1.10221410     -0.539359868       1.22710431
However, in the ``Copenhagen'' case when $M_1 = M_2$ = 1/2 (right-hand edge),
$|s|$ and $i$ both peak at $8\sqrt{3}\tau\epsilon/5 \approx$ 2.771 281 $\tau\epsilon$, 
while $c$ vanishes entirely.  

\begin{figure}
%\centerline{\psfig{figure=inc.eps,width=3.25in,height=3.25in}}
\begin{center}
\includegraphics[width=3.25in,height=3.25in]{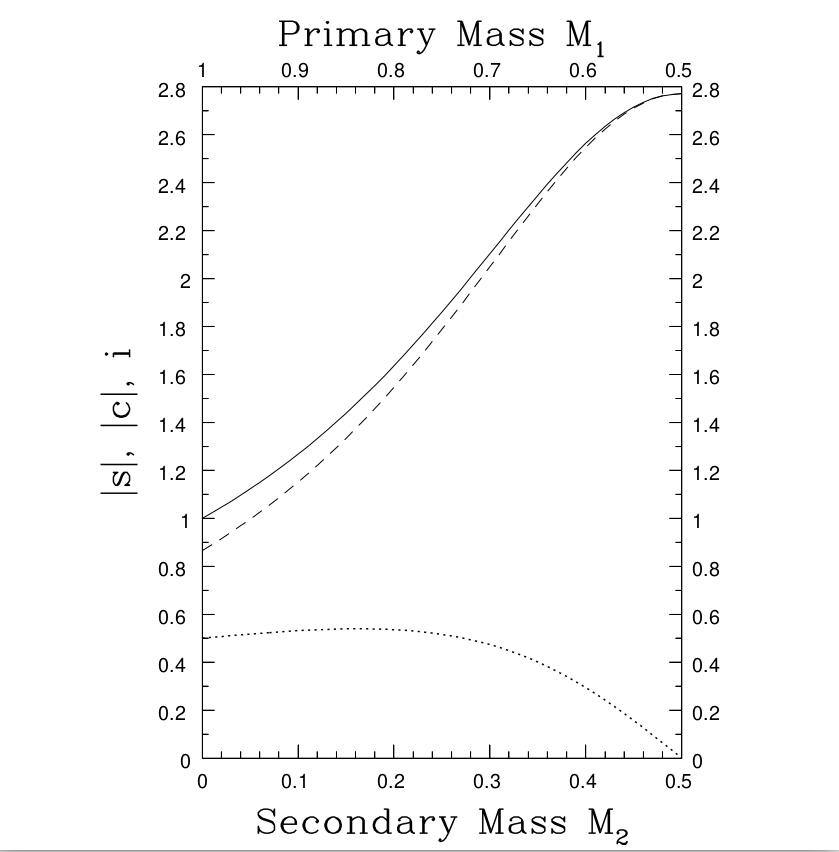}
\end{center}
\caption{ $|s|$ (dashed curve), $|c|$ (dotted curve), 
and $i$ (solid curve), each normalized by $\tau\epsilon$, 
as functions of $M_1$ or of $M_2$.  }
\end{figure}

\subsection{Free solution}

For the free solution, we assume the form $z = H\exp(\nu t)$, 
where $H$ is a complex amplitude of vertical motion 
and $\nu$ is its complex rate constant.  
Then substituting this into Eq. (25) gives the quadratic equation 
\begin{equation}
		\nu^2 = -1 -\kappa[1 -11M_1 M_2/4] -\kappa[1 -M_1 M_2]\nu\tau , 
\end{equation}
% Note that the infinitesimal amplitude $C$ turns out to be arbitrary.  
%
% Eq. (44) above has two roots:  
with the two roots 
\[
				\nu = -\kappa\tau[1 -M_1 M_2]/2 
		\pm\sqrt{\kappa^2\tau^2[1 -M_1 M_2]^2/4 -1 -\kappa[1 -11M_1 M_2/4]} 
\] \begin{equation}
		\approx -\kappa\tau[1 -M_1 M_2]/2 \pm\sqrt{-1 -\kappa[1 -11M_1 M_2/4]} 
\end{equation}
to first degree in $\tau$, where for the moment $\pm$ has its usual meaning.  
% Here we have neglected $\tau^2$, for consistency with the development of the Appendix.  

Note that the discriminant $-1 -\kappa[1 -11M_1 M_2/4]$ in Formula (45) above is always negative, 
because $[1 -11M_1 M_2/4]$ is always positive.  Therefore $\nu$ can be written as $\zeta \pm j\eta$, 
where $\zeta = -\kappa\tau[1 -M_1 M_2]/2$ is the e-folding rate of the free solution, 
$\eta = \sqrt{1 +\kappa[1 -11M_1 M_2/4]}$ is the angular frequency of its oscillations,
and $j$ is the imaginary unit ($j^2 = -1$).  In strictly real terms, 
the complete free solution to Eq. (38) can be written as 
\begin{equation}
		z_{free} = \exp(\zeta t)[S\sin(\eta t) +C\cos(\eta t)] , 
\end{equation}
where $S$ and $C$ are real constants.  
Because $\zeta$ is always negative, the vertical motion is always stable.  

It is of interest to consider special cases of Formula (46) above.  
When $\kappa$ vanishes, Formula (45) reduces to just $\nu = +j$ or $\nu = -j$; 
so $\zeta$ = 0 and $\eta$ = 1, and Formula (46) becomes $z_{free} = S\sin(t) +C\cos(t)$.  
Then the Trojan executes harmonic oscillations above and below the plane $z = 0$ 
with angular frequency unity (or $n$, in dimensional units), as expected; 
and their absolute amplitude $|H| = \sqrt{S^2 +C^2}$ effectively becomes 
the inclination of the Trojan orbit to the $XY$ plane.  

When the time lag $\tau$ vanishes, but $\kappa$ does not, 
the Trojan still oscillates vertically about its preferred plane, 
again with effective inclination $|H| = \sqrt{S^2 +C^2}$, 
but now with slightly increased angular frequency $\eta$ = 
$\sqrt{1 +\kappa[1 -11M_1 M_2/4]} \approx 1 +\kappa[1/2 -11M_1 M_2/8]$ 
due to the enhanced attraction of $M_1$ and $M_2$ on the tidal bulges of $m_o$.  

When neither $\kappa$ nor $\tau$ vanishes, the Trojan executes {\it damped} harmonic oscillations 
above and below the preferred plane, with essentially the same angular frequency $\eta$ as above, 
but decaying to the preferred plane with e-folding time $-1/\zeta$ (or $-1/(2\pi\zeta)$ in orbital periods).  
% However, this solution was obtained assuming that $\boldsymbol \omega$ was fixed in the sidereal frame; 
% in reality, the equator plane of $\mu$ would damp to its orbital plane much faster than {\it vice versa}.  

% In any case, we conclude that the vertical motion of $m_0$ is stable, 
% and henceforth we confine ourselves to its horizontal motion.  

% WAIT - WHEN $m_2 = 0 \ne \kappa$, WE HAVE THE 2-BODY PROBLEM WITH TIDES; 
% WHY DOESN'T THE DAMPING VANISH ?  IS IT BECAUSE THE SPIN IS PERP. TO THE XY PLANE ?

% $\sqrt{1 +\kappa[1 -11m_1 m_2/4] -\kappa^2[1 -m_1 m_2]^2\tau^2}$ 
% (provided that this frequency is real; in the unlikely event that it is imaginary, 
% the Trojan decays monotonically toward the plane $z$ = 0).  

% When $m_2$ vanishes (as in the two-body problem), 
% but $\kappa$, $\tau$, and $m_1$ do not, Formula (35) reduces to 
% \begin{equation}
%		\zeta = -\kappa m_1^2 \tau/2 \pm\sqrt{\kappa^2 m_1^4 \tau^2/4 -1 -\kappa m_1^2} . 
% \end{equation}
% When $m_1 = m_2$ (as in the Copenhagen problem), Formula (35) reduces to 
% \begin{equation}
%		\zeta = -3\kappa m_1^2\tau/2 \pm\sqrt{9\kappa^2 m_1^4 \tau^2/4 - 1 -5\kappa m_1^2/4} . 
% \end{equation}

\subsection{Complete solution}

In order to evaluate the coefficients of the free solution (46), 
first write out the general solution to Eq. (38):  
\begin{equation}
		z = z_{forced} +z_{free} = s\sin(t) +c\cos(t) +\exp(-\zeta t)[S\sin(\eta t) +C\cos(\eta t)] , 
\end{equation}
where $s$ and $c$ are given by Formulae (42).  
Then differentiating solution (47) above gives 
\begin{equation}
\dot{z} = s\cos(t) -c\sin(t) +\exp(-\zeta t)[(S\eta -C\zeta)\cos(\eta t) -(C\eta +S\zeta)\sin(\eta t)] . 
\end{equation}

Next, set Formulae (47) and (48) above equal to their respective 
initial conditions $z = z_0$ and $\dot{z} = \dot{z}_0$ at time $t$ = 0.  
Then solving the resulting two simultaneous equations gives 
\begin{equation}
			C = z_0 -c \; \; {\rm and} \; \; S = (\dot{z}_0 -s +C\zeta)/\eta . 
\end{equation}
In the dissipationless case when $\tau$ vanishes, 
then $s$, $c$, and $\zeta$ all vanish as well from Formulae (42) and (45); 
while Formulae (49) above reduce to $C = z_0$ and $S = \dot{z}_0$, 
and the general solution to Eq. (38) becomes just $z = \dot{z}_0\sin(\eta t) +z_0\cos(\eta t)$.  

However, all of the above solutions are unrealistic, because they were obtained 
assuming that $\boldsymbol \omega$ was fixed in the sidereal frame.  
In reality, the equator plane of $m_o$ would move, 
and would damp to its orbital plane much faster than {\it vice versa},  
because the spin angular momentum of $m_o$ is much less than its orbital angular momentum.  
Nevertheless, we conclude that the vertical motion of $m_o$ is stable, 
and henceforth we confine ourselves to its horizontal motion.  

% CHANGE A TO S, B TO C, and C to H

\section{Horizontal motion}

The two-dimensional solution for the horizontal motion is generally more complicated 
than the one-dimensional solution for the vertical motion, 
because the $x$ and $y$ components are strongly coupled.  
Like the vertical solution, the horizontal solution consists of a free part and a forced part, 
but the forced part of the horizontal solution is constant rather than sinusoidal in time.

\subsection{Shifted equilibria}

To solve for the forced part of the horizontal motion, 
we assume that $x = x'$ and $y = y'$, where $x'$ and $y'$ are constants.  
Then Eqs. (23) and (24) respectively reduce to 
\[
				0 = 3x'/4 \pm3\sqrt{3}(1-2M_2)y'/4 
%		+\kappa[(m_2 -m_1)/2 +(1 -15m_1 m_2/8)x' \pm2\sqrt{3}(m_1 -m_2)y'] 
		+\kappa[(M_2 -M_1)/2 +(1  -9M_1 M_2/8)x' \pm2\sqrt{3}(M_1 -M_2)y'] 
\] \begin{equation}
			\approx 3x'/4 \pm3\sqrt{3}[M_1-M_2]y'/4 -\kappa[M_1 -M_2]/2 
\end{equation}
					and 
\[
				0 = 9y'/4 \pm3\sqrt{3}(1-2M_2)x'/4 
%	+\kappa[\pm\sqrt{3}(-1/2 +11m_1 m_2/8) \pm2\sqrt{3}(m_1 -m_2)x' +(5 -7m_1 m_2)y']  
	+\kappa[\pm\sqrt{3}(-1/2 +11M_1 M_2/8) \pm2\sqrt{3}(M_1 -M_2)x' +(5-16M_1 M_2)y']  
\] \begin{equation}
		\approx 9y'/4 \pm3\sqrt{3}[M_1-M_2]x'/4 \mp\kappa\sqrt{3}[1/2 -11M_1 M_2/8] , 
\end{equation}
where we neglect terms of order $\kappa^2$ and higher powers of $\kappa$. 

Solving System (50) and (51) above gives 
\begin{equation}
		x' \approx 11\kappa[M_1 -M_2]/24 \approx 0.458 \; 333 \; \kappa[M_1 -M_2] 
\end{equation}
					and 
\begin{equation}
		y' \approx \pm5\sqrt{3}\kappa/72 \approx \pm0.120 \; 281 \; \kappa , 
\end{equation}
provided that $M_1 M_2$ does not vanish.  
Formulae (52) and (53) above represent a constant shift of the equilibrium points 
from the equilateral points L4 and L5 in the CR3BP without tides 
to the new equilibria L$'4$ and L$'5$, respectively, 
due to the enhanced attraction of $M_1$ and $M_2$ on the tidal bulges of $m_o$.  

Note that $x'$ vanishes when $M_1 = M_2$, 
and is positive when $M_1 > M_2$.  
% but changes sign when $M_1$ and $M_2$ are switched.  
The shift in $y$ from the leading L4 point is non-negative, 
while that from the trailing L5 point is non-positive.  
Thus these equilibrium points lie 
farther from the origin with increasing $\kappa$, 
to compensate for the increased attraction of the tides.  
Note also that $x'$ and $y'$ both vanish when $\kappa$ vanishes 
(provided that $M_2 \ne 0$), consistent with the classic CR3BP.  

When $M_2$ vanishes, 
% In the special case when $\kappa$ and $m_2$ both vanish, 
% the situation reduces to the classic two-body problem, while 
Eqs. (50) and (51) both reduce to $0 = x' \pm\sqrt{3}y' -2\kappa/3$.  
This means that the equilibrium points may lie anywhere 
along a pair of infinitesimal arcs through L$'4$ and L$'5$ 
at a distance of $1 +\kappa/3$ from $M_1$.  
% perpendicular to ${\bf r}_1$.  
This is consistent with the two-body problem including tides, 
and thus provides another valuable check on our results.  
Note that solutions (52) and (53) lie on these arcs.  
When $\kappa$ vanishes as well, these arcs run through L4 and L5.

\subsection{Characteristic equation}

To find the free part of the solution to System (23) and (24), 
we assume solutions of the form $x = D\exp(\lambda t)$ and $y = E\exp(\lambda t)$, 
where $D$ and $E$ are complex amplitudes 
and $\lambda$ is their shared complex rate constant.  
Then substituting these into Eqs. (23) and (24) gives 
\begin{equation}
			\lambda^2 D = 2\lambda E +3D/4 \pm3\sqrt{3}[M_1 -M_2]E/4 
\end{equation} \[
%			+\kappa[(1 -15m_1 m_2/8)A +(-3/2 +93m_1 m_2/16)A\nu\tau 
%			+\kappa[(1 -9 m_1 m_2/8)A +(-3/2 +63m_1 m_2/16)A\nu\tau 
			+\kappa[(1 -9 M_1 M_2/8)D +(-3/2 +93M_1 M_2/16)D\lambda\tau 
			\pm2\sqrt{3}(M_1 -M_2)E \mp\sqrt{3}(M_1 -M_2)E\lambda\tau/2] 
\]
					and
\begin{equation}
			\lambda^2 E = -2\lambda D +9E/4 \pm3\sqrt{3}(M_1 -M_2)D/4
\end{equation} \[
			+\kappa[\pm2\sqrt{3}(M_1 -M_2)D \mp\sqrt{3}(M_1 -M_2)D\lambda\tau/2
%			+(5 -7M_1 M_2)B +(-5   +121M_1 M_2/16)B\nu\tau] . 
			+(5-16M_1 M_2)E +(-5/2 +121M_1 M_2/16)E\lambda\tau] . 
\]

System (54) and (55) above consists of two homogeneous linear equations in the two unknowns, 
$D$ and $E$.  In matrix form, this system can be written as 
\[
%	\left[ \begin{array}{cc} 3/4 -\nu^2 +\kappa[(1 -15m_1 m_2/8) -(3/2 -93m_1 m_2/16)\nu\tau)] 
%	\left[ \begin{array}{cc} 3/4 -\nu^2 +\kappa[(1 -9 m_1 m_2/8) -(3/2 -63m_1 m_2/16)\nu\tau] 
	\left[ \begin{array}{cc} 3/4 -\lambda^2 +\kappa[(1 -9 M_1 M_2/8) -(3/2 -93M_1 M_2/16)\lambda\tau] 
		 & +2\lambda \pm3\sqrt{3}(M_1 -M_2)/4 \pm\kappa\sqrt{3}(M_1 -M_2)[2 -\lambda\tau/2] 
		\\ -2\lambda \pm3\sqrt{3}(M_1 -M_2)/4 \pm\kappa\sqrt{3}(M_1 -M_2)[2 -\lambda\tau/2] 
%		& 9/4 -\nu^2 +\kappa[(5 -7m_1 m_2) -(5   -121m_1 m_2/16)\nu\tau] \end{array} \right] 
		& 9/4 -\lambda^2 +\kappa[(5-16M_1 M_2) -(5/2 -121M_1 M_2/16)\lambda\tau] \end{array} \right] 
\] \begin{equation}
			\bullet \left[ \begin{array}{c} D \\ E \end{array} \right] 
			      = \left[ \begin{array}{c} 0 \\ 0 \end{array} \right] . 
\end{equation}

System (56) above is self-consistent only if the determinant of the coefficient matrix vanishes.  
To first degree in $\kappa$, this yields 
\[
%	0 = \nu^4 +\nu^2 +27m_1 m_2/4 +\kappa\nu^3[13/2 -107m_1 m_2/8]\tau -\kappa\nu^2[6 -71m_1 m_2/8] 
%	0 = \nu^4 +\nu^2 +27m_1 m_2/4 +\kappa\nu^3\tau[4 -23m_1 m_2/2] +\kappa\nu^2[-6 +137m_1 m_2/8] 
	0 = \lambda^4 +\lambda^2 +27M_1 M_2/4 +\kappa\lambda^3\tau[4-107M_1 M_2/8] +\kappa\lambda^2[-6 +137M_1 M_2/8]
\] \begin{equation}
%		+\kappa\nu[(-5/8 +75m_1 m_2/4)\tau +27/4 -27m_1 m_2] -\kappa[3 +319m_1 m_2/32] . 
%		+\kappa\nu\tau[-3 +177m_1 m_2/32] +\kappa[-3 +465m_1 m_2/32] . 
%		+\kappa\nu\tau[-3 +177m_1 m_2/32] +\kappa[-3 +687m_1 m_2/32] . 
		+\kappa\lambda\tau[-3  +39M_1 M_2/4 ] +\kappa[-3 +687M_1 M_2/32] . 
\end{equation}

Eq. (57) above can be solved easily as a linear equation for $M_1 M_2$ 
(or as a quadratic equation for $M_2$) as a function of $\kappa$ and $\lambda$; 
for example, $M_2 \approx M_1 M_2 \approx 4\kappa/9$ when $\lambda$ = 0. 
However, we wish to find $\lambda$ as a function of $M_2$ and $\kappa$; 
then Eq. (57) is a quartic equation, analytically solvable in principle, 
but cumbersome in practice.  Its four roots $\lambda_N$ may be real or complex 
(where the index $N$ runs from 1 through 4); 
but any complex roots must come in conjugate pairs, 
because all of the coefficients in Eq. (57) are real.

\subsection{Perturbation method}

Consistent with our neglect of $\kappa^2$, 
we approximate the roots of Eq. (57) by a method of perturbation.  
We assume that each root can be represented as $\lambda_N = \gamma_N +\delta_N$, 
where $\lambda_N$ reduces to $\gamma_N$ in the absence of tides.  
When tides are absent, $\kappa$ vanishes, and Eq. (54) reduces to just 
\begin{equation}
			0 = \lambda^4 +\lambda^2 +27M_1 M_2/4 , 
\end{equation}
equivalent to Eq. (3.141) of Murray and Dermott (1999).  
Eq. (58) above is a {\it biquadratic} equation; 
still quartic, but now a quadratic equation in $\lambda^2$, 
and much easier to solve for all four roots $\gamma_N$.  

Solving Eq. (58) by the quadratic formula gives 
$\gamma_1^2 = \gamma_2^2 = -1/2 -\sqrt{1 -27M_1 M_2}/2$ for the first two roots, and 
$\gamma_3^2 = \gamma_4^2 = -1/2 +\sqrt{1 -27M_1 M_2}/2$ for the other two roots.  Then all four roots are 
\[
%			\gamma_1 = +j\sqrt{1 +\sqrt{1 -27m_1 m_2}}/\sqrt{2} , 
			\gamma_1 = \frac{+j}{\sqrt{2}}\sqrt{1 +\sqrt{1 -27M_1 M_2}} , 
\] \[
%			\gamma_2 = -j\sqrt{1 +\sqrt{1 -27m_1 m_2}}/\sqrt{2} , 
			\gamma_2 = \frac{-j}{\sqrt{2}}\sqrt{1 +\sqrt{1 -27M_1 M_2}} , 
\] \[
%			\gamma_3 = +j\sqrt{1 -\sqrt{1 -27m_1 m_2}}/\sqrt{2} , 
			\gamma_3 = \frac{+j}{\sqrt{2}}\sqrt{1 -\sqrt{1 -27M_1 M_2}} , 
\] \begin{equation}
%			\gamma_4 = -j\sqrt{1 -\sqrt{1 -27m_1 m_2}}/\sqrt{2} , 
			\gamma_4 = \frac{-j}{\sqrt{2}}\sqrt{1 -\sqrt{1 -27M_1 M_2}} , 
\end{equation}
equivalent to Formulae (3.143) and (3.144) of Murray and Dermott (1999).  
Note that $\gamma_2 = -\gamma_1$ and $\gamma_4 = -\gamma_3$, 
while $\gamma_1^2 +\gamma_3^2 = -1$.  
% and $\gamma_1^2 \gamma_3^2 = 27m_1 m_2/4$.  

If $1 -27M_1 M_2 < 0$ 
(so $M_2 > (1 -\sqrt{23/27})/2 \approx 0.03852 \approx 1/25.96$; Dobrovolskis, 2013), 
$\gamma_2$ and $\gamma_3$ have positive real parts, and the motion is unstable.  
Otherwise, all four roots are pure imaginary, 
and $\gamma_1$ and $\gamma_2$ correspond to harmonic epicyclic motion 
with a period slightly longer than the orbital period of $M_2$ around $M_1$, 
while $\gamma_3$ and $\gamma_4$ correspond to a harmonic libration 
of still longer period (Murray and Dermott, 1999).  

\begin{figure}
%\centerline{\psfig{figure=gammas.eps,width=3.3in,height=3.3in}}
\begin{center}
\includegraphics[width=3.3in,height=3.3in]{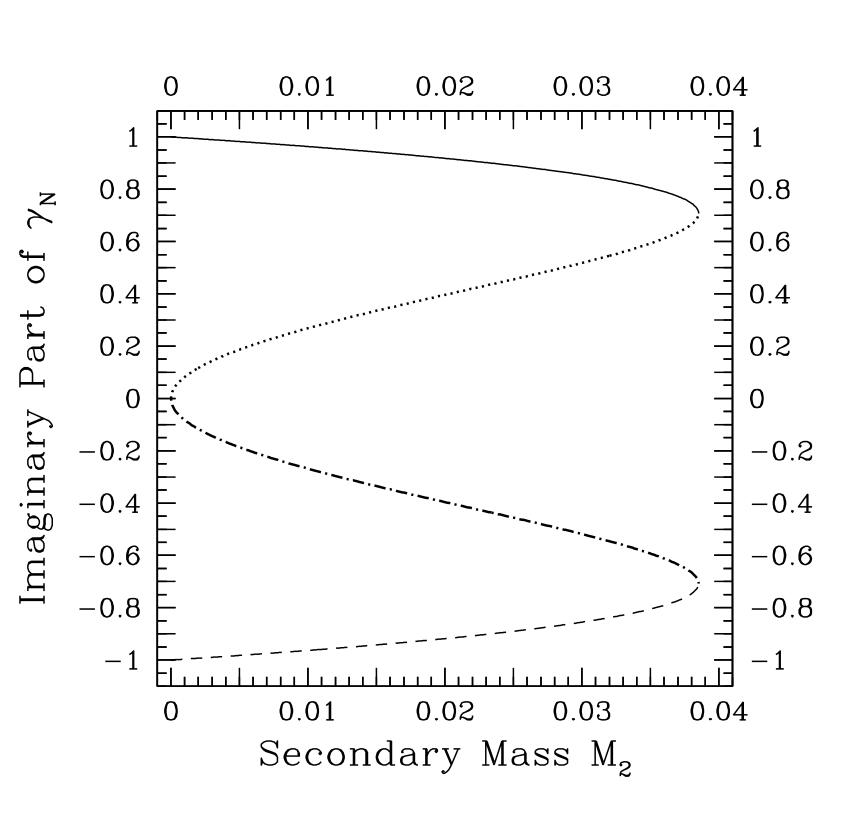}
\end{center}
\caption{ Imaginary parts of $\gamma_1$, $\gamma_2$, $\gamma_3$, 
and $\gamma_4$ as functions of $M_2$, the mass of the secondary.  
Solid curve:  Im($\gamma_1$).  Dashed curve:  Im($\gamma_2$) = --Im($\gamma_1$).  
Dotted curve:  Im($\gamma_3$).  Dot-dashed curve:  Im($\gamma_4$) = --Im($\gamma_3$).  }
\end{figure}

Figure 2 graphs the imaginary parts of $\gamma_1$, $\gamma_2$, $\gamma_3$, and $\gamma_4$ 
as functions of $M_2$, for $M_1 M_2 \le 1/27$, from Formulae (59).  For $M_2 < \sim0.02$, 
$\gamma_1 \approx +j\sqrt{1 -27M_2/4} \approx +j[1 -27M_2/8]$, $\gamma_2 \approx -j[1 -27M_2/8]$, 
$\gamma_3 \approx +\sqrt{-27M_2/4} = +3j\sqrt{3M_2}/2$, and $\gamma_4 \approx -3j\sqrt{3M_2}/2$.  

In the two-body case when $M_2$ vanishes, $\gamma_3$ and $\gamma_4$ both vanish as well, 
while $\gamma_1 = +j$ and $\gamma_2 = -j$.  (In this special case, 
the exponential solutions $\exp(\gamma_3 t)$ and $\exp(\gamma_4 t)$ are replaced 
by secular solutions of the form $D +Et$, where $E$ is now a constant velocity.)  
At $M_1 M_2$ = 1/27 ($M_2 \approx$ 0.03852), 
$\gamma_1$ and $\gamma_3$ both equal $+j/\sqrt{2}$, while 
$\gamma_2$ and $\gamma_4$ both equal $-j/\sqrt{2}$.  
(In this special case, the solutions take the form $[D +Et]\exp(\gamma_N t)$.)  

Now when tides are present, $0 < \kappa << 1$, and we assume that 
each root of Eq. (54) is perturbed to $\lambda_N = \gamma_N +\delta_N$.  
For the time being, we also presume that $\delta_N$ is of the same order as $\kappa$.  
Then Eq. (57) can be expanded as 
\[
                0 = [\gamma_N^4 +4\gamma_N^3\delta_N] +[\gamma_N^2 +2\gamma_N\delta_N] +27M_1 M_2/4
\] \begin{equation}
%		+\kappa\gamma_N^3[13/2 -107m_1 m_2/8]\tau -\kappa\gamma_N^2[6 -71m_1 m_2/8]
%		+\kappa\gamma_N^3\tau[4 -23m_1 m_2/2] +\kappa\gamma_N^2[-6 +137m_1 m_2/8] 
		+\kappa\gamma_N^3\tau[4-107M_1 M_2/8] +\kappa\gamma_N^2[-6 +137M_1 M_2/8] 
\end{equation} \[
%		+\kappa\gamma_N[(-5/8 +75m_1 m_2/4)\tau +27/4 -27m_1 m_2 -\kappa[3 +319m_1 m_2/32] 
%		+\kappa\gamma_N\tau[-3 +177m_1 m_2/32] +\kappa[-3 +465m_1 m_2/32] , 
%		+\kappa\gamma_N\tau[-3 +177m_1 m_2/32] +\kappa[-3 +687m_1 m_2/32] , 
		+\kappa\gamma_N\tau[-3  +39M_1 M_2/4 ] +\kappa[-3 +687M_1 M_2/32] , 
\]
to first degree in $\delta_N$ and $\kappa$.  

Subtracting Eq. (58) from Eq. (60) above and solving for $\delta_N$ then gives 
\begin{equation}
%		\delta_N = \kappa \{ -\gamma_N^3[13/2 -107m_1 m_2/8]\tau +\gamma_N^2[6 -71m_1 m_2/8] 
%		\delta_N = \kappa \{ \gamma_N^3\tau[-4 +23m_1 m_2/2] +\gamma_N^2[6 -137m_1 m_2/8] 
		\delta_N = \kappa \{ \gamma_N^3\tau[-4+107M_1 M_2/8] +\gamma_N^2[6 -137M_1 M_2/8] 
\end{equation} \[
%		-\gamma_N[49/8 -33m_1 m_2/4]\tau +[3 +319m_1 m_2/32] \} /[4\gamma_N^3 +2\gamma_N] . 
%		+\gamma_N\tau[3 -177m_1 m_2/32] +[3 -465m_1 m_2/32] \} /[4\gamma_N^3 +2\gamma_N] . 
%		+\gamma_N\tau[3 -177m_1 m_2/32] +[3 -687m_1 m_2/32] \} /[4\gamma_N^3 +2\gamma_N] . 
		+\gamma_N\tau[3  -39M_1 M_2/4 ] +[3 -687M_1 M_2/32] \} /[4\gamma_N^3 +2\gamma_N] . 
\]

Note that Eqs. (60) and (61) above become invalid when $\gamma_N$ vanishes, 
as $\gamma_3$ and $\gamma_4$ do when $M_2$ vanishes; 
we avoid this difficulty by finding only $\delta_1$ and $\delta_2$ this way.  
Eqs. (60) and (61) also become invalid when $\gamma_N^2$ approaches --2, 
as all four roots do when $M_1 M_2$ approaches 1/27 ($M_2 \approx$ 0.03852). 
We avoid this issue by treating only $M_2 < \sim 0.03$ henceforth; 
then all four $\gamma_N$ are strictly imaginary.

\subsection{Epicycles}

At this point it is useful to consider the real and imaginary parts of $\delta_N$ separately.  
Because each $\gamma_N$ is imaginary, the entire denominator of Formula (61) also is imaginary, 
while the constant, linear, quadratic, and cubic terms of its numerator 
alternate between real and imaginary.  Then the real part of $\delta_N$ becomes 
\begin{equation}
%			{\rm Re}(\delta_N) = \kappa\tau \{ -\gamma_N^3[13/2 -107m_1 m_2/8] 
%			{\rm Re}(\delta_N) = \kappa\tau \{ \gamma_N^3[-4 +23m_1 m_2/2] 
			{\rm Re}(\delta_N) = \kappa\tau \{ \gamma_N^3[-4+107M_1 M_2/8] 
%			-\gamma_N[49/8 -33m_1 m_2/4] \} / [4\gamma_N^3 +2\gamma_N] , 
%			+\gamma_N[3 -177m_1 m_2/32] \} / [4\gamma_N^3 +2\gamma_N] , 
			+\gamma_N[3  -39M_1 M_2/4 ] \} / [4\gamma_N^3 +2\gamma_N] , 
\end{equation}
				while its imaginary part becomes 
\begin{equation}
%			{\rm Im}(\delta_N) = -j\kappa \{ \gamma_N^2[6 -71m_1 m_2/8] 
			{\rm Im}(\delta_N) = -j\kappa \{ \gamma_N^2[6-137M_1 M_2/8] 
%			+[3 +319m_1 m_2/32] \} / [4\gamma_N^3 +2\gamma_N] . 
%			+[3 -465m_1 m_2/32] \} / [4\gamma_N^3 +2\gamma_N] . 
			+[3 -687M_1 M_2/32] \} / [4\gamma_N^3 +2\gamma_N] . 
\end{equation}

Note from Formula (62) that the real part of $\delta_N$ vanishes when the tidal time lag $\tau$ vanishes, 
as expected for the case with no dissipation.  
Note also that $4\gamma_1^2 +2 = 4\gamma_2^2 +2= -2\sqrt{1 -27M_1 M_2}$.  
Then substituting $\gamma_1$ into Formula (62) gives 
\[
%		{\rm Re}(\delta_1) = \kappa\tau \{ [1/2 +\sqrt{1 -27m_1 m_2}/2][13/2 -107m_1 m_2/8] 
%		{\rm Re}(\delta_1) = \kappa\tau \{ [1/2 +\sqrt{1 -27m_1 m_2}/2][4 -23m_1 m_2/2] 
		{\rm Re}(\delta_1) = \kappa\tau \{ [1/2 +\sqrt{1 -27M_1 M_2}/2][4-107M_1 M_2/8] 
%				-49/8 +33m_1 m_2/4 \} / [-2\sqrt{1 -27m_1 m_2}] 
%				+[3-177m_1 m_2/32] \} / [-2\sqrt{1 -27m_1 m_2}] 
				+[3 -39M_1 M_2/4 ] \} / [-2\sqrt{1 -27M_1 M_2}] 
\] \[
%	= \kappa\tau \{ -23/8 +25m_1 m_2/16 +[13/2 -107m_1 m_2/8] \sqrt{1 -27m_1 m_2}/2 \} / [-2\sqrt{1 -27m_1 m_2}]  
%	= \kappa\tau \{ [5 -361m_1 m_2/32] +[2 -23m_1 m_2/4] \sqrt{1 -27m_1 m_2} \} / [-2\sqrt{1 -27m_1 m_2}] 
	= \kappa\tau \{ [5 -263M_1 M_2/16] +[2-107M_1 M_2/16]\sqrt{1 -27M_1 M_2} \} / [-2\sqrt{1 -27M_1 M_2}] 
\] \begin{equation}
%		= \kappa\tau \{ -13/8 +107m_1 m_2/32 +[23/16 -25m_1 m_2/32]/\sqrt{1 -27m_1 m_2} \} . 
%		= \kappa\tau \{ -1 +23m_1 m_2/8 +[-5/2 +361m_1 m_2/64]/\sqrt{1 -27m_1 m_2} \} . 
		= \kappa\tau \{ -1+107M_1 M_2/32+[-5/2 +263M_1 M_2/32] /\sqrt{1 -27M_1 M_2} \} . 
\end{equation}

Likewise, substituting $\gamma_1$ into Formula (63) gives 
\[ 
%		{\rm Im}(\delta_1) = -\kappa \{ [1/2 +\sqrt{1 -27m_1 m_2}/2][6 -71m_1 m_2/8] 
		{\rm Im}(\delta_1) = \kappa \{ [-1/2 -\sqrt{1 -27M_1 M_2}/2][6-137M_1 M_2/8] 
%		-[3 +319m_1 m_2/32] \} / \sqrt{ 2[1 -27m_1 m_2][1 +\sqrt{1 -27m_1 m_2}] } 
%		+[3 -465m_1 m_2/32] \} / \sqrt{ 2[1 -27m_1 m_2][1 +\sqrt{1 -27m_1 m_2}] } 
		+[3 -687M_1 M_2/32] \} / \sqrt{ 2[1 -27M_1 M_2][1 +\sqrt{1 -27M_1 M_2}] } 
\] \[
%			= \kappa \{ 461m_1 m_2/32 -[3 -71m_1 m_2/16]\sqrt{1 -27m_1 m_2} \} 
			= \kappa \{-413M_1 M_2/32+[-3+137M_1 M_2/16]\sqrt{1 -27M_1 M_2} \} 
			/ \sqrt{ 2[1 -27M_1 M_2][1 +\sqrt{1 -27M_1 M_2}] } 
\] \begin{equation}
%	= \kappa \{ -3 +71m_1 m_2/16 +461m_1 m_2[1 -27m_1 m_2]^{-1/2}/32 \} / \sqrt{ 2 +2\sqrt{1 -27m_1 m_2} } . 
	= \kappa \{ -3+137M_1 M_2/16 -413M_1 M_2[1 -27M_1 M_2]^{-1/2}/32 \} / \sqrt{ 2 +2\sqrt{1 -27M_1 M_2} } . 
\end{equation}
Substituting $\gamma_2 = -\gamma_1$ into Formulae (62) and (63) gives just 
Re($\delta_2$) = Re($\delta_1$), but Im($\delta_2$) = --Im$(\delta_1)$.  
Thus $\delta_1$ and $\delta_2$ are complex conjugates, 
% like $\gamma_1$ and $\gamma_2$, or $\gamma_3$ and $\gamma_4$.  
and so are $\lambda_1 = \gamma_1 +\delta_1$ and $\lambda_2 = \gamma_2 +\delta_2$, as anticipated.  

Figure 3 graphs the real and imaginary parts of $\delta_1$ and $\delta_2$ 
as functions of $M_2$ from Formulae (64) and (65) above.  
Note that the vertical scale of Fig. 3 is ten times greater than that of Fig. 2, 
but that the real parts of $\delta_1$ and $\delta_2$ are normalized by $\kappa\tau$, 
while their imaginary parts are normalized by just $\kappa$.  
The imaginary parts of $\delta_1$ and $\delta_2$ are not very important, 
because they only change the epicylic frequency slightly; 
but their real parts are very important, 
because they may determine the stability or instability of the epicycles.  
Note that the real parts of $\delta_1$ and $\delta_2$ 
are negative for all $M_1 M_2 <$ 1/27 ($M_2 < \sim0.03852$), 
% are positive for $\sim0.007447 < m_2 < \sim0.03852$,  
% so that the epicycles grow exponentially 
% with an $e$-folding time of $1/[n {\rm Re}(\delta_1)]$, 
% and the Trojan is unstable.  
% For $m_2 < \sim0.007447 \approx$ 1/134.3, 
% the real parts of $\delta_1$ and $\delta_2$ are negative, 
so the epicycles decay exponentially 
with an e-folding time of $-1/[n {\rm Re}(\delta_1)]$.  

\begin{figure}
%\centerline{\psfig{figure=deltas.eps,width=3.1in,height=3.1in}}
\begin{center}
\includegraphics[width=3.1in,height=3.1in]{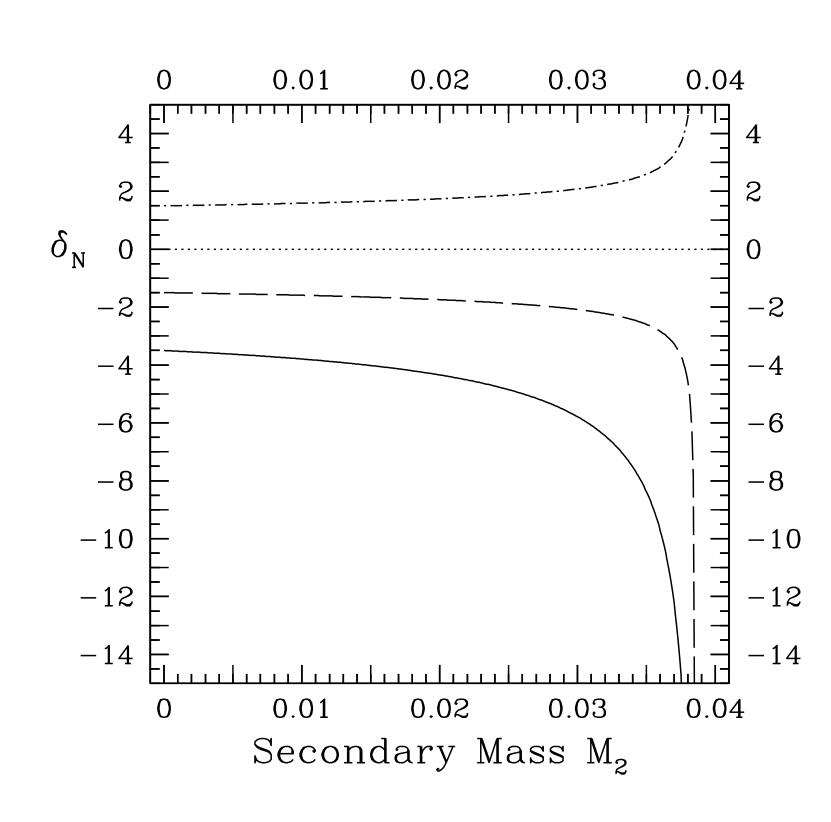}
\end{center}
\caption{ $\delta_1$ and $\delta_2$ as functions of $M_2$, the mass of the secondary.  
Solid curve:  Re($\delta_1$) = Re($\delta_2$), normalized by $\kappa\tau$.  
Dashed curve:  Im($\delta_1$), normalized by $\kappa$.  
Dot-dashed curve:  Im($\delta_2$) = --Im($\delta_1$).  
The horizontal dotted line denotes the zero level. }
\end{figure}

When $M_2$ vanishes, as in the two-body problem with tides, 
Re($\delta_1$) and Re($\delta_2$) reduce to $-7\kappa\tau/2$ from Formula (64), 
while Im($\delta_1$) = --Im($\delta_2$) reduces to $-3\kappa/2$ from Formula (65).  
The former agrees with the damping time $\tau_{\rm AL}$ 
from Formula (61) of Couturier {\it et al.} (2021), 
when $Q$ of the secondary is set to infinity; 
our Re($\delta_1$) also agreees with the eccentricity damping time $\tau_e$ 
from Formula 4.198 of Murray and Dermott (1999), where their $C_S$, $\tilde{\mu}_S$, 
and $Q_S$ are identified with our $R$, $3/(2k)$, and $1/(n\tau)$, respectively.  
Murray and Dermott (1999) also determined 
that 3/7 of the damping rate arises from radial tides, and 4/7 from transverse tides; 
however, they used an energy argument inapplicable to problems with more than two bodies.  
Note that Formulae (64) and (65) cannot be represented as any simple combination 
of two two-body problems.  

% Then the Trojan is stable provided that the long-term librations also are stable; 
% but if the long-term librations grow,  the Trojan is still unstable.  

% MENTION Lissauer, J. J., and J. E. Chambers, 2008.  
% Solar and planetary destabilization of the 
% Earth-Moon triangular Lagrangian points.  
% Icarus 195, 16--27.  

\newpage

\subsection{Librations}

The above perturbation procedure fails for $\lambda_3$ and $\lambda_4$ 
as $M_2$ approaches zero, because $\gamma_3$ and $\gamma_4$ both vanish then.  
However, now that we know two roots $\lambda_1 = \gamma_1 +\delta_1$ 
and $\lambda_2 = \gamma_2 +\delta_2$ of Eq. (57), 
we can use them to find its last two roots $\lambda_3$ and $\lambda_4$.  
The quartic Eq. (54) is of the form $0 = \prod_1^4 (\lambda -\lambda_N)$.  
In principle, we could divide Eq. (57) by its factors $(\lambda -\lambda_1)$ 
and $(\lambda -\lambda_2)$ to deflate it into a quadratic, namely 
\begin{equation}
			0 = (\lambda -\lambda_3)(\lambda -\lambda_4) 
		= \lambda^2 -\nu[\lambda_3 +\lambda_4] +\lambda_3\lambda_4 . 
\end{equation}

% Then solving Eq. (62) above by the quadratic formula recovers 
% \begin{equation}
%	\lambda = [\lambda_3 +\lambda_4]/2 \pm\sqrt{[\lambda_3 +\lambda_4]^2/4 -\lambda_3\lambda_4} 
%		= [\lambda_3 +\lambda_4]/2 \pm\sqrt{[\lambda_3 -\lambda_4]^2/4} 
%			= \lambda_3 \; {\rm or} \; \lambda_4 , 
% \end{equation}
% where now and henceforth we use the standard meaning of $\pm$.  

The coefficents of quadratic equation (66) above 
are just the sum and product of $\lambda_3$ and $\lambda_4$; 
so we can use a short-cut to find these coefficients.  
The constant term on the right-hand side of the quartic Eq. (57) 
is just $\prod_1^4 \lambda_N = \lambda_1\lambda_2\lambda_3\lambda_4$, 
while its cubic term is $-\nu^3\sum_1^4 \lambda_N$.  Thus from Eq. (57), 
the sum of all four roots $\lambda_N$ is $\kappa\tau[-4+107M_1 M_2/8]$, 
while their product is $27M_1 M_2/4 +\kappa[-3 +687M_1 M_2/32]$.  
When $\kappa$ vanishes, note from Formulae (59) how 
the roots $\gamma_N$ of the biquadratic Eq. (58) satisfy the above, 
because their sum vanishes, while their product is just $27M_1 M_2/4$.  

When $\kappa \ne 0$, since we know $\lambda_1$ and $\lambda_2$, 
and we also know $\sum_1^4 \lambda_N$ and $\prod_1^4 \lambda_N$, 
then we can find the sum and product of $\lambda_3$ and $\lambda_4$:  
\begin{equation}
                \lambda_3 +\lambda_4 = \kappa\tau[-4+107M_1 M_2/8] -[\lambda_1 +\lambda_2] , 
\end{equation}
			while 
\begin{equation}
                \lambda_3 \lambda_4 = \{ 27M_1 M_2/4 +\kappa[-3 +687M_1 M_2/32] \} / [\lambda_1 \lambda_2] . 
\end{equation}

Because $\lambda_1$ and $\lambda_2$ are complex conjugates, their sum is just $\lambda_1 +\lambda_2$ = 
Re($\lambda_1$) +Re($\lambda_2$) = Re($\delta_1$) +Re($\delta_2$) = 2 Re$(\delta_1)$ = 2 Re$(\delta_2)$, 
while their product is just the square of their magnitudes:  
\[                      \lambda_1 \lambda_2 = {\rm Re}^2(\lambda_1) +{\rm Im}^2(\lambda_1) 
                        = {\rm Re}^2(\gamma_1 +\delta_1) +{\rm Im}^2(\gamma_1 +\delta_1) 
			= [{\rm Re}(\gamma_1) +{\rm Re}(\delta_1)]^2 
			+ [{\rm Im}(\gamma_1) +{\rm Im}(\delta_1)]^2 
\] \begin{equation}
= {\rm Re}^2(\delta_1) +{\rm Im}^2(\gamma_1) +2 \; {\rm Im}(\gamma_1){\rm Im}(\delta_1) +{\rm Im}^2(\delta_1) 
                \approx {\rm Im}^2(\gamma_1) +2 \; {\rm Im}(\gamma_1){\rm Im}(\delta_1) , 
\end{equation}
%		\approx 1/2 +\sqrt{1 -27m_1 m_2}/2 +\kappa\{-3 +137m_1 m_2/16 -413m_1 m_2[1 -27m_1 m_2]^{-1/2}/32\} , 
% \end{equation}
to first degree in $\kappa$.  

Then the sum of $\lambda_3$ and $\lambda_4$ is just 
\begin{equation}
                	\lambda_3 +\lambda_4 = \kappa\tau[-4 +107M_1 M_2/8] -2 \; {\rm Re}(\delta_1) 
\end{equation} \[
	\approx \kappa\tau[-4 +107M_1 M_2/8] -\kappa\tau\{-2 +107M_1 M_2/16 +[-5 +263M_1 M_2/16]/\sqrt{1 -27M_1 M_2}\} 
\] \[
		= \kappa\tau\{-2 +107M_1 M_2/16 +[5 -263M_1 M_2/16]/\sqrt{1 -27M_1 M_2} \} , 
\]
while their product is 
\begin{equation}
                	\lambda_3 \lambda_4 \approx \frac{27M_1 M_2/4 +\kappa[-3 +687M_1 M_2/32]}
			{{\rm Im}^2(\gamma_1) +2 \; {\rm Im}(\gamma_1){\rm Im}(\delta_1)} 
\end{equation} \[
				\approx \{27M_1 M_2/4 +\kappa[-3 +687M_1 M_2/32]\} 
		[{\rm Im}^2(\gamma_1) -2 \; {\rm Im}(\gamma_1){\rm Im}(\delta_1)]/{\rm Im}^4(\gamma_1) 
\] \[
		= [1 +\sqrt{1 -27M_1 M_2} -\kappa\{-6 +137M_1 M_2/8 -413M_1 M_2[1 -27M_1 M_2]^{-1/2}/16\}]
\] \[
		\{ 27M_1 M_2/4 +\kappa[-3 +687M_1 M_2/32] \}/[1 -27M_1 M_2/2 +\sqrt{1 -27M_1 M_2}]
\] \[
				\approx \{ 27M_1 M_2[1 +\sqrt{1 -27M_1 M_2}]/4 
		+\kappa [27M_1 M_2 \{ 6 -137M_1 M_2/8 -413M_1 M_2[1 -27M_1 M_2]^{-1/2}/16 \}/4] 
\] \[				+\kappa[-3 +687M_1 M_2/32][1 +\sqrt{1 -27M_1 M_2}] 
				\} /[1 -27M_1 M_2/2 +\sqrt{1 -27M_1 M_2}] , 
\]
again to first degree in $\kappa$.  

Note that $\lambda_3 +\lambda_4$ from Formula (70) is proportional to $\kappa\tau$, 
while $\lambda_3 \lambda_4$ from Formula (71) above is linear in $\kappa$.  
Thus to first degree in $\kappa$, Formula (66) reduces to just
\begin{equation}
			\lambda \approx [\lambda_3 +\lambda_4]/2 \pm\sqrt{-\lambda_3 \lambda_4} 
			\equiv \kappa\tau\Delta \pm\sqrt{\gamma_3^2 +\kappa\Gamma} , 
\end{equation}
where 
\begin{equation}
			\Delta = -1 +107M_1 M_2/32 +[5/2 -263M_1 M_2/32]/\sqrt{1 -27M_1 M_2} 
\end{equation}
and
\begin{equation}
		\Gamma = ( 27M_1 M_2 \{-6 +137M_1 M_2/8 +413M_1 M_2[1 -27M_1 M_2]^{-1/2}/16 \}/4 
\end{equation} \[
		+[3 -687M_1 M_2/32][1 +\sqrt{1 -27M_1 M_2}] )/(1 -27M_1 M_2/2 +\sqrt{1 -27M_1 M_2}) . 
\]
% and
% \begin{equation}
%		\beta = \{ 27m_1 m_2[1 +\sqrt{1 -27m_1 m_2}]/4 \}/[1 -27m_1 m_2/2 +\sqrt{1 -27m_1 m_2}] . 
% \end{equation}

Figure 4 graphs $\Gamma$ and $\Delta$ as functions of $M_2$.  
Note how $\Delta$ reduces to 3/2 for the two-body case $M_2$ = 0, 
but increases monotonically with $M_2$, until it becomes infinite 
at $M_1 M_2$ = 1/27 $\approx$ 0.037 037 ($M_2 \approx$ 0.03852).  
In contrast, $\Gamma$ reduces to 3 for $M_2$ = 0, 
decreases to a shallow minimum of $\sim$2.5215 at $M_2 \approx$ 0.0278, 
and then increases again to infinity at $M_1 M_2$ = 1/27.  

\begin{figure}
%\centerline{\psfig{figure=kappaC.eps,width=3.1in,height=3.1in}}
\begin{center}
\includegraphics[width=3.1in,height=3.1in]{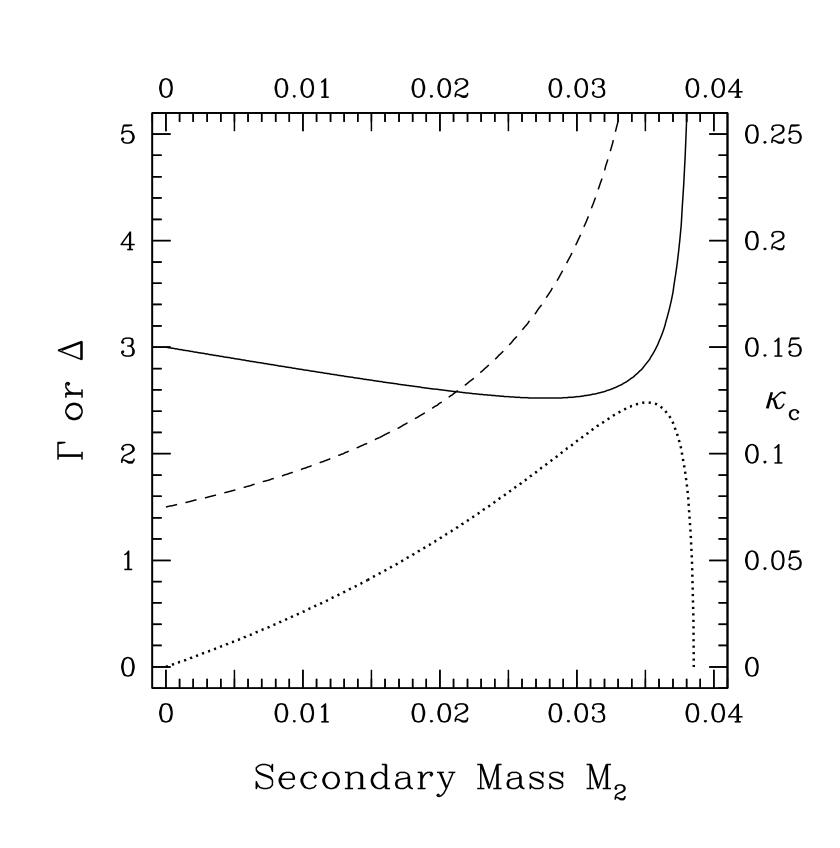}
\end{center}
\caption{ Left-hand scale: $\Gamma$ (solid curve) and $\Delta$ (dashed curve) 
as functions of $M_2$, the mass of the secondary. 
Right-hand scale: $\kappa_c$ (dotted curve) as a function of $M_2$. }
\end{figure}

Because $\gamma_3^2$ is negative, while $\Gamma$ and $\Delta$ both are positive, 
$\gamma_3^2$ represents the stabilizing influence of the secondary mass $M_2$ 
in Formula (72) for $\lambda$, while $\kappa\Gamma$ and $\kappa\tau\Delta$ represent 
the destabilizing influences of the tidal force and of tidal dissipation, respectively.  

Note that $\sqrt{\gamma_3^2 +\kappa\Gamma}$ in Formula (72) vanishes 
when $\kappa$ reaches a critical value $\kappa_c \equiv -\gamma_3^2/\Gamma$.  
Figure 4 also graphs $\kappa_c$ as a function of $M_2$, 
on a vertical scale 20 times smaller than $\Gamma$ and $\Delta$.  
Note how $\kappa_c$ rises from zero at $M_2$ = 0 
to a peak of $\sim$0.1241 at $M_2 \approx$ 0.0351, 
but then falls to zero again at $M_1 M_2$ = 1/27.  

For $\kappa > \kappa_c$, the square root in Formula (72) is positive, 
and the Lagrange points L$'4$ and L$'5$ are unstable.  
% For example, in the two-body limit when the secondary mass $M_2$ is negligible, 
% $\gamma_3$ also vanishes, and Formula (72) reduces to $\nu = 3\kappa\tau/2 \pm\sqrt{3\kappa}$.  
% Then L$'4$ and L$'5$ become unstable with a relatively short e-folding time of $\sim1/\sqrt{3\kappa}$.  
For $\kappa < \kappa_c$, the square root in Formula (72) becomes imaginary, 
and $\lambda$ represents a slowly-growing libration 
with a period of $2\pi/\sqrt{-\gamma_3^2 -\kappa\Gamma}$ 
and an e-folding time of $1/(\kappa\tau\Delta)$.  
% Then L$'4$ and L$'5$ remain unstable, but with a long lifetime, 
% inversely proportional to both $\kappa$ and $\tau$.  
When tides from the secondary are neglected, $\Delta$ reduces to 3/2, 
so the e-folding time of the libration becomes $2/(3\kappa\tau)$.  
This agrees with the e-folding time $\tau_{\rm lib} = 7\tau_{\rm AL}/3$ 
from Formula (61) of Couturier {\it et al.} (2021), 
when tides in the secondary are neglected.

\subsection{Complete solution}

The general solution to System (23) and (24) is of the form
\begin{equation}
x = x' +\alpha_1\exp(\lambda_1 t) +\alpha_2\exp(\lambda_2 t) +\alpha_3\exp(\lambda_3 t) +\alpha_4\exp(\lambda_4 t) , 
\end{equation}
\begin{equation}
y = y' +\beta_1\exp(\lambda_1 t) +\beta_2\exp(\lambda_2 t) +\beta_3\exp(\lambda_3 t) +\beta_4\exp(\lambda_4 t) , 
\end{equation}
where the constant coefficients $\alpha_N$ and $\beta_N$ are complex amplitudes.  

To evaluate these constants, first differentiate solutions (75) and (76) above:  
\begin{equation}
		\dot{x} = \alpha_1\lambda_1\exp(\lambda_1 t) +\alpha_2\lambda_2\exp(\lambda_2 t) 
			 +\alpha_3\lambda_3\exp(\lambda_3 t) +\alpha_4\lambda_4\exp(\lambda_4 t) , 
\end{equation}
\begin{equation}
		\dot{y} =  \beta_1\lambda_1\exp(\lambda_1 t) + \beta_2\lambda_2\exp(\lambda_2 t) 
			 + \beta_3\lambda_3\exp(\lambda_3 t) + \beta_4\lambda_4\exp(\lambda_4 t) . 
\end{equation}
Then setting Formulae (75) through (78) above respectively equal to the initial conditions 
$x = x' +x_0$, $y = y' +y_0$, $\dot{x} = \dot{x}_0$, and $\dot{y} = \dot{y}_0$ at time $t$ = 0 
gives $\sum\alpha_N = x_0$, $\sum\beta_N = y_0$, $\sum\alpha_N\lambda_N = \dot{x}_0$, 
and $\sum\beta_N\lambda_N = \dot{y}_0$, as in Formulae (3.127) of Murray and Dermott (1999).  

This appears to leave us with four linear equations in eight unknowns, 
but substituting Formulae (75) through (78) 
(along with their next derivatives $\ddot{x}$ and $\ddot{y}$) into Eqs. (23) and (24) 
gives two relations between $\alpha_N$ and $\beta_N$ for each $N$:  
\begin{equation}
	\beta_N/\alpha_N = \frac{\lambda_N^2 -3/4 -\kappa[1-9M_1 M_2/8] -\kappa\tau[93M_1 M_2/16 -3/2]\lambda_N}
	{2\lambda_N \pm3\sqrt{3}(M_1-M_2)/4 \pm\kappa\sqrt{3}(M_1-M_2)(2-\lambda_N\tau/2)} \equiv q_N 
\end{equation}
and
\begin{equation}
	\beta_N/\alpha_N = \frac{-2\lambda_N \pm3\sqrt{3}(M_1-M_2)/4 \pm\kappa\sqrt{3}(M_1-M_2)(2-\lambda_N\tau/2)}
	{\lambda_N^2 -9/4 -\kappa(5-16M_1 M_2) -\kappa(121M_1 M_2 -5/2)\lambda_N\tau} \equiv q'_N .  
\end{equation}
Actually the complex quotients $q_N$ and $q'_N$ from Formulae (79) and (80) above are equal by Eqs. (56) and (57).  

Then the final system for the initial conditions takes the pleasing form 
\begin{equation}
					\left[ \begin{array}{cccc}

	1		&	1		&	1		&	1		\\

	\lambda_1       &       \lambda_2       &       \lambda_3       &       \lambda_4       \\

	q_1		&	q_2		&	q_3		&	q_4		\\

	q_1\lambda_1	&	q_2\lambda_2	&	q_3\lambda_3	&	q_4\lambda_4	\\

					\end{array}\right] 
	\bullet \left[ \begin{array}{c} \alpha_1 \\ \alpha_2 \\ \alpha_3 \\ \alpha_4 \end{array} \right]
		= \left[ \begin{array}{c} x_0 \\ \dot{x}_0 \\ y_0 \\ \dot{y}_0 \\ \end{array} \right] . 
\end{equation}
System (81) above can be solved numerically for any set of initial conditions 
$x_0$, $y_0$, $\dot{x}_0$, $\dot{y}_0$, and parameters $M_2$, $\kappa$, $\tau$.  

Furthermore, it is possible to express $x$ and $y$ entirely in terms of real quantities. 
Note that $\lambda_1$ and $\lambda_2$, $q_1$ and $q_2$, $\alpha_1$ and $\alpha_2$, 
$\beta_1$ and $\beta_2$ all are complex conjugate pairs; 
likewise, $\lambda_3$ and $\lambda_4$, $q_3$ and $q_4$, $\alpha_3$ and $\alpha_4$, 
$\beta_3$ and $\beta_4$ all are complex conjugate pairs as well.  
Thus $\alpha_1 = A_1 +ja_1$, $\alpha_2 = A_1 -ja_1$, $\alpha_3 = A_3 +ja_3$, 
and $\alpha_4 = A_3 -a_3$, where $A_1$ = Re($\alpha_1$), $a_1$ = Im($\alpha_1$), 
$A_3$ = Re($\alpha_3$), and $a_3$ = Im($\alpha_3$).  Similarly, we write $\beta_1 = B_1 +jb_1$, 
$\beta_2 = B_1 -jb_1$, $\beta_3 = B_3 +jb_3$, and $\beta_4 = B_3 -jb_3$, where 
$B_1$ = Re($\beta_1$), $b_1$ = Im($\beta_1$), $B_3$ = Re($\beta_3$), and $b_3$ = Im($\beta_3$).  

Finally, $\lambda_1 = \Lambda_1 +j\ell_1$, $\lambda_2 = \Lambda_1 -j\ell_1$,
$\lambda_3 = \Lambda_3 +j\ell_3$, and $\lambda_4 = \Lambda_3 -j\ell_3$.  
Here $\Lambda_1$ = Re($\lambda_1$) = Re($\delta_1$) is the decay rate of $\mu$'s eccentric epicyles, 
while $\ell_1$ = Im($\lambda_1$) = Im($\gamma_1 +\delta_1$) is their angular frequency.  Likewise, 
$\Lambda_3$ = Re($\lambda_3$) = $\kappa\tau\Delta$ is the growth rate of $\mu$'s tadpole librations, 
while $\ell_3$ = Im($\lambda_3$) = $\sqrt{-\gamma_3^2 -\kappa\Gamma}$ is their angular frequency.  

Formulae (75) and (76) can be expressed in terms of real quantities as 
\[
		x = x'	+\exp(\Lambda_1 t)[(\alpha_1+\alpha_2)\cos(\ell_1 t) +(\alpha_1-\alpha_2)j\sin(\ell_1 t)] 
			+\exp(\Lambda_3 t)[(\alpha_3+\alpha_4)\cos(\ell_3 t) +(\alpha_3-\alpha_4)j\sin(\ell_3 t)]
\] \begin{equation}
			= x' +\exp(\Lambda_1 t)[2A_1\cos(\ell_1 t) -2a_1\sin(\ell_1 t)]
			     +\exp(\Lambda_3 t)[2A_3\cos(\ell_3 t) -2a_3\sin(\ell_3 t)] 
\end{equation}
		and
\[
		y = y'	+\exp(\Lambda_1 t)[(\beta_1+\beta_2)\cos(\ell_1 t) +(\beta_1-\beta_2)j\sin(\ell_1 t)] 
			+\exp(\Lambda_3 t)[(\beta_3+\beta_4)\cos(\ell_3 t) +(\beta_3-\beta_4)j\sin(\ell_3 t)]
\] \begin{equation}
			= y' +\exp(\Lambda_1 t)[2B_1\cos(\ell_1 t) -2b_1\sin(\ell_1 t)]
			     +\exp(\Lambda_3 t)[2B_3\cos(\ell_3 t) -2b_3\sin(\ell_3 t)] . 
\end{equation}

% For example, Case 0 in Table 4 lists the values of $\lambda_N$, $q_N$, 
% $\alpha_N$, and $\beta_N$ given by Murray and Dermott (1999, p. 92) for the case 
% $m_2$ = .01 without tides, when $\kappa$ vanishes and $\tau$ is irrelevant, 
% for $x_0 = y_0 = 10^{-5}$ and $\dot{x}_0 = \dot{y}_0 = 0$ (in our notation).  
% For comparison, Case 1 gives the corresponding values from this work 
% (as well as $t_z$ and $\zeta$) for the same parameters; 
% note that the values agree to the given precision.  
% In both cases, only values for $N$ = 1 and $N$ = 3 are given, 
% because the values for $N$ = 2 are just the complex conjugates of those for $N$ = 1, 
% while the values for $N$ = 4 are just the complex conjugates of those for $N$ = 3.  

% 1.8586799848897801 2.7891275858275084

For example, Table 4 lists the numerical results for the above quantities 
for several different parameter sets, but all with a fairly heavy 
secondary mass $M_2$ = 0.01, comparable to the Moon/Earth mass ratio.  
Then $\Delta \approx$ 1.858 680, 
$\Gamma \approx$ 2.789 128, $\kappa_c \approx$ 0.025 818 3, 
and the secondary's dimensionless Hill radius is $(M_2/3)^{1/3} \approx$ 0.149 380 . 
Furthermore, we chose initial conditions $x_0 = y_0 = 10^{-5}$ 
and $\dot{x}_0 = \dot{y}_0 = 0$ in each case.  

Case 1 is the standard CR3BP without tides, 
when $\kappa$ vanishes and $\tau$ is irrelevant.  
Murray and Dermott (1999, p. 92) did the same case; 
for comparison, Table 4 lists the values they reported as ``Case 0''.  
% for comparison, Case 0 lists the values they reported 
% for the two angular frequencies $\ell_1$ and $\ell_3$, 
% both e-folding rates $\Lambda_1$ and $\Lambda_3$, and the eight real coefficients 
% $2A_1$, $-2a_1$, $2A_3$, $-2a_3$, $2B_1$, $-2b_1$, $2B_3$, and $-2b_3$.  
Note that Cases 0 and 1 agree to the given precision.  

Case 2 in Table 4 lists the results for the same parameters, 
except now with tides of large dimensionless magnitude $\kappa$ = 0.000 1, 
but still no dissipation.  Case 3 lists the results for tides of the same magnitude, 
but now with weak dissipation, for a dimensionless time lag $\tau$ = 0.01 
($Q \approx$ 100), appropriate for a ``dry'' planet like the Moon or Mars.  
Case 4 lists the results for tides of the same magnitude again, 
but now with strong dissipation, for a dimensionless time lag $\tau$ = 0.10 
($Q \approx$ 10), appropriate for a ``wet'' planet like the Earth.  
In each of these cases, the tabulated $\lambda_N$ found by perturbing $\gamma_N$  
agree with those found more directly by solving the quartic equation (57) numerically.  

The e-folding rates $\zeta$ and $\Lambda_N$ all vanish in Cases 1 and 2 
without dissipation, so $m_o$ maintains quasi-periodic vertical oscillations, 
eccentric epicycles, and tadpole librations indefinitely.  In Case 3 with weak dissipation, 
the vertical oscillations decay slowly with an e-folding time of $\sim$320 000 orbital periods, 
while the epicycles decay quickly with an e-folding time of only $\sim$42 000 orbits; however, 
the tadpole librations grow, with an intermediate e-folding time of $\sim$86 000 orbits.  
In Case 4 with strong dissipation, the same things happen, but ten times faster.  
Thus Cases 3 and 4 with tidal dissipation both are unstable.  

% 0.26834774854251275       0.96332210908509952        1.0004862692710979

\begin{center}
			Table 4.  Numerical results.  In all cases, $M_2$ = 0.01, 
			$x_0 = y_0 = 10^{-5}$, and $\dot{x}_0 = \dot{y}_0 = 0$.  
\vspace*{0.1in}

\begin{tabular}{|c||c||c|c|c|c|}

\hline
		Case		&	0	&	1	&	2	&	3	&	4	\\
\hline
		$\kappa$	&	0	&	0	&	0.000 1	&	0.000 1	&	0.000 1	\\
		$\tau$		&	0    	&	0	&	0	&	0.01	&	0.10	\\
		$Q$		&   $\infty$	&   $\infty$	&   $\infty$	&   $\sim$100	&   $\sim$10	\\
\hline
\hline
		$1/(2\pi\zeta)$&	.	&    $\infty$	&    $\infty$	&   -321 493.	&    -32 149.3	\\
		$\zeta$		&	.	&	0	&	0	&-0.000 000 495 050&-0.000 004 950 50	\\
		$\eta$		&	.	&	1	&1.000 048 637 6&1.000 048 637 6&1.000 048 637 6\\
\hline
\hline
                $x'$            &       .       &       0       &0.000 044 916 7&0.000 044 916 7&0.000 044 916 7\\
                $y'$            &       .       &       0       &0.000 012 028 1&0.000 012 028 1&0.000 012 028 1\\
\hline
\hline
		$\delta_1$	&	.	&	0	&	&-0.000 003 792 47	& -0.000 037 924 7	\\
				&		&	& -0.000 159 059 $j$ & -0.000 159 059 $j$ & -0.000 159 059 $j$	\\
\hline
	$1/(2\pi\Lambda_1)$	&	.	&   $\infty$	&   $\infty$	&   -41 966.0	&   -4 196.60	\\
		$\Lambda_1$	&	.	&	0	&	0	&-0.000 003 792 47 & -0.000 037 924 7	\\
		$\ell_1$	&	0.963	&   0.963 322	&   0.963 163	&   0.963 163	&   0.963 163	\\
\hline
	$1/(2\pi\Lambda_3)$	&	.	&   $\infty$	&   $\infty$	&    85 627.9	&    8 562.79	\\
		$\Lambda_3$	&	.	&	0	&	0	& 0.000 001 858 68 &  0.000 018 586 8	\\
		$\ell_3$	&	0.268	&   0.268 348	&   0.267 828	&   0.267 828	&   0.267 828	\\
\hline
\hline
        	$q_1$		&	.	& -0.400 586	& -0.400 670	& -0.400 672	& -0.400 685	\\
				&		&+0.606 246 $j$	&+0.606 112 $j$	&+0.606 113 $j$	&+0.606 121 $j$	\\
\hline
		$q_3$		&	.	& -0.548 257	& -0.548 355	& -0.548 353	& -0.548 340	\\
				&		&+0.231 134 $j$	&+0.230 666 $j$	&+0.230 665 $j$	&+0.230 663 $j$	\\
\hline
\hline
		$q_1\lambda_1$	&	.	& -0.584 010	& -0.583 785	& -0.583 784	& -0.583 778	\\
				&		&-0.385 893 $j$	&-0.385 911 $j$	&-0.385 914 $j$	&-0.385 948 $j$	\\
\hline
		$q_3\lambda_3$	&	.	& -0.062 024 3	& -0.061 778 6	& -0.061 779 6	& -0.061 788 1	\\
				&		&-0.147 123 $j$	&-0.146 865 $j$	&-0.146 864 $j$	&-0.146 856 $j$	\\
\hline
\hline
		$10^5\alpha_1$	&	.	&  -1.223 77	&  -1.223 92	&  -1.223 89	&  -1.223 35	\\
				&		& +4.272 53 $j$	& +4.274 37 $j$	& +4.274 44 $j$	& +4.274 47 $j$	\\
\hline
		$10^5\alpha_3$	&	.	&   1.723 77	&   1.723 92	&   1.723 89	&   1.723 35	\\
				&		& -15.337 6 $j$	& -15.371 5 $j$	& -15.371 7 $j$	& -15.371 6 $j$	\\
\hline
\hline
		$10^5\beta_1$	&	.	&  -2.099 98	&  -2.100 36	&  -2.100 42	&  -2.100 67	\\
				&		& -2.453 42 $j$	& -2.454 44 $j$	& -2.454 46 $j$	& -2.454 21 $j$	\\
\hline
		$10^5\beta_3$	&	.	&   2.599 97	&   2.600 36	&   2.600 41	&   2.600 68	\\
				&		& +8.807 37 $j$	& +8.826 69 $j$	& +8.826 76 $j$	& +8.826 38 $j$	\\
\hline
\hline
	 $2A_1\times10^5$	&	-2.45	&   -2.447 54	&   -2.447 84	&  -2.447 78	&  -2.446 70	\\
	$-2a_1\times10^5$	&	-8.55	&   -8.545 06	&   -8.548 74	&  -8.548 88	&  -8.548 94	\\
\hline
	 $2A_3\times10^5$	&	3.45	&   3.447 54	&    3.447 84	&   3.447 78	&   3.446 70	\\
	$-2a_3\times10^5$	&	30.7	&   30.675 2	&    30.743 0	&   30.743 4	&   30.743 2	\\
\hline
\hline
	 $2B_1\times10^5$	&	-4.20	&   -4.199 96	&   -4.200 72	&  -4.200 84	&  -4.201 34	\\
	$-2b_1\times10^5$	&	 4.90	&    4.906 84	&    4.908 88	&   4.908 92	&   4.908 42	\\
\hline
	 $2B_3\times10^5$	&	 5.20	&    5.199 94	&    5.200 72	&   5.200 82	&   5.201 36	\\
	$-2b_3\times10^5$	&	-17.6	&  -17.614 74	&  -17.653 38	& -17.653 52	& -17.652 76	\\
\hline

\end{tabular}

\end{center}

To illustrate, the loopy trajectory in Fig. 5 plots our analytic solution for Case 1 
during the first 12.5 orbital periods $P$ of $M_2$ around $M_1$.  
Compare Fig. 5 with Fig. 3.15 of Murray and Dermott (1999, p. 96) for Case 0; 
they agree in essentially every detail.  For this length of time,
the solutions for Cases 2, 3, and 4 with tides 
are almost indistinguishable from Case 1 without tides; 
slight differences on the order of $10^{-5}$ or less 
appear to be due mainly to higher-degree terms in $\kappa$.  

\begin{figure}
%\centerline{\psfig{figure=compare.eps,width=6.0in,height=6.0in}}
\includegraphics[width=6.0in,height=6.0in]{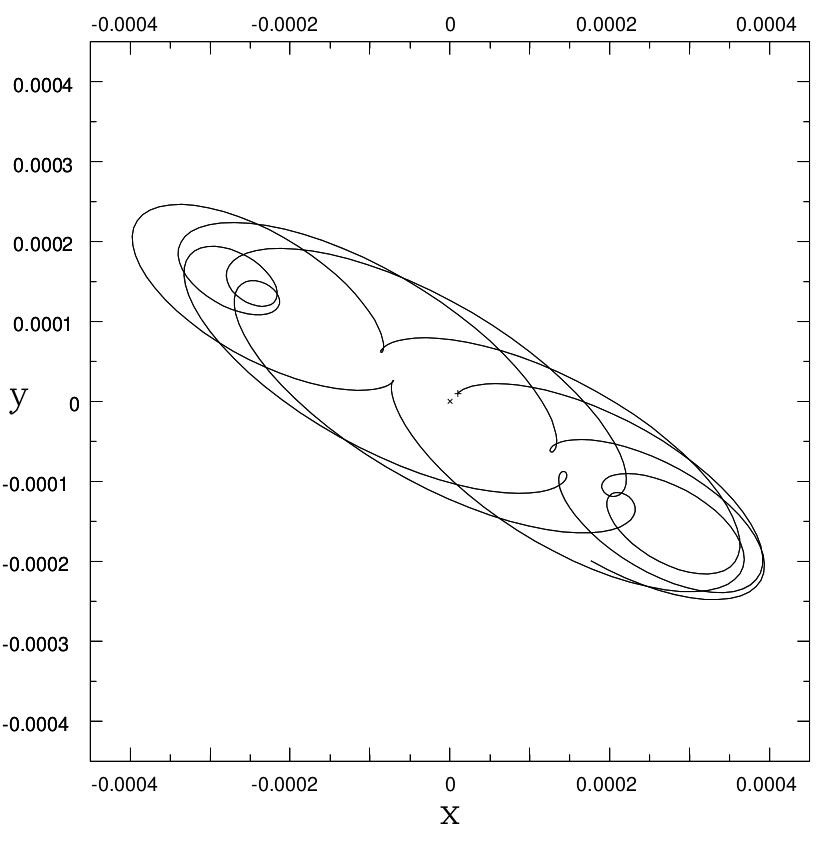}
\caption{ Our analytic solution for Case 1 without tides 
during the first 12.5 orbital periods of the secondary.  
$x$ and $y$ are the horizontal deviations 
from the leading equilateral Lagrange point L4.  
The $\times$ marks the origin, at L4, 
while the + marks the intial location of $m_o$.  
Compare with Fig. 3.15 of Murray and Dermott (1999, p. 96).  }
\end{figure}

\newpage

\section{Numerical Simulations of the Non-Linearized Problem }

In order to overcome some limitations of our analytic approach 
(particularly the assumption of small $x$ and $y$), 
we numerically integrated the non-linearized System (1) through (3) 
(with Formulae 104 through 107 from the Appendix for the tidal forces) 
for Cases 1 through 4 of Table 4.  
In each case, we neglected the obliquity $\epsilon$ of $m_o$, for simplicity; 
and we set $z_0 = 10^{-6}$ and $\dot{z}_0$ = 0, 
just to verify the stability of the vertical oscillations; 
these oscillations are too small to test their coupling with the horizontal motion.  
% Furthermore, we neglected the obliquity $\epsilon$ of $m_o$, for simplicity.  

We used the Bulirsch-Stoer integrator ``bsstep'' (Press {\it et al.}, 1992) 
for each simulation, with a tolerance ``eps'' of $10^{-9}$ 
and an initial stepsize of 1/40 of the orbit period $P$ of $M_2$ 
($9^\circ$ of its mean anomaly).  We ran each simulation 
for one million orbits of $M_2$, or until a close encounter occurred.  

Our numerical simulation for the first 12.5 periods of Case 1 (no tides) 
is plotted in Fig. 5, on top of our analytic solution.  
On this scale, the two solutions are indistinguishable.  As a further check, 
we computed the Jacobi constant $C_J$ from Formula (9) at each time step; 
its initial value of 3.000 000 000 555 361 was conserved 
up to the final digit for the entire million periods.  

Panel A of Fig. 6 plots our numerical simulation of Case 1 
during the first 100 orbital periods $P$ of $M_2$, 
while Panel B plots it for the final 100 $P$ (periods 999 900 through 1 000 000).  
Although Panels A and B differ in details, the general character 
and boundary of the trajectory have not changed over a million periods.  

\begin{figure}
%\centerline{\psfig{figure=case12.eps,width=6.0in,height=6.0in}}
\includegraphics[width=6.0in,height=6.0in]{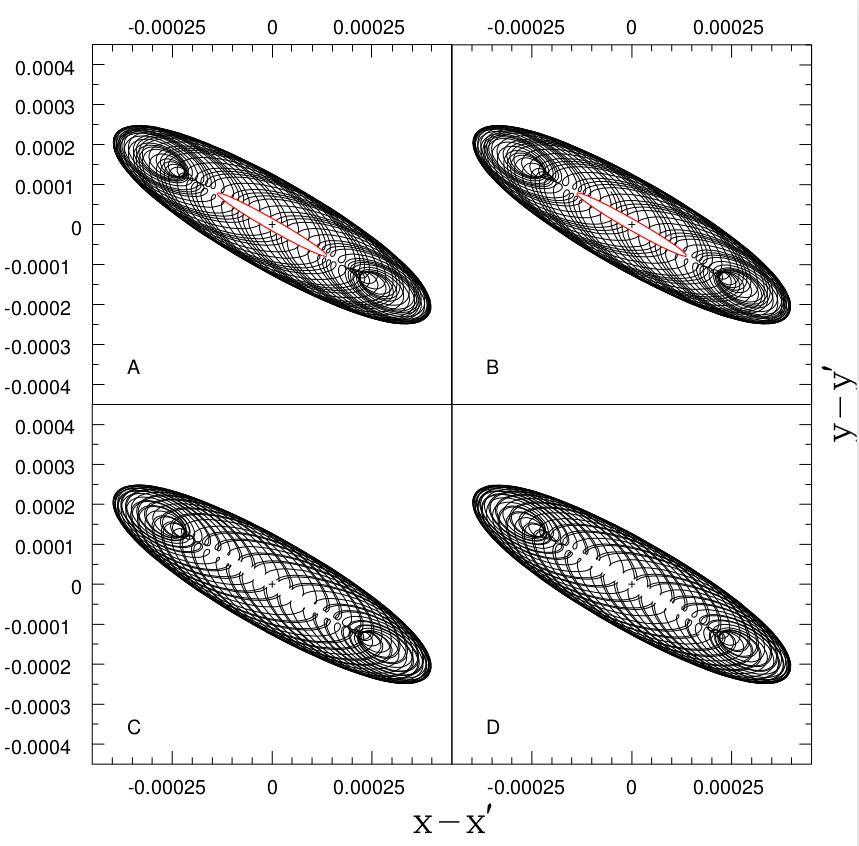}
\caption{ Panel A:  Numerical simulation of Case 1 
during the first 100 orbital periods $P$ of the secondary.  
Panel B:  Numerical solution for Case 1 during the final 100 $P$.  
The plus signs in both A and B mark the origin, at L4, 
while the red ellipses plot the zero-velocity curves 
corresponding to the Jacobi constant $C_J$. 
Panel C:  Numerical solution of Case 2 during the first 100 $P$.  
Panel D:  Numerical solution of Case 2 during the final 100 $P$.  
$x -x'$ and $y -y'$ are the horizontal deviations 
from the leading equilateral Lagrange point L$'4$.  
The plus signs in both C and D mark the origin, at L$'4$.  }
\end{figure}

In both Panels A and B of Fig. 6, the red ellipses with semi-major axis 
0.000 157 200 8, semi-minor axis 0.000 013 647 8, and slope --29\dotdeg748 130 graph 
an analytic approximation to the zero-velocity curve $C_J = 2U +X^2 +Y^2 +M_1 M_2$, 
defined implicitly by setting $\dot{X} = \dot{Y} = \dot{Z} = 0$ 
in Formula (9) for the Jacobi constant.  Note that 
$\mu$ can never cross a zero-velocity curve, or its speed would become imaginary!  
Note also how the trajectory in Panels A and B never crosses the red ellipses, 
although it ``bounces'' off of them at times, when its speed momentarily vanishes.  

For comparison, Panel C of Fig. 6 plots our numerical simulation of Case 2 
(dissipationless tides) 
during the first 100 $P$, while Panel D plots it for the final 100 $P$. 
Again, Panels C and D differ in details, but the general character and 
boundary of the trajectory still do not change over a million periods, 
because in this case the tides are conservative ($\tau$ = 0); 
even though the Jacobi constant $C_J$ and corresponding zero-velocity curve 
do not exist, at least not in the simple form given by Formula (9).  
The main effect of conservative tides is to shift the equilibria 
slightly away from the equilateral points L4 and L5, to L$'4$ and L$'5$.  

Figure 7 plots our numerical simulation of Case 3 with weak tidal dissipation 
($\tau$ = 0.01) in six stages.  Panel A plots the first 100 $P$, 
in a format like Fig. 6; and in fact the results look very similar.  
Panel B plots periods  20 000 through  20 100, on the same scale as Panel A; 
note how the eccentric epicyles have damped considerably, 
as expected, while the tadpole librations have grown.  
Panel C plots periods 200 000 through 200 100, on a scale ten times larger, while 
Panel D plots periods 400 000 through 400 100, on a scale ten times larger still; 
in both cases, the eccentric epicycles have damped away to insignificance, 
while the tadpole librations have grown exponentially.  

\begin{figure}
%\centerline{\psfig{figure=case3rev.eps,width=6.0in,height=6.0in}}
\includegraphics[width=6.0in,height=6.0in]{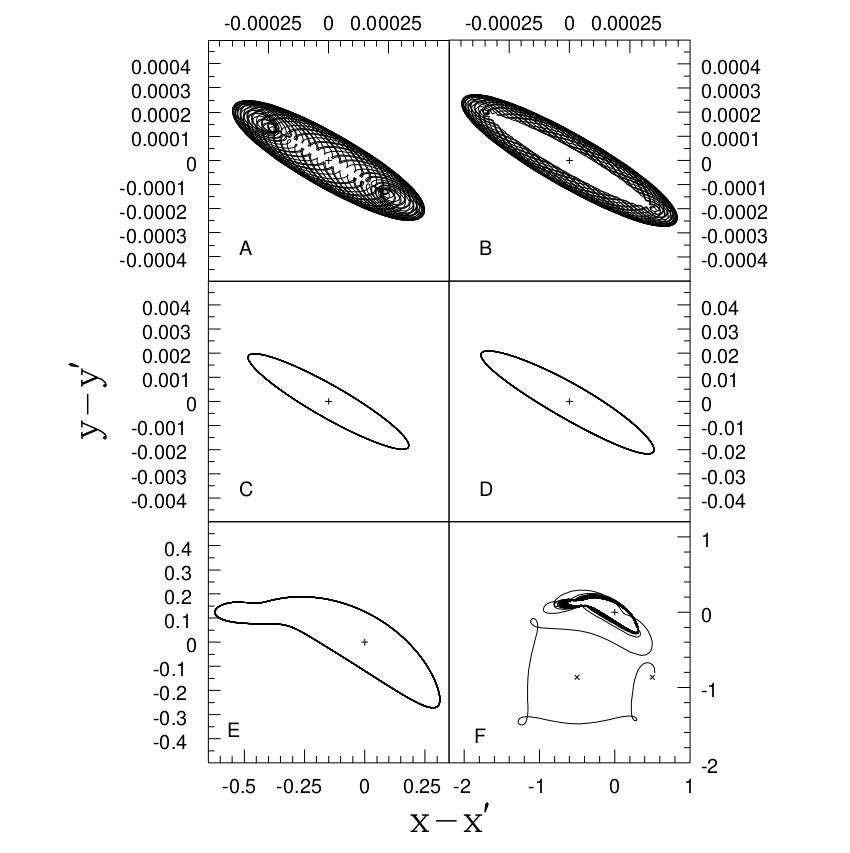}
\caption{ Numerical simulation of Case 3 (weak tidal dissipation).  
Panel A:  The first 100 orbital periods $P$ of $M_2$.  
Panel B:  $t$ =  20 000 $P$ through  20 100 $P$.  
Panel C:  $t$ = 200 000 $P$ through 200 100 $P$.  
Panel D:  $t$ = 400 000 $P$ through 400 100 $P$.  
Panel E:  $t$ = 600 000 $P$ through 600 100 $P$.  
Panel F:  $t$ = 600 100 $P$ through 602 999 $P$.  
$x -x'$ and $y -y'$ are the horizontal deviations 
from the leading equilateral Lagrange point L$'4$.  
Note different scales in each panel.  In Panel F, 
the $\times$ on the left-hand side marks the location of $M_1$, 
while the $\times$ on the right marks the location of $M_2$.  
The plus signs in each panel mark the origin, at L$'4$.  } 
\end{figure}

% Panel D of Fig. 7 plots our numerical simulation 
% of Case 3 for $t$ = 610 000 $P$ through 615 000 $P$.  
% Note that the scale of Panel D is now 100 times larger than in Panel C, 
% and that the primary mass $m_1$ resides at $x-x' \approx$ -0.500 044 916 7, 
% $y-y' \approx$ -0.866 037 431 9, marked by the cross on the left, 
% while the secondary mass $m_2$ resides at $x-x' \approx$ 0.499 955 083 3, 
% $y-y' \approx$ -0.866 037 431 9, marked by the cross on the right.   

Panel E of Fig. 7 plots our numerical simulation 
of Case 3 for $t$ = 600 000 $P$ through 600 100 $P$, 
on a shifted scale again ten times larger than Panel D.  
By this time, the tadpole librations have grown by another order of magnitude,
and taken on a peculiar shape which resembles a banana more than a polliwog.
Finally, Panel F plots periods 600 100 through 602 999, 
on a scale about three times larger than Panel E.  
Note that the primary mass $M_1$ resides at $x-x' \approx$ --0.500 044 916 7, 
$y-y' \approx$ -0.866 037 431 9, marked by the $\times$ on the left, 
while the secondary mass $M_2$ resides at $x-x' \approx$ 0.499 955 083 3, 
$y-y' \approx$ --0.866 037 431 9, marked by the $\times$ on the right.   
At $t \approx$ 602 996 $P$, $m_o$ escapes the banana orbit and assumes an eccentric orbit inferior to $M_2$.  
(Recall that $M_2$ = 0.01, so horseshoe orbits are unstable even if dissipation can be neglected; 
Cuk {\it et al.}, 2012.)  Only about three periods later, at $t \approx$ 602 999 $P$, 
$m_o$ has a close encounter with the secondary $M_2$ ($r_2 \approx$ 0.000 861 403; 
the final time-step is not plotted), and the integrator stops.  

For comparison, Fig. 8 plots our numerical simulation of Case 4 with strong tidal dissipation
($\tau$ = 0.10) in the same format as Fig. 7, and indeed the results do not look very different.  
Panels A through E look very similar to those of Fig. 7, 
except that everthing happens about ten times faster; 
Panel A plots periods 0 through 100, while 
Panel B plots periods  2 000 through  2 100, 
Panel C plots periods 20 000 through 20 100, 
Panel D plots periods 40 000 through 40 100, and 
Panel E plots periods 60 000 through 60 100.  By $t$ = 60 000 $P$, 
the tadpole librations have taken the peculiar banana-like shape previously seen in Fig. 7.  

\begin{figure}
%\centerline{\psfig{figure=case4rev.eps,width=6.0in,height=6.0in}}
\includegraphics[width=6.0in,height=6.0in]{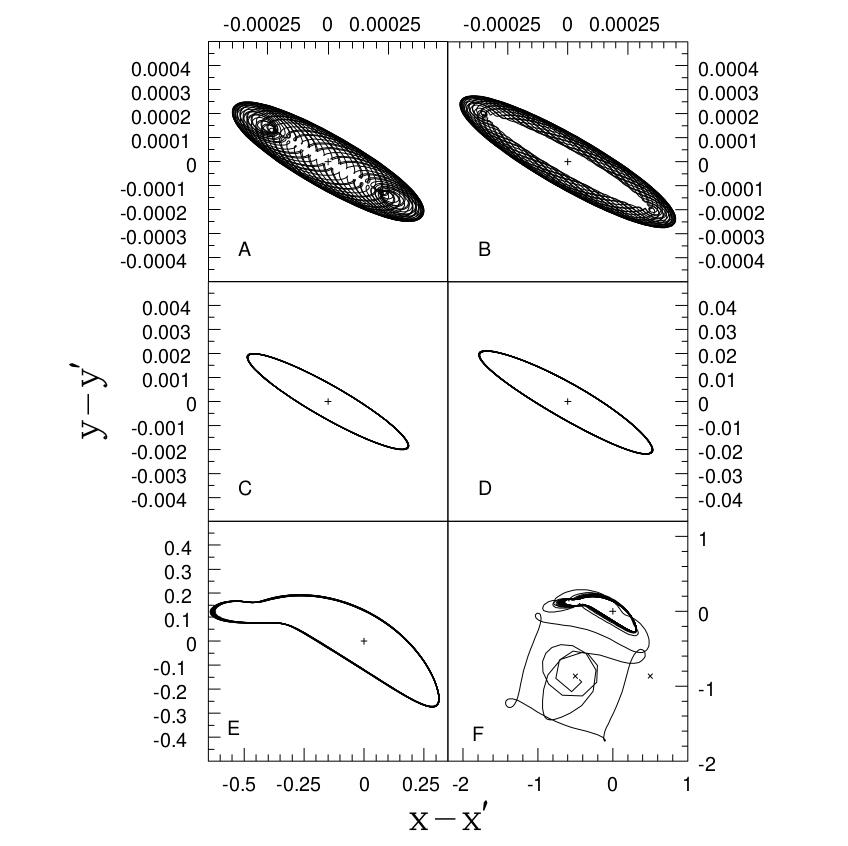}
\caption{ Numerical simulation of Case 4 (strong tidal dissipation).  
Panel A:  The first 100 orbital periods $P$ of $M_2$.  
Panel B:  $t$ =  2 000 $P$ through  2 100 $P$.  
Panel C:  $t$ = 20 000 $P$ through 20 100 $P$.  
Panel D:  $t$ = 40 000 $P$ through 40 100 $P$.  
Panel E:  $t$ = 60 000 $P$ through 60 100 $P$.  
Panel F:  $t$ = 60 100 $P$ through 61 327 $P$.  
$x -x'$ and $y -y'$ are the horizontal deviations 
from the leading equilateral Lagrange point L$'4$.  
Note different scales in each panel.  In Panel F, the polygonal appearance 
of the final orbit is not due to the integrator, but to the discrete output steps; 
and the $\times$ on the left-hand side marks the location of $M_1$, 
while the $\times$ on the right marks the location of $M_2$.  
The plus signs in each panel mark the origin, at L$'4$.  }
\end{figure}

% Panel D of Fig. 8 plots Case 4 for orbits 60 000 through 61 500; by $t$ = 60 000 $P$, 
% the tadpole librations have taken the peculiar banana shape previously seen in Fig. 7.  
% At $t \approx$ 61 393 $P$, $m_0$ escapes the tadpole orbit and adopts an orbit passing $m_2$; 
% but only about four orbits later, at $t \approx$ 61 396.6 $P$, $m_0$ has a close encounter 
% with $m_2$ ($r_2 \approx$ 0.063 167 0), and is immediately ejected from the system.  

Panel F of Fig. 8 plots Case 4 for periods 60 100 through 60 327.  At $t \approx$ 61 323 $P$, 
$m_o$ escapes the banana orbit and assumes a tidally decaying eccentric orbit inferior to $M_2$.  
Only about four periods later, at $t \approx$ 61 327 $P$, $m_o$ has a close encounter 
with the primary $M_1$ this time ($r_1 \approx$ 0.047 711 1; the final time-step is not plotted), 
and the integrator stops.  The polygonal appearance of the final orbit in Fig. 8 
is not due to the integrator, but is an artefact of the discrete output steps.  

Certain processes lead to asymmetry 
between the leading and trailing Trojan points L$'4$ and L$'5$, while others do not; 
for example, radiation pressure is symmetric, but Poynting-Robertson drag is not.  
Our formulation of tides is symmetric, so that L$'4$ and L$'5$ should be equivalent.  
When $Y$ is replaced by $-Y$ in Formulae (1) through (8), (23) through (25), 
and (35) through (37), then Formulae (2), (7), (24), and (36) for $\ddot{Y}$, 
$\partial U/\partial Y$, $\ddot{y}$, and $f_Y$ change sign; 
but their $X$ and $Z$ counterparts are unaffected.  
In order to demonstrate this symmetry, we also simulated an extra Case 5, identical to Case 4, 
except starting with $x_0 = -y_0 = 10^{-5}$ from the shifted trailing Lagrange point L$'5$, 
rather than $x_0 = y_0 = 10^{-5}$ from the shifted leading Lagrange point L$'4$.  

Figure 9 plots the results of Case 5, in a similar format to Fig. 8 for Case 4, except for changes of scale.  
Except for the flip of the $y$ axis, panels A through E of Fig. 9 are nearly identical to those of Fig. 8; 
but as shown in Panel F of Fig. 9, after escaping the banana orbit at $t \approx$ 60 324 $P$, 
now $m_o$ assumes an eccentric orbit superior to $M_2$.  About nine periods later, at $t \approx$ 60 333 $P$, 
$m_o$ has a close encounter with the secondary $M_2$ ($r_2 \approx$ 0.015 045 7; 
the final time-step is not plotted), and the integrator stops.  Except for the flip of the $y$ axis, 
we attribute the minor differences between Cases 4 and 5 to numerical noise and/or chaos.  

\begin{figure}
%\centerline{\psfig{figure=case5rev.eps,width=6.0in,height=6.0in}}
\includegraphics[width=6.0in,height=6.0in]{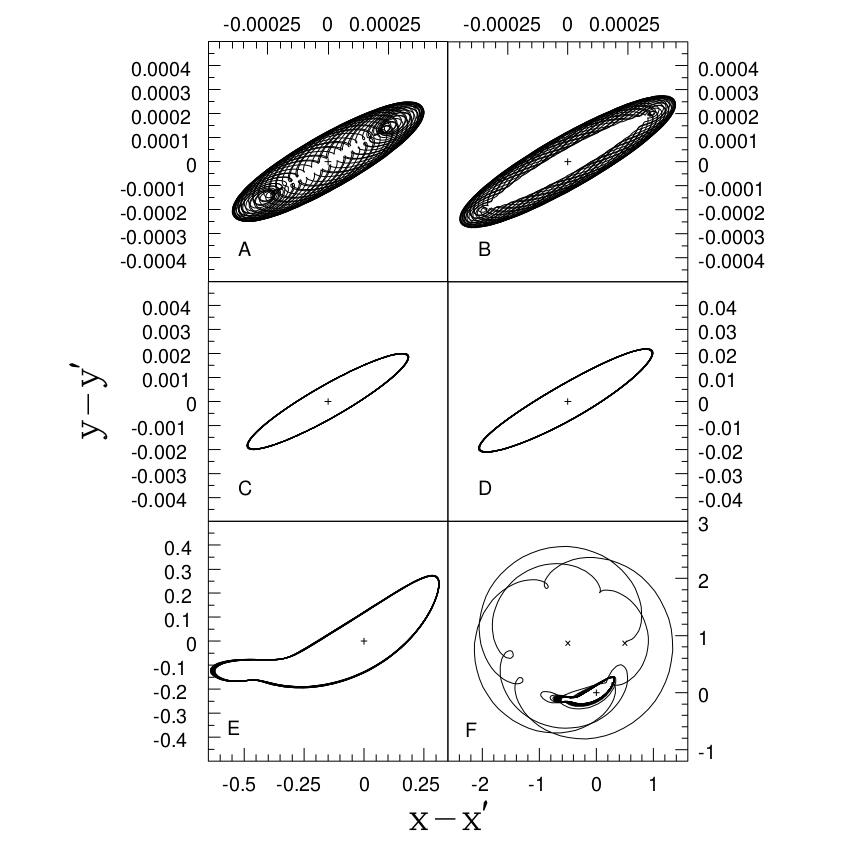}
\caption{ Numerical simulation of Case 5.  
Panel A:  The first 100 orbital periods $P$ of $M_2$.  
Panel B:  $t$ =  2 000 $P$ through  2 100 $P$.  
Panel C:  $t$ = 20 000 $P$ through 20 100 $P$.  
Panel D:  $t$ = 40 000 $P$ through 40 100 $P$.  
Panel E:  $t$ = 60 000 $P$ through 60 100 $P$.  
Panel F:  $t$ = 60 100 $P$ through 60 333 $P$.  
$x -x'$ and $y -y'$ are the horizontal deviations 
from the leading equilateral Lagrange point L$'4$.  
Note different scales in each panel.  In Panel F, 
the $\times$ on the left-hand side marks the location of $M_1$, 
while the $\times$ on the right marks the location of $M_2$.  
The plus signs in each panel mark the origin, at L$'5$.  }
\end{figure}

Finally, in Cases 1 and 2, the $Z$ component of motion 
showed nearly harmonic oscillations for the entire simulation, 
with a period nearly equal to unity, as expected.  
In contrast, Cases 3 through 5 experienced a simple exponential decay 
of the vertical oscillations until the end, 
with the time constants approximately as given in Table 4, 
practically uncoupled from the horizontal motion in $X$ and $Y$.

\section{Discussion}

% symmetry between L4 and L5

% Apply to Telesto, Calypso, Helene, and Polydeuces ? 
% No - dominated by tides in Saturn 

% frequency dependence of Q

% What about non-synchronous rotation ? 

% How about tides in the primary and secondary ? 

% What if the secondary is eccentric ? 

We have found analytic solutions for the motion of a small planet 
near the Trojan point L4 or L5 of a more massive planet, 
including the effects of tides in the small planet.  
Our analytic results include Formula (64) 
for the decay rate of the small planet's eccentric epicycles, 
and Formula (73) for the growth rate of its tadpole librations, 
which agree with those of Couturier {\it et al.} (2021).  
Thus we confirm their conclusion that dissipation of energy 
by tides in a Trojan planet tends to destabilize its orbit.  
We also confirm this analytic conclusion by several numerical simulations.  

The vast majority of known exoplanets that have been observed 
transiting their host star have orbital periods of a few months or less, 
so they and any co-orbital companions may well be subjected to strong tidal forces. 
Therefore such planets may not survive for long times, which may explain 
why no Trojan planets have yet been discovered by {\it Kepler}, {\it TESS}, or other searches.  

If tidal forces are responsible for removing co-orbital companions from these planets 
(rather than such configurations being absent at the end of the planet formation era), 
then smaller Trojan planets and those with longer orbital periods 
% co-orbital planets are likely to be more common for longer-period planets, 
may begin to turn up over longer periods of observation.  
The tidal properties of water-rich sub-Neptune exoplanets 
may differ substantially from those predicted by dry gas-rich models.  
Thus, future detection of a population of Trojan companions 
may help elucidate the internal structure of exoplanets 
that are unlike any of their Solar System cousins.  

Our conclusion that tides destabilize the orbits of Trojan planets 
may be altered by relaxing some of the assumptions in our model.  For example, 
we have neglected any eccentricity of the secondary's orbit around the primary.
Many known exoplanets are in quite eccentric orbits; 
but note that Trojans are even less stable in such systems.  

% Our results are not applicable to exoplanets exhibiting large eccentricities
% *** aren't trojans less stable under such circumstances?.  SLIGHTLY

We also have neglected tides in the primary and secondary.  
Tides in the primary are certainly important in many planet/satellite systems; 
but as distances between bodies are generally larger in the context of exoplanets, 
and stars tend to be less tidally dissipative than even giant planets, tides in stars are 
less likely to be important, except for planets on very eccentric or very short-period orbits.

We have neglected the mass $m_o$ of the Trojan as well; 
its influence on the primary and secondary might lead 
to interesting and possibly observable behaviors.  
We may revisit this question in future work.  

Similarly, a significant obliquity $\epsilon$ of Trojans 
or their non-synchronous rotation might affect our results.  
Of still greater concern, the Appendix discusses the possible 
frequency dependence of $\tau$ or $Q$; in other contexts, 
different frequency dependences can lead to different behaviors. 

% Sargasso quip ? - IN INTRO

\newpage

\section{Appendix: Generalized Tidal Formulation}

Here we derive the form of the tidal forces on a body of mass $m_o$ 
in a right-handed Cartesian coordinate system $X,Y,Z$, retaining dimensional units.  
First, we must consider the tidal potential.  

% but so far, we make no assumptions about $m_1$, $m_2$, $r_1$, $r_2$, or $\Theta$, 
% the angular distance between $m_1$ and $m_2$ as measured from the center of $m_3$.  

\subsection{Potential}

To begin, we assume that $m_o$ is approximately spherically symmetric, with surface radius $R$. 
		Then to lowest degree in $R$, $m_o$ experiences a tidal potential 
\begin{equation}
			V_1 = GM_1 R^2 r_1^{-3} [ 3\cos^2(\theta_1) -1 ]/2 
%			    = Gb^2 m_1 r_1^{-3} [ 3\cos(2\theta_1) +1 ]/4 
\end{equation}
		at its surface due to the primary $M_1$, and an analogous tidal potential 
\begin{equation}
			V_2 = GM_2 R^2 r_2^{-3} [ 3\cos^2(\theta_2) -1 ]/2 
%			    = Gb^2 m_2 r_2^{-3} [ 3\cos(2\theta_2) +1 ]/4 
\end{equation}
			at its surface due to the secondary $M_2$.  

			Here 
\begin{equation}
			r_1 = \sqrt{ (X_1-X_0)^2 +(Y_1-Y_0)^2 +(Z_1-Z_0)^2} 
\end{equation}
is the distance between the centers of $m_o$ at $(X_o,Y_o,Z_o)$ 
and of $M_1$ at $(X_1,Y_1,Z_1)$, while  
\begin{equation}
			r_2 = \sqrt{ (X_2-X_o)^2 +(Y_2-Y_o)^2 +(Z_2-Z_o)^2} 
\end{equation}
is the distance between the centers of $m_o$ and of $M_2$, at $(X_2,Y_2,Z_2)$.  
Similarly, $\theta_1$ is the angular distance 
along the surface of $m_o$ from the point nearest to $M_1$, 
while $\theta_2$ is the analogous angular distance from the sub-$M_2$ point.  

Actually, Formulae (84) through (87) above apply to any number of massive bodies 
$M_1$, $M_2$, $M_3$, ... $M_N$, as desired.  For our present purposes, 
we confine ourselves to just $M_1$ and $M_2$ in the following; 
but our methods can readily be generalized for N-body applications.  

Each of the tidal potentials $V_1$ and $V_2$ from Formulae (84) and (85) raises 
a bulge on the surface of $m_o$, of height $hV_1/g$ and $hV_2/g$, respectively.  
Here $g = Gm_o/R^2$ is the acceleration of gravity on the surface of $m_o$, 
while $h$ is a dimensionless constant depending on the internal structure of $m_o$, 
called its height Love number of the second degree (usually written $h_2$).  
For large, homogeneous fluid bodies, $h$ = 5/2; but for a small solid object, 
$h$ is smaller and proportional to its squared radius $R^2$, all else being equal.  

However, due to dissipation of the energy of deformation in the body of $m_o$, 
the tidal bulges do not align exactly with the peaks of the tidal potentials 
$V_1$ or $V_2$, but rather with the {\it lagged} potentials $V'_1$ and $V'_2$; 
these are equal to $V_1$ and $V_2$ at the surface of $m_o$, 
but at a certain time $t-\tau$ {\it before} the current time $t$ 
(somewhat analogous to the retarded potential of electrodynamics).  

			These lagged potentials can be written as 
\begin{equation}
			V' \approx GM R^2 r'^{-3} [ 3\cos^2(\theta') -1 ]/2 , 
\end{equation}
where $V'$, $r'$, $\theta'$, and $M$ all are subscripted 1, as in Formula (84); 
or else they all are subscripted 2, as in Formula (85).  Here $r'$ is the distance 
between $m_o$ and the tide-raising body $M$ at the previous time $t -\tau$, 
while $\theta'$ is the angular distance along the surface of $m_o$ 
from the location of the sub-perturber point at the previous time $t -\tau$.  

It is common in many applications to replace $\tau$ with $P/(2\pi Q)$, 
where $P$ is the period of the tides 
and $Q$ is a dimensionless parameter known as the tidal ``quality factor''.  
Note that the dimensionless value of $\tau$ is just $1/Q$.  
For example, the tidal time lag $\tau$ is on the order of ten minutes 
for semi-diurnal tides ($P \approx$ 12 hours) in the Earth, mostly due to the oceans; 
then the dimensionless value of $\tau$ is on the order of 0.1, so $Q \approx$ 10.  
Using the same $Q$ for annual tides ($P \approx$ 1 year) would give $\tau \approx$ 5 days.  

Assuming that $Q >> 1$, or equivalently, 
that $\tau$ is short compared to the tidal periods of interest 
(the ``weak friction'' approximation), then 
the lagged distance can be approximated as $r' \approx r -\dot{r}\tau$, 
while the lagged angular distance can be approximated as 
$\theta' \approx \theta -\dot{\theta}\tau$, 
both to first degree in $\tau$.  

% Then each lagged potential can be expanded as 
% \begin{equation}
%			V' \approx Gb^2 m [r^{-3} +3r^{-4} \dot{r} \tau] 
%		[ 3\cos^2(\theta) -1 +6\sin(\theta)\cos(\theta) \dot{\theta}\tau ]/2 , 
% \end{equation}
%			again to first degree in $\tau$.  
% Note that the product term in $\tau^2$ is neglected, for consistency; 
% then radial and transverse tides may be treated separately.  

Strictly speaking, the above expansion is formally appropriate 
only when the time lag $\tau$ is a constant, independent of the period $P$; 
this corresponds to ``viscous''-type tides, 
where the quality factor $Q$ is directly proportional to $P$.  However, 
the above treatment is questionable if $\tau$ depends on period (as in ``constant-Q'' tides, 
where $\tau$ is directly proportional to $P$, more analogous to friction between solids).  
Still worse, if $\tau$ and $Q$ depend on the amplitude of the tides, 
the tides are not even linear, and different components cannot simply be added together.  

Assuming that the tides are linear, 
each tidal bulge generates its own quadrupole (inverse cube) potential 
\begin{equation}
				W = kV'R^3/r_o^3 . 
\end{equation}
Here $r_o$ is the distance of the field point from the center of $m_o$, 
while $k$ is another dimensionless constant depending on the internal structure of $m_o$, 
called its potential Love number of the second degree (usually written $k_2$); 
$k_2$ is also known to astrophysicists as the ``apsidal motion constant'' 
({\it e.g.}, Alexander, 1973).  
For large, homogeneous fluid bodies, $k$ = 3/2; but for a small solid object, 
$k$ is smaller and proportional to its squared radius $R^2$, like the height Love number $h$.

\subsection{Forces}

The quadrupole potentials $W_1$ and $W_2$ from Formula (89) above are responsible 
for the tidal forces and torques between $m_o$ and the tide-raising bodies.  
The lag terms involving $\dot{r} \tau$ and $\dot{\theta} \tau$ in Formula (88) for $V'$ 
are the reasons why those tidal forces and torques are not conservative in general.  

For convenience, we decompose the forces into radial and transverse components.  
For example, the quadrupole potential $W$ exerts a radial force of strength 
\begin{equation}
			m \partial W/\partial r_o = -3mkV'R^3/r_o^4 
		= -3kGMmR^5 r'^{-3} r_o^{-4} [3\cos^2(\theta') -1]/2 , 
\end{equation}
as well as a {\it transverse} force of strength 
\begin{equation}
			m r_o^{-1} \partial W/\partial\theta' 
		= -3kGMmR^5 r'^{-3} r_o^{-4} \sin(\theta')\cos\theta' , 
\end{equation}
on any mass $m$ at a distance $r_o$ from $m_o$, 
and at an angular distance $\theta'$ from the axis of the tidal bulge.  

If the perturbed body $m$ is located at $(X_m, Y_m, Z_m)$, 
the radial force on $m$ from Formula (90) 
is directed along ${\bf r}_o = (X_m-X_o, Y_m-Y_o, Z_m-Z_o)$, 
the vector location of $m$ relative to the center of $m_o$; 
but the transverse force from Formula (91) above lies in the direction 
of the part of ${\bf r}_0 -{\bf r}'$ perpendicular to ${\bf r}_0$, 
where ${\bf r}'$ is the time-lagged vector location of the tide-raising body 
relative to $m_o$.  By Newton's third law, 
$m$ exerts an equal force on $m_o$, but in the opposite direction.  

% Note that if either Formula (90) or (91) turns out to be negative, 
% the direction of the corresponding force is reversed.  POOR 
It is interesting to note that in an equilateral configuration 
where $\theta_1$ and $\theta_2$ are each $60^\circ$, the radial force of $M_1$ 
on the tidal bulge induced in $m_o$ by $M_2$ turns out to be positive, 
corresponding to a slight repulsion; and the same applies to 
the radial force of $M_2$ on the tidal bulge induced in $m_o$ by $M_1$.  

In order to evaluate $\cos\theta'$ in Formulae (90) and (91), 
we take the scalar product of the unit vectors (designated by carats) 
in the directions of ${\bf r}_o$ and ${\bf r}'$:  
\begin{equation}
			\cos\theta' = \hat{r}_o \bullet \hat{r}' 
			= \frac{ {\bf r}_o \bullet {\bf r}'}{r_o r'} . 
%		= \frac{ {\bf r}_\mu \bullet [{\bf r} -{\bf \dot{r}}\tau]}{r_\mu [r -\dot{r}\tau] } , 
\end{equation}
However, Formula (92) above is not adequate to determine $\theta'$ when it is small, 
such as when the tide-raising and perturbed bodies are one and the same.  
Therefore we use the {\it cross} product of $\hat{r}_o$ and $\hat{r}'$ 
to evaluate $\sin\theta'$ in Formula (91):  
\begin{equation}
                        \sin\theta' = |\hat{r}_o \times \hat{r}'|
                	= \frac{|{\bf r}_o \times {\bf r}'|}{r_o r'} .  
\end{equation}

Now we separate ${\bf r}'$ itself into radial and transverse vectors, 
respectively parallel and perpendicular to ${\bf r}_o$.  
The radial part of ${\bf r}'$ is 
\begin{equation}
			{\bf r}'_R = \hat{r}_o r'\cos\theta' 
			= \hat{r}_o ({\bf r}_o \bullet {\bf r}')/r_o 
			= {\bf r}_o ({\bf r}_o \bullet {\bf r}')/r_o^2 . 
\end{equation}
Then subtracting this radial part from ${\bf r}'$ 
leaves the transverse part of ${\bf r}'$:  
\begin{equation}
                        {\bf r}'_T = {\bf r}' -{\bf r}'_R 
		= {\bf r}' -({\bf r}_o \bullet {\bf r}'){\bf r}_o/r_o^2 . 
\end{equation}
It is easy to verify that ${\bf r}'_T$ above is orthogonal to ${\bf r}_o$:  
${\bf r}_o \bullet {\bf r}'_T$ = 
${\bf r}_o \bullet [{\bf r}' -{\bf r}_o({\bf r}_o \bullet {\bf r}')/r_o^2]$ = 
${\bf r}_o \bullet {\bf r}' - ({\bf r}_o \bullet {\bf r}_o)({\bf r}_o \bullet {\bf r}')/r_o^2$ 
= ${\bf r}_o \bullet {\bf r}' - {\bf r}_o \bullet {\bf r}'$ = 0.  

This decomposition of ${\bf r}'$ into radial and transverse parts 
makes it easier to evaluate the cross product in Formula (93):  
\begin{equation}
			\sin\theta' = \frac{|{\bf r}_o \times {\bf r}'|}{r_o r'} 
			= \frac{|{\bf r}_o \times ({\bf r}'_R +{\bf r}'_T)|}{r_o r'} 
		= \frac{|({\bf r}_o \times {\bf r}'_R) +({\bf r}_\mu \times {\bf r}'_T)|}{r_\mu r'} 
			= \frac{|{\bf 0} +{\bf r}_o \times {\bf r}'_T|}{r_o r'} 
			= \frac{|{\bf r}_0||{\bf r}'_T|}{r_o r'} = |{\bf r}'_T|/r' . 
% \] \begin{equation}
%			= \sqrt{{\bf r}'_T \bullet {\bf r}'_T}/r' = 
%	\sqrt{{\bf r}' \bullet {\bf r}' -2{\bf r}' \bullet {\bf r}'_R +{\bf r}'_R \bullet {\bf r}'_R}/r'
%			= \sqrt{r'^2 -2r'^2_R +r'^2_R}/r' = \sqrt{1 -r'^2_R/r'^2} . 
\end{equation}

Now we can write the radial part of the tidal force on $m_o$ as 
\begin{equation}
	{\bf F}_R = 3kGMmR^5 r'^{-5}r_o^{-7}[3({\bf r}_o\bullet{\bf r}')^2 -r_o^2 r'^2]{\bf r}_o/2 
\end{equation}
from Formulae (90) and (92), and the transverse part of the tidal force on $m_o$ as 
\begin{equation}
		{\bf F}_T = -3kGMmR^5 r'^{-5}r_o^{-5}[{\bf r}_o \bullet {\bf r}']{\bf r}'_T
			= -3kGMmR^5 r'^{-5}r_o^{-7}[{\bf r}_o \bullet {\bf r}'] 
			[r_o^2{\bf r}' -({\bf r}_o \bullet {\bf r}'){\bf r}_o] 
\end{equation}
from Formulae (91) and (95).  Finally, re-combining these gives the total tidal force on $m_o$:  
\begin{equation}
			{\bf F} = {\bf F}_R +{\bf F}_T = 3kGMmR^5 r'^{-5}r_o^{-7} 
			\{ [5({\bf r}_o \bullet {\bf r}')^2 -r_o^2r'^2]{\bf r}_o/2 
			-[{\bf r}_o \bullet {\bf r}']r_o^2{\bf r}' \} . 
\end{equation}

\subsection{Lagged location}

The next issue is to evaluate ${\bf r}'$, the effective value of ${\bf r}$ 
at a time $t -\tau$ before the current time $t$.  If $m_o$ were not spinning at all, 
we could approximate ${\bf r}' \approx {\bf r} -\dot{\bf r}\tau$, to first degree in $\tau$.  
However, if $m_o$ is spinning rigidly with angular velocity vector $\boldsymbol \omega$,
its rotation drags the tidal bulge away from the sub-perturber point, 
as if the tide-raising body $M$ had an additional transverse velocity 
\begin{equation}
                        {\boldsymbol \varpi} = {\bf r} \times {\boldsymbol \omega} 
\end{equation}
in a reference frame fixed in the body of $m_o$.  
% with respect to the surface of $m_0$.  

	Note that ${\boldsymbol \varpi}$ is orthogonal to ${\bf r}$, 
so ${\bf r} \bullet {\boldsymbol \varpi}$ = 0.  Then we may approximate 
\begin{equation}
			{\bf r}' \approx {\bf r} -[\dot{\bf r} +{\boldsymbol \varpi}]\tau , 
\end{equation}
and 
\begin{equation}
				r'^2 = {\bf r}' \bullet {\bf r}' 
		\approx r^2 -2[{\bf r} \bullet \dot{\bf r}]\tau = r^2 -2r\dot{r}\tau, 
\end{equation}
both to first degree in $\tau$.  

In Formula (102) above, we have replaced ${\bf r} \bullet \dot{\bf r}$ 
with $\frac{d}{dt}({\bf r} \bullet {\bf r})/2 = \frac{d}{dt}r^2/2 = r\dot{r}$.  
The time derivative of the distance $r$ appearing in Formula (102) is 
\begin{equation}
	\dot{r} = ([X-X_o][\dot{X}-\dot{X}_o] +[Y-Y_o][\dot{Y}-\dot{Y}_o] +[Z-Z_o][\dot{Z}-\dot{Z}_o])/r , 
\end{equation}
from Formula (84) or (85).  Note that Formula (103) above for $\dot{r}$ is not the same as the speed 
$|{\bf \dot{r}}|$ = \\ $\sqrt{ [\dot{X}-\dot{X}_o]^2 +[\dot{Y}-\dot{Y}_o]^2 +[\dot{Z}-\dot{Z}_o]^2 }$ 
of the tide-raising body $M$ relative to $m_o$; 
instead, $\dot{r}$ is just the {\it radial component} of its vector velocity ${\bf \dot{r}}$.

\subsection{Primary}

Now consider the case when the primary $M_1$ 
is both the tide-raising body $M$ and the perturbed body $m$.  
Then ${\bf r}_0 = {\bf r} = {\bf r}_1$, ${\bf r}' = {\bf r}'_1$, 
${\bf s} = {\bf s}_1$, and Formula (59) for the tidal force on $m_o$ becomes 
\[
				{\bf F}_{11} = 3kGM_1^2 R^5 r_1'^{-5} r_1^{-7} 
			\{ [5({\bf r}_1 \bullet {\bf r}'_1)^2 -r_1^2 r_1'^2]{\bf r}_1/2 
			-[{\bf r}_1 \bullet {\bf r}'_1]r_1^2 {\bf r}'_1 \} 
\] \[
				\approx 3kGM_1^2 R^5 r_1'^{-5} r_1^{-7} 
\{ [5r_1^4/2 -5r_1^2({\bf r}_1 \bullet \dot{\bf r}_1)\tau -r_1^4/2 +r_1^2({\bf r}_1 \bullet \dot{\bf r}_1)\tau]{\bf r}_1 
			-[{\bf r}_1 \bullet {\bf r}'_1]r_1^2{\bf r}'_1 \} 
\] \[
				= 3kGM_1^2 R^5 r_1'^{-5} r_1^{-5} 
			\{ [2r_1^2 -4({\bf r}_1 \bullet \dot{\bf r}_1)\tau]{\bf r}_1 
			-[{\bf r}_1 \bullet {\bf r}'_1]{\bf r}'_1 \}
\] \[
				\approx 3kGM_1^2 R^5 r_1'^{-5} r_1^{-5} 
                        \{ [2r_1^2 -4({\bf r}_1 \bullet \dot{\bf r}_1)\tau]{\bf r}_1 
	-[r_1^2 -({\bf r}_1 \bullet \dot{\bf r}_1)\tau][{\bf r}_1 -(\dot{\bf r}_1 +{\boldsymbol \varpi}_1)\tau] \}
\] \[
				\approx 3kGM_1^2 R^5 r_1'^{-5} r_1^{-5} 
                        \{ [r_1^2 -3({\bf r}_1 \bullet \dot{\bf r}_1)\tau]{\bf r}_1 
				+r_1^2[\dot{\bf r}_1 +{\boldsymbol \varpi}_1]\tau \}
\] \[
				\approx 3kGM_1^2 R^5 r_1'^{-5} r_1^{-5} 
                        	\{ [r_1^2 -3r_1\dot{r}_1\tau]{\bf r}_1 
                        +r_1^2[\dot{\bf r}_1 +{\boldsymbol \varpi}_1]\tau \} 
\] \[
				\approx 3kGM_1^2 R^5 r_1'^{-5} r_1^{-3}
                                \{ [1 -3\dot{r}_1\tau/r_1]{\bf r}_1
                        +[\dot{\bf r}_1 +{\boldsymbol \varpi}_1]\tau \}
\] \[
				\approx 3kGM_1^2 R^5 r_1^{-8}[1 +5\dot{r}_1\tau/r_1]
                                \{ [1 -3\dot{r}_1\tau/r_1]{\bf r}_1 
                        +[\dot{\bf r}_1 +{\boldsymbol \varpi}_1]\tau \}
\] \begin{equation}
				\approx 3kGM_1^2 R^5 r_1^{-8}
                                \{ [1 +2\dot{r}_1\tau/r_1]{\bf r}_1
                        +[\dot{\bf r}_1 +{\boldsymbol \varpi}_1]\tau \} , 
% \[
%		= 3kGb^5 m_1^2 r_1'^{-5} r_1^{-5} [r_1^2 -r_1\dot{r}_1\tau]{\bf r}'_1 
% \] \[
%	\approx 3kGb^5 m_1^2 r_1^{-8}[1 +5\dot{r}_1\tau/r_1][1 -\dot{r}_1\tau/r_1]{\bf r}'_1 
% \] \[
%		\approx 3kGb^5 m_1^2 r_1^{-8}[1 +4\dot{r}_1\tau/r_1]{\bf r}'_1 
% \] \[
%	= 3kGb^5 m_1^2 r_1^{-8}[1 +4\dot{r}_1\tau/r_1][{\bf r}_1 -(\dot{\bf r}_1 +{\bf s}_1)\tau]
% \] \begin{equation}
%	\approx 3kGb^5 m_1^2 r_1^{-8}\{[1 +4\dot{r}_1\tau/r_1]{\bf r}_1 -(\dot{\bf r}_1 +{\bf s}_1)\tau \} . 
\end{equation}
again to first degree in $\tau$.  

The first term inside the braces $\{\}$ in Formula (104) above represents 
the enhanced radial attraction of $M_1$ on $m_o$ due to its tidal bulge, 
independent of dissipation or of the time lag $\tau$.  
The three remaining terms all are proportional to $\tau$; 
the second term $+2\dot{r}_1\tau{\bf r}_1/r_1$ gives a small modification 
of the first term due to just the radial component of $\dot{\bf r}_1$.  
The third term $+\dot{\bf r}_1\tau$ represents 
an additional small force on $m_o$ proportional to $\dot{\bf r}_1$, 
while the last term $+{\boldsymbol \varpi}_1\tau$ 
gives an additional transverse force on $m_o$ due to its rotation.  
It is gratifying to verify that the latter three terms 
are equivalent to Formula (5) of Mignard (1979) in the two-body case.

\subsection{Secondary}

In a two-body problem, ${\bf F}_{11}$ above would be the only tidal force to include; 
but a three-body problem introduces three more tidal forces to consider.  
Of course, the attraction of the secondary $M_2$ 
on the tidal bulge it raises in $m_o$ also produces its own force 
\begin{equation}
				{\bf F}_{22} \approx 3kGM_2^2 R^5 r_2^{-8} 
                                \{ [1 +2\dot{r}_2\tau/r_2]{\bf r}_2 
                                +[\dot{\bf r}_2 +{\boldsymbol \varpi}_2]\tau \} 
%				ABOVE AND BELOW AGREE - YAY ! 
%		{\bf F}_{22} \approx 3kGb^5 m_2^2 \{ [r_2^{-8} +2r_2^{-9}\dot{r}_2 \tau]{\bf r}_2 
%			+r_2^{-8} \tau [{\bf s}_2 +{\bf \dot{r}}_2] \} 
\end{equation}
on $m_o$, analogous to Formula (104) for ${\bf F}_{11}$.  

%		To find the Cartesian components of ${\bf R}_{22}$, 
% just project it onto all three Cartesian axes $X$, $Y$, $Z$.  Then
% \begin{equation}
%                R_{X22} = R_{22} [X_2 -X_3]/r_2 , \; \; 
%                R_{Y22} = R_{22} [Y_2 -Y_3]/r_2 , \; \; {\rm and} \; \; 
%                R_{Z22} = R_{22} [Z_2 -Z_3]/r_2 , 
% \end{equation}
%		analogous to Eqs. (26) through (28).  

%		Likewise, the attraction of the secondary $m_2$ 
% on the tidal bulge it raised in $m_0$ also exerts its own transverse force 
% \begin{equation}
%		{\bf T}_{22} \approx 3kGb^5 m_2^2 r_2^{-8} \tau [{\bf s}_2 +{\bf t}_2] 
% \end{equation}
% on $m_0$, analogous to Formula (69) for ${\bf T}_{11}$.  

% ${\bf T}_{22}$ on $m_3$ in the direction of ${\bf s}_2 +{\bf t}_2$, 
% opposite to the relative angular displacement $\dot{\theta}_2 \tau$, of strength 
% \begin{equation}
%                T_{22} = 3kGb^5 m_2^2 r_2^{-7} \dot{\theta}_2 \tau ,  
%\end{equation}
%			analogous to Formula (26) for $T_{11}$.  

Note that the secondary force ${\bf F}_{22}$ 
is weaker than the corresponding primary force ${\bf F}_{11}$ 
by a factor of $(M_2/M_1)^2$, all else being equal.  
However, there are also two cross-interactions to consider as well, 
each of intermediate order $M_2/M_1$.

\subsection{Mixed}

Now consider the case when the tide-raising body $M$ is the primary $M_1$, 
but the perturbed body $m$ is the secondary $M_2$.  
Then ${\bf r}_o = {\bf r}_2$, but ${\bf r} = {\bf r}_1$, 
${\bf r}' = {\bf r}_1'$, ${\boldsymbol \varpi} = {\boldsymbol \varpi}_1$, 
and Formula (99) for the tidal force on $m_o$ becomes 
\[
		{\bf F}_{12} = 3kGM_1 M_2 R^5 r_1'^{-5} r_2^{-7} 
                \{ [5({\bf r}_2 \bullet {\bf r}'_1)^2 -r_2^2 r_1'^2]{\bf r}_2/2 
                -[{\bf r}_2 \bullet {\bf r}'_1]r_2^2 {\bf r}'_1 \} 
\] \[
			\approx 3kGM_1 M_2 R^5 r_1'^{-5} r_2^{-7} 
		\{ [5({\bf r}_2 \bullet [{\bf r}_1 -(\dot{\bf r}_1 +{\boldsymbol \varpi}_1)\tau])^2 
		-r_2^2 [r_1^2 -2({\bf r}_1 \bullet \dot{\bf r}_1)\tau]]{\bf r}_2/2 
\] \[
		-[{\bf r}_2 \bullet ({\bf r}_1 -[\dot{\bf r}_1 +{\bf s}_1]\tau)] 
			r_2^2 [{\bf r}_1 -(\dot{\bf r}_1 +{\boldsymbol \varpi}_1)\tau] \} 
\] \[
	\approx 3kGM_1 M_2 R^5 r_1'^{-5} r_2^{-7} \{ [5({\bf r}_2 \bullet {\bf r}_1)^2 
	-10({\bf r}_2 \bullet {\bf r}_1)[{\bf r}_2 \bullet (\dot{\bf r}_1 +{\boldsymbol \varpi}_1)]\tau
		-r_1^2 r_2^2 +2r_2^2({\bf r}_1 \bullet \dot{\bf r}_1)\tau]{\bf r}_2/2 
\] \[
		-[{\bf r}_2 \bullet ({\bf r}_1 -[\dot{\bf r}_1 +{\bf s}_1]\tau)]r_2^2 {\bf r}_1 
			+[{\bf r}_2 \bullet {\bf r}_1]r_2^2(\dot{\bf r}_1 +{\boldsymbol \varpi}_1)\tau \}
\] \[
	\approx 3kGM_1 M_2 R^5 r_1^{-5} r_2^{-7} [1 +5\dot{r}_1\tau/r_1] \{ [5({\bf r}_2 \bullet {\bf r}_1)^2 
        	-10({\bf r}_2 \bullet {\bf r}_1)[{\bf r}_2 \bullet (\dot{\bf r}_1 +{\boldsymbol \varpi}_1)]\tau 
                	-r_1^2 r_2^2 +2r_2^2 r_1 \dot{r}_1\tau]{\bf r}_2/2 
\] \[
		-[{\bf r}_2 \bullet ({\bf r}_1 -[\dot{\bf r}_1 +{\bf s}_1]\tau)]r_2^2 {\bf r}_1 
                	+[{\bf r}_2 \bullet {\bf r}_1]r_2^2(\dot{\bf r}_1 +{\boldsymbol \varpi}_1)\tau \} 
\] \[
		\approx 3kGM_1 M_2 R^5 r_1^{-5} r_2^{-7} \{ [5({\bf r}_2 \bullet {\bf r}_1)^2 
        	-10({\bf r}_2 \bullet {\bf r}_1)[{\bf r}_2 \bullet (\dot{\bf r}_1 +{\boldsymbol \varpi}_1)]\tau 
                        -r_1^2 r_2^2 +2r_2^2 r_1 \dot{r}_1\tau]{\bf r}_2/2 
\] \[
		-[{\bf r}_2 \bullet ({\bf r}_1 -[\dot{\bf r}_1 +{\bf s}_1]\tau)]r_2^2 {\bf r}_1 
                        +[{\bf r}_2 \bullet {\bf r}_1]r_2^2(\dot{\bf r}_1 +{\boldsymbol \varpi}_1)\tau 
			+5([5({\bf r}_2 \bullet {\bf r}_1)^2 -r_1^2 r_2^2]{\bf r}_2/2 
			-[{\bf r}_2 \bullet {\bf r}_1]r_2^2{\bf r}_1)\dot{r}_1\tau/r_1 \} 
\] \[
			= 3kGM_1 M_2 R^5 r_1^{-5} r_2^{-7} \{ [5({\bf r}_2 \bullet {\bf r}_1)^2 
                -10({\bf r}_2 \bullet {\bf r}_1)[{\bf r}_2 \bullet (\dot{\bf r}_1 +{\boldsymbol \varpi}_1)]\tau 
                        -r_1^2 r_2^2 -3r_2^2 r_1 \dot{r}_1\tau]{\bf r}_2/2 
\] \begin{equation}
		-[{\bf r}_2 \bullet ({\bf r}_1 -[\dot{\bf r}_1 +{\bf s}_1]\tau)]r_2^2 {\bf r}_1 
                        +[{\bf r}_2 \bullet {\bf r}_1]r_2^2(\dot{\bf r}_1 +{\boldsymbol \varpi}_1)\tau 
                        	+5[5({\bf r}_2 \bullet {\bf r}_1)^2 {\bf r}_2/2 
                        -({\bf r}_2 \bullet {\bf r}_1)r_2^2{\bf r}_1]\dot{r}_1\tau/r_1 \} , 
\end{equation}
again to first degree in $\tau$.  

Formula (106) above is equivalent to the intermediate result (4.3) of Mignard (1979), 
except that his formula does not include the terms without $\tau$.  
It also comes as some relief to verify that Formula (106) reduces 
to Formula (104) or (105) when $M_1$ and $M_2$ are the same object.  
In the case when the tide-raising body $M$ is the secondary $M_2$, 
but the perturbed body $m$ is the primary $M_1$, 
the tidal force ${\bf F}_{21}$ on $m_o$ is found 
by swapping the subscripts 1 and 2 in Formula (106) above.  

Finally, the net tidal force on $m_o$ is just the sum of the four individual terms:  
\begin{equation}
		{\bf F} = {\bf F}_{11} +{\bf F}_{12} +{\bf F}_{21} +{\bf F}_{22} . 
\end{equation}
Unfortunately, adding ${\bf F}_{12}$ and ${\bf F}_{21}$ together 
does not seem to lead to any important simplifications.

\newpage

\begin{center}
                                ACKNOWLEDGEMENTS
\end{center}

Support for this work was provided by NASA's PSD ISFM program.  
We thank two anonymous reviewers for their service and helpful suggestions, 
as well as Jeff Cuzzi and Jos\'{e} Alvarellos for constructive critiques of a draft.

\setlength{\parindent}{-0.1in}

\begin{center}
                                REFERENCES
\end{center}

Alexander, M. E., 1973.  The weak friction approximation and tidal evolution 
in close binary systems.  {\it Astrophysics and Space Science} {\bf 23}, 459--510.  

% Barnett, C. T., 1976. Theoretical modeling of the magnetic and gravitational fields \\
% of an arbitrarily shaped three-dimensional body. {\it Geophysics} {\bf 41}, 1353--1364.

Beaug\'e, C., Zs. S\'andor, B. \'Erdi, and \'A. S\"uli, 2007.  
Co-orbital terrestrial planets in exoplanetary systems: \\ 
a formation scenario.  {\it Astronomy \& Astrophysics} {\bf 463}, 359--367.  

% Beyer, R. A., {\it et al.}, 2019.  Potential mapping schemes and reference systems for MU69.  
% \\ {\it LPSC Abstracts} {\bf 50}, 2258.pdf.  

% Binzel, R. P., {\it et al.}, 2019.  Highly localized seasonal cold-trapping 
% in the neck of 2014 MU69 'Ultima Thule'.  \\ {\it LPSC Abstracts} {\bf 50}, 2933.pdf.  

% Blitzer, L., 1982. Dynamical stability and potential energy.
% {\it Am. J. Physics} {\bf 50}, 431--434.

% Brooker, R. A., and T. W. Olle, 1955.  Apsidal-motion constants for polytropic models.  
% {\it M.N.R.A.S.} {\bf 115}, 101--106.  

% Broucke, R. A., 1995. Closed form expressions for some gravitational potentials: \\
% Triangle, rectangle, pyramid and polyhedron. Pp. 1293--1309 in {\it Spaceflight Mechanics 1995:
% \\ Proceedings of the AAS/AIAA Spaceflight Mechanics Conference}, R. J. Proulx, J. J. F. Liu,
% \\ P. K. Seidelmann, and S. Alfano, editors. Univelt, Inc. Mostly republished in 1999
% \\ as Pp. 321--340 in {\it The Dynamics of Small Bodies in the Solar System},
% \\ B. A. Steves and A. E. Roy, editors. Kluwer, Inc.

% Burov, A. A., A. D. Guerman, I. I. Kosenko, and V. I. Nikonov, 2017.  
% On the gravity of dumbbell-like bodies represented by a pair of intersecting balls 
% (in Russian).  {\it Rus. J. Nonlin. Dyn.} {\bf 13}, 243--256.  

% Carbognani, A., 2017.  The spin-barrier ratio for S and C-type main belt asteroids. \\
% {\it Planetary and Space Science} {\bf 147}, 1--5.  

Caton, D. B., S. A. Davis, and B. D. Walls, 1999. A search for Trojan planets:  
\\ A novel approach for looking for transits of extrasolar planets (abstract).  
{\it B.A.A.S.} {\bf 31}, 1534.  

Caudal, G. V., 2013.  The role of tidal torques on the evolution of the system of 
Saturn's co-orbital satellites Janus and Epimetheus.  {\it Icarus} {\bf 223}, 733--740.  

% Chandrasekhar, S., 1969. {\it Ellipsoidal Figures of Equilibrium}. Yale U. Press.

% Charnoz, S., A. Brahic, P. C. Thomas, and C. C. Porco, 2007.
% The equatorial ridges of Pan and Atlas: \\ Terminal accretionary ornaments?
% {\it Science} {\bf 318}, 1622-1624.

Colombo, G., D. A. Lautman, and I. I. Shapiro, 1966.  The Earth's dust belt:  Fact or fiction?  
\\ 2.  Gravitational focussing and Jacobi capture.  {\it JGR} {\bf 71}, 5705--5717.  

% Countermarsh, B., 2007. Velocity effect of vehicle rolling resistance in sand. \\
% {\it J. of Terramechanics} {\bf 44}, 275--291.

Couturier, J., P. Robutel, and A. C. M. Correia, 2021.  An analytical model 
for tidal evolution in co-orbital systems.  I.  Application to exoplanets.  
{\it Cel. Mech. Dyn. Astron.} 133:37. 

Couturier, J., P. Robutel, and A. C. M. Correia, 2022.  
Dynamics of co-orbital exoplanets in a first order resonance chain with tidal dissipation.  
Submitted to {\it Astronomy \& Astrophysics}.  arXiv:2204.08074v1.  

\'{C}uk, M., D. P. Hamilton, and M. J. Holman, 2012.  
Long-term stability of horseshoe orbits.  
{\it M.N.R.A.S} {\bf 426}, 3051--3056.  

% Danby, J. M. A., 1962. {\it Fundamentals of Celestial Mechanics}. MacMillan.

% Davis, D. R., K. R. Housen, and R. Greenberg, 1981. \\ The unusual
% dynamical environment of Phobos and Deimos. {\it Icarus} {\bf 47}, 220--233.

Davis, S. A., D. B. Caton, K. A. Klutz, K. D. Wohlman, R. J. Stamilio, and K. B. Hix, 2001.  
\\ The search for extrasolar Trojan planets: An update (abstract).  
{\it B.A.A.S.} {\bf 33}, 1303.  

de la Barre, C. M., W. M. Kaula, and F. Varadi, 1996.  \\ 
A study of orbits near Saturn's triangular Lagrangian points.  
{\it Icarus} {\bf 121}, 88--113.  

% Dermott, S. F., and P. C. Thomas, 1988. The shape and internal structure of Mimas.
% {\it Icarus} {\bf 73}, 25--65. 

% Dobrovolskis, A. R., and J. A. Burns, 1980. Life near the Roche limit: \\
% Behavior of ejecta from satellites close to planets. {\it Icarus} {\bf 42}, 422--441.

% Dobrovolskis, A. R., 1990. Tidal disruption of solid bodies.
% {\it Icarus} {\bf 88}, 24--38.

% Dobrovolskis, A. R., 1995. Chaotic rotation of Nereid?
% {\it Icarus} {\bf 118}, 181--198.

% Dobrovolskis, A.R., 1996.  Inertia of any polyhedron.  {\it Icarus} {\bf 124}, 698--704.  

% Dobrovolskis, A. R., 2019.  Classification of ellipsoids by shape and surface gravity.  
% {\it Icarus} {\bf 321}, 891--928.  \\ (Paper III)  

% Dobrovolskis, A. R., and D. G. Korycansky, 2013. \\
% The quadrupole model for rigid-body gravitational simulations.
% {\it Icarus} {\bf 225}, 623--635.  (Paper I)  

% Dobrovolskis, A. R., and D. G. Korycansky, 2018. \\
% Internal gravity, self-energy, and disruption of comets and asteroids.  
% {\it Icarus} {\bf 303}, 234--250.  (Paper II)  

Dobrovolskis, A. R., 2007.  Spin states and climates of eccentric explanets.  
{\it Icarus} {\bf 192}, 1--23.  

Dobrovolskis, A. R., 2012.  Counter-orbitals:  another class of co-orbitals.  \\
{\it AAS Division for Planetary Sciences Meeting} {\bf 44}, abstract 112.22 .  

Dobrovolskis, A. R., 2013.  Effects of Trojan exoplanets on the reflex motions of their parent stars.  
\\ {\it Icarus} {\bf 26}, 1635--1641.  

Dobrovolskis, A. R., 2015. Radial velocities of stars with multiple co-orbital planets.  \\ 
{\it Astrophysics and Space Science} {\bf 356}, 241--249.  See also arXiv :1404.5377v1.  

% Ferguson, C. C., 1979. Intersections of ellipsoids and planes of arbitrary
% orientation and position. \\ {\it Mathematical Geology} {\bf 11}, 329--336. 

Ferraz-Mello, F., 2022.  On tides and exoplanets.  
In {\it Multi-scale dynamics of space objects}.  Proceedings of IAU Symposium No. 364, 
ed. A. Celletti, C. Beaug\'{e}, C. Gale\c{s}, and A. Lemaitre.  arXiv:2111.01984v2.  

Ford, E. B., and B. S. Gaudi, 2006.  Observational constraints on Trojans of transiting extrasolar planets.  
\\ {\it Ap. J. Letters} {\bf 652}, L137--L140.  

Ford, E. B., and M. J. Holman, 2007.  Using transit timing observations to search for Trojans 
\\ of transiting extrasolar planets.  {\it Ap. J. Letters} {\bf 664}, L51--L54.  

% Fujiwara, A., and 21 co-authors, 2006. The rubble-pile asteroid Itokawa
% as observed by Hayabusa. \\ {\it Science} {\bf 312}, 1330--1334. 

% Gendzwill, D. J., and M. R. Stauffer, 1981. Analysis of triaxial ellipsoids:
% \\ Their shapes, plane sections, and plane projections.
% {\it Mathematical Geology} {\bf 13}, 135--152.

Giuppone, C. A., P. Ben\'itez-Llambay, and C. Beaug\'e, 2012.  Origin and detectability 
\\ of co-orbital planets from radial velocity data.  {\it M.N.R.A.S.} {\bf 421}, 356--368.  

Go\'zdziewski, K., and M. Konacki, 2006.  Trojan pairs in the HD128311 and HD82943 planetary systems?  
\\ {\it Ap. J} {\bf 647}, 573--586.  

Greenberg, R. (1978).  Orbital resonance in a dissipative medium.  {\it Icarus} {\bf 48}, 12--22.  

Guerrero, N. M., and 104 co-authors, 2021.  The TESS objects of interest catalog 
from the TESS prime mission.  {\it Ap. J. Supplements} {\bf 254}, 39.  

% Guibout, V., and D. J. Scheeres, 2003. Stability of surface motion on a rotating ellipsoid.
% \\ {\it Cel. Mech. Dyn. Astron.} {\bf 87}, 263--290.

% Hamelin, M., 2011. Motion of blocks on the surface of Phobos:
% \\ New constraints for the formation of grooves.
% {\it Planetary \& Space Science} {\bf 59}, 1293--1307.

H'{e}non, M., and M. Guyot, 1970.  Stability of periodic orbits in the restricted problem.  
Pp. 349--374 in {\it Periodic Orbits, Stability and Resonances}, ed. G. E. O. Giacaglia.  
Dodrecht-Holland: D. Reidel Publishing Company.  

Hippke, M., and D. Angerhousen, 2015.  
A statistical search for a population of exo-Trojans 
\\ in the Kepler data set.  {\it Ap. J.} {\bf 811}:1.  

% Horedt, G. P., and G. Neukum, 1984.  Cratering rate over the surface of a synchronous satellite.  
% \\ {\it Icarus} {\bf 60}, 710--717.  

% Holsapple, K. A., and P. Michel, 2008. Tidal disruptions II.
% A continuum theory for solid bodies with strength,
% with applications to the Solar System. {\it Icarus} {\bf 193}, 283--301.

Hut, P., 1981.  Tidal evolution in close binary systems.  
{\it Astronomy \& Astrophysics} {\bf 99}, 126--140.  

Ingersoll, A. P., and A. R. Dobrovolskis, 1978.  
Venus' rotation and atmospheric tides.  
{\it Nature} {\bf 275}, 37--38.  

% Ingersoll, A. P., T. Svitek, and B. C. Murray, 1992.  
% Stability of polar frosts in spherical bowl-shaped craters \\ 
% on the Moon, Mercury, and Mars.  {\it Icarus} {\bf 100}, 40--47.  

% Ivory, J., 1809. On the attractions of homogeneous ellipsoids.
% {\it Phil. Trans. Roy. Soc.} {\bf 99}, 345--372.

Janson, M., 2013.  A systematic search for Trojan planets in the Kepler data.  {\it Ap. J.} {\bf 774}:156.  

Jeffreys, H., 1929.  {\it The Earth}, 2d ed.  Cambridge U. Press.  

% Keane, J. T., {\it et al.}, 2019.  Gravity, rotation, and hill slopes of 2014 MU69.  
% {\it LPSC Abstracts} {\bf 50}, 3145.pdf.  

% Kellogg, O. D., 1929. {\it Foundations of Potential Theory}.
% Springer, reprinted by Dover, 1953.

Kipping, D., 2020.  An independent analysis of the six recently claimed exomoon candidates.  
{\it Ap. J. Letters} {\bf 900}: L44.

% Kondratev, B. P., 1989.  Dinamika Ellipsoidal'nykh Gravitiruiushchikh Figure \\ 
% (Dynamics of Ellipsoidal Gravitating Figures).  272 pp. (in Russian).  Nauka, Moscow (in NASA ADS)

% Kondratev, B. P., and V. A. Antonov, 1993.  New methods in the theory of the Newtonian potential.  
% \\ Potential energy of homogenous lens-shaped bodies and segments of spheres.  \\ 
% {\it Astron. Zh.} {\bf 70}, 594--609 (in Russian).  Translated in {\it Astron. Rep.} {\bf 37}, 300--307.  

% Kondratev, B. P., 2003.  The Theory of Potential and Equilibrium Figures.  
% \\ 624 pp. (in Russian).  Institute for Computer Research, Moscow-Izhevsk.  

% Kondratev, B. P., 2007.  Potential Theory:  New Methods and Problems with Solutions.  515 pp. 
% (in Russian). \\ MIR, Moscow.  http://vixri.com/d/Kondratev\%20B.\%20\_TEORIJa\%20POTENCIALA.pdf 

Lainey, V., and 10 co-authors, 2020.  Resonance locking in giant planets 
indicated by the rapid orbital expansion of Titan.  {\it Nature Astronomy} {\bf 4}, 1053--1058.  

% Lakdawalla, E., 2010. Hartley 2 compared to other comets, and in motion 3D. \\
% http://www.planetary.org/blogs/emily-lakdawalla/2010/2763.html.

% Lakdawalla, E., 2014. New Rosetta view of the comet -- and a comparison to other comets. \\ 
% http://www.planetary.org/blogs/emily-lakdawalla/2014/07311300-new-rosetta-view-of-the-comet.html.

Laughlin, G., and J. E. Chambers, 2002.  Extrasolar Trojans:  \\ The viability  
and detectability of planets in the 1:1 resonance.  {\it Astron. J.} {\bf 124}, 592--600.  

% Lebovitz, N. R., 1998. The mathematical development of the classical ellipsoids. \\
% {\it Int. J. of Engineering Science} {\bf 36}, 1407--1420.

Leleu, A., G. A. L. Coleman, and S. Ataiee, 2019.  Stability of the co-orbital resonance 
under dissipation.  Application to its evolution is protoplanetary discs.  
{\it Astronomy \& Astrophysics} {\bf 631}, A6.  

Lillo-Box, J., and 7 co-authors, 2018a.  The TROY project: Searching for co-orbital bodies to known planets.  
\\ I.  Project goals and first results from archival radial velocity.  
{\it Astronomy \& Astrophysics} {\bf 609}, A96.  

Lillo-Box, J., and 12 co-authors, 2018b.  
The TROY project: Multi-technique constraints on exotrojans \\ 
in nine planetary systems.  {\it Astronomy \& Astrophysics} {\bf 618}, A42.  

Lissauer, J. J., P. Goldreich, and S. Tremaine, 1985.  
Evolution of the Janus-Epimetheus coorbital resonance \\ 
due to torques from Saturn's rings.  {\it Icarus} {\bf 64}, 425--434.  

Lissauer, J. J, and 24 co-authors, 2011.  
Architecture and dynamics of {\it Kepler}'s candidate multiple transiting planet systems.  
{\it Ap. J.} {\bf 197}:8.  

% Ma, S. D., and X. Chen, 1994. Reconstruction of quadric surface from occluding contour. \\
% Pp. 27--31 in {\it Proceedings of the 12th International Conference on Pattern Recognition.}

% Ma, S. D., and L. Li, 1996. Ellipsoid reconstruction from three perspective views. \\
% Pp. 344--348 in {\it Proceedings of the 13th International Conference on Pattern Recognition.}

% MacMillan, W. D., 1930. {\it The Theory of the Potential}.
% McGraw-Hill, reprinted by Dover, 1958.

Madhusudhan, N., and J. N. Winn, 2009.  Empirical constraints on Trojan companions \\ 
and orbital eccentricities in 25 transiting exoplanetary systems.  {\it Ap. J.} {\bf 693}, 784--793.  

% Marchis, F.,  M. Kaasalainen, E.F.Y. Hom, J. Berthier,
% J. Enriquez, D. Hestroffer, D. Le Mignant, and I. de Pater, 2006.
% Shape, size and multiplicity of main-belt asteroids: \\ 
% I. Keck Adaptive Optics survey. {\it Icarus} {\bf 185}, 39--63.

% Meech, K. J., and 17 co-authors, 2017.  A brief visit from a red 
% and extremely elongated interstellar asteroid.  {\it Nature} {\bf 552}, 378--381.  

Mignard, F., 1979.  The evolution of the lunar orbit revisited.  
I.  {\it The Moon and the Planets} {\bf 20}, 301--315.  

Morais, M. H. N., and F. Namouni, 2013.  Retrograde resonance in the planar three-body problem.  
\\ {\it Cel. Mech. Dyn. Astron.} {\bf 125}, 91--106.  

Morais, M. H. N., and F. Namouni, 2016.  A numerical investigation of coorbital stability and libration 
\\ in three dimensions.  {\it Cel. Mech. Dyn. Astron.} {\bf 125}, 91--106.  

Murray, C. D., 1994. Dynamical Effects of Drag in the Circular Restricted Three-Body Problem: \\
I. Location and Stability of the Lagrangian Equilibrium Points. {\it Icarus} {\bf 112}, 465--484.

Murray, C. D., and S. F. Dermott, 1999. {\it Solar System Dynamics}. Cambridge U. Press.

Narita, N., {\it et al.}, 2007.  Measurement of the Rossiter-McLaughlin effect \\ 
in the transiting exoplanetary system TrES-1.  {\it Pub. Astron. Soc. Japan} {\bf 59}, 763--770.  

Nauenberg, M., 2002.  Stability and eccentricity for two planets in a 1:1 resonance,
\\ and their possible occurrence in extrasolar planetary systems.
{\it Astron. J.} {\bf 124}, 2332--2338.  

% Neutsch, W., 1979.  On the gravitational energy of ellipsoidal bodies and some related functions.  
% \\ {\it Astronomy \&  Astrophysics} {\bf 72}, 339-347.  

% Newton, I., 1687. {\it Philosophi{\ae} Naturalis Principia Mathematica}. Streater, London.

Noyelles, B., J. Frouard, V. V. Makarov, and M. Efroimsky, 2014.  
Spin-orbit evolution of Mercury revisited.  {\it Icarus} {\bf 241}, 26--44.  

% Okabe, M., 1979. Analytical expressions for gravity anomalies due to homogeneous polyhedral bodies 
% \\ and translations into magnetic anomalies.  {\it Geophysics} {\bf 44}, 730--741.  

% Ostro, S.J., R. S. Hudson, M. C. Nolan, J.-L. Margot, D. J. Scheeres, D. K. Campbell,
% C. Magri, J. D. Giorgini, and D. K. Yeomans, 2000.
% Radar observations of asteroid 216 Kleopatra. \\ 
% {\it Science} {\bf 288}, Issue 5467, 836--839 + cover.

% Paul, M. K., 1974. The gravity effect of a homogeneous polyhedron for 
% three-dimensional interpretation. \\ {\it Pure Appl. Geophys.} {\bf 112}, 553--561.  

Peale, S. J. (1993).  The effect of the nebula on the Trojan precursors.  
{\it Icarus} {\bf 106}, 308--322.  

% Poh\'{a}nka, B., 1988. Optimum expression for computation of the gravity field of 
% a homogeneous polyhedral body. \\ {\it Geophys. Prospecting} {\bf 36}, 733--751.  

% Porco, C. C., P. C. Thomas, J. W. Weiss, and D. C. Richardson, 2007. \\
% Saturn's small inner satellites: Clues to their origins.
% {\it Science} {\bf 318}, 1602--1607.

% Pravec, P., and A. W. Harris, 2000.  Fast and slow rotation of asteroids. 
% {\it Icarus} {\bf 148}, 12--20.  

Press, W. H., S. A. Teukolsky, W. T. Vetterling, and B. P. Flannery, 1992. \\
{\it Numerical Recipes in Fortran, Second Edition}. Cambridge U. Press.

% Ramsey, A. S., 1940. {\it An Introduction to the Theory of Newtonian Attraction}.  Cambridge U.

% Richardson, D. C., P. Elankumaran, and R. E. Sanderson, 2005. \\
% Numerical experiments with rubble piles: equilibrium shapes and spins.  
% {\it Icarus} {\bf 173}, 349--361.  

Rodr\'{i}guez, A., C. A. Giuppone, and T. A. Michtchenko, 2013.  
Tidal evolution of close-in exoplanets in co-orbital configurations.  
{\it Cel. Mech. Dyn. Astron.} {\bf 117}, 59--74.  

Rowe, J. F., and 28 co-authors, 2014.  Validation of Kepler's multiple planet candidates.  
III.  Light curve analysis and announcement of hundreds of new multi-planet systems.  
{\it Ap. J.} {\bf 784}, 45.  

Schuerman, D. (1980).  The restricted three-body problem including radiation pressure.  
{\it Ap. J.} {\bf 238}, 337--322.  

% Shepard, M. K., and 10 co-authors, 2018.  
% A revised shape model of asteroid (216) Kleopatra.  
% \\ {\it Icarus} {\bf 311}, 197--209.  

Sheppard, S. S., and C. A. Trujillo, 2006.  
A thick cloud of Neptune Trojans and their colors.  
{\it Science} {\bf 313}, 511--514.  

% Shi, X., J. Oberst, and K. Wilner, 2016.  
% Mass wasting on Phobos triggered by an evolving tidal environment.  
% {\it Geophysical Research Letters} {\bf 43}, 12371--12379.  

Simmons, J. F. L., A. J. C. McDonald, and J. C. Brown, 1985.  
\\ The restricted 3-body problem with radiation pressure.  
{\it Cel. Mech.} {\bf 35}, 145--187.  

% Slyuta, E. N., 2012a. Shape distribution of ordinary chondrites, iron meteorites, and 
% metallic asteroids. \\ {\it Lunar and Planetary Science Conference} {\bf 43}, abstract 1088.

% Slyuta, E. N., 2012b.
% Shape of ordinary chondrites, iron meteorites, metallic, S- and C-asteroids. \\
% {\it Asteroids, Comets, Meteors 2012}, abstract 6091.

Smith, A. M., and J. J. Lissauer, 2010.  Orbital stability of systems of closely-spaced planets, 
II:  \\ configurations with coorbital planets.  {\it Cel. Mech. Dyn. Astron.} {\bf 107}, 487--500.  

% Stern, S. A., {\it et al.}, 2019.  Initial results from the New Horizons exploration of 2014 MU69, 
% \\ a small Kuiper belt object.  {\it Science} {\bf 364}, eaaw9771.

% Thomas, P. C., 2010. Sizes, shapes, and derived properties of the saturnian satellites \\
% after the Cassini nominal mission. {\it Icarus} {\bf 208}, 395--401. 

% Thomas, P. C., J. A. Burns, M. S. Tiscareno, M. M. Hedman, and P. Helfenstein, 2013a.
% \\ Saturn's mysterious arc-embedded moons: Recycled fluff? \\
% {\it Lunar and Planetary Science Conference} {\bf 44}, abstract 1598.

% Thomas, P. C., J. A. Burns, M. Hedman, P. Helfenstein, S. Morrison, M. Tiscareno,
% and J. Veverka, 2013b. \\ The inner small satellites of Saturn: 
% A variety of worlds.  {\it Icarus} {\bf 226}, 999--1019.

% Waldvogel, J., 1979. The Newtonian potential of homogeneous polyhedra. \\
% {\it Z. Angew. Mathe. Phys.} {\bf 30}, 388--398.

% Washabaugh, P. D., and D. J. Scheeres, 2002.
% Energy and stress distributions in ellipsoids.
% {\it Icarus} {\bf 159}, 314--321.

% Weaver, H. A., {\it et al.}, 2016.  The small satellites of Pluto as observed by New Horizons.  
% \\ {\it Science} {\bf 351}, aae0030.

% Werner, R. A., 1994. The gravitational potential of a homogeneous polyhedron or don't cut corners.  
% \\ {\it Cel. Mech. Dyn. Astron.} {\bf 59}, 253--278.  

% Werner, R. A., and D. J. Scheeres, 1995. Exterior gravitation of a polyhedron derived
% and compared with harmonic and mascon gravitation representations of asteroid 4769 Castalia.
% {\it Cel. Mech. Dyn. Astron.} {\bf 65}, 313--344.

Wiegert, P., M. Connors, and C. Veillet, 2017.  A retrograde co-orbital asteroid of Jupiter.  
\\ {\it Nature} {\bf 543}, 687--689.  

% Wilson, L., and J. W. Head, 1989. Dynamics of groove formation on Phobos by ejecta from Stickney 
% (abstract). \\ {\it Lunar and Planetary Science Conference} {\bf XX}, 1211--1212.

Yoder, C. F., 1979. Notes on the origin of the Trojan asteroids. {\it Icarus} {\bf 40}, 341--344.

Yoder, C. F., G. Colombo, S. P. Synnott, and K. A. Yoder, 1983. \\
Theory of motion of Saturn's coorbiting satellites. {\it Icarus} {\bf 53}, 431--443.

\end{document}